\documentclass[12pt,tightenlines,eqsecnum,floats,aps,amsmath,amssymb,nofootinbib,prd,superscriptaddress]{revtex4}

\usepackage{setspace}
\usepackage{amsmath,amssymb,amsfonts,amsthm,mathrsfs}
\usepackage{graphicx}
\usepackage{enumerate} 

\usepackage[pdftex]{hyperref}

\def\be{\begin{equation}}
\def\ee{\end{equation}}
\def\ba{\begin{eqnarray}}
\def\ea{\end{eqnarray}}
\def\bi{\begin{itemize}}
\def\ei{\end{itemize}}
\def\nn{\nonumber}

\def\integers{\mathbb{Z}}
\def\reals{\mathbb{R}}
\def\complex{\mathbb{C}}

\def\aut{\text{Aut}}
\def\autq{\text{Aut}_{ \text{sa}}}

\def\diff{\text{Diff}}

\def\autEo{\text{Aut}^{\Eo}}
\def\autqEo{\text{Aut}^{\Eo}_{ \text{sa}}}

\def\tr{\text{Tr}}

\def\bra{\langle}
\def\ket{\rangle}

\def\id{\text{Id}}

\def\idtwo{\mathbf{1}}
\def\w{\omega}
\def\rk{\text{rank}\,}
\def\t{\tau}

\def\lqg{\text{\tiny LQG}}

\def\Hks{\mathcal{H}_{\text{KS}}}
\def\P{\mathcal{P}}
\def\E{\mathcal{E}}
\def\A{\mathcal{A}}
\def\Abar{\bar{\mathcal{A}}}
\def\hba{\mathcal{HBA}}
\def\hbabar{\overline{\mathcal{HBA}}}

\def\Ah{\bar{\mathcal{A}}_\text{H}}
\def\Ab{\bar{\mathcal{A}}_\text{B}}

\def\Lh{\mathcal{L}_\text{H}}
\def\Lb{\mathcal{L}_\text{B}}
\def\L{\mathcal{L}}
\def\spec{\Delta}
\def\muks{\mu_{\rm{KS}}}

\def\cyl{\text{Cyl}}
\def\hom{\text{Hom}}

\def\p{\tilde{p}}
\def\Eb{\bar{E}}
\def\pol{\text{Pol}}

\def\et{\tilde{e}}

\def\Eo{\mathring{E}}
\def\xh{\hat{x}}
\def\hb{\bar{h}}

\def\eo{\mathring{e}}
\def\qo{\mathring{q}}
\def\bE{\beta_{\Eb}}

\def\LR{\Lambda_R}

\def\e{\epsilon}
\def\D{\mathcal{D_{\rm KS}}}
\def\odd{{\rm odd}}
\def\even{{\rm even}}
\def\Haut{\mathcal{H}_{\text{Aut}}}
\def\Lo{\mathring{\Lambda}}
\def\Ph{\text{Ph}}
\def\sym{\text{Sym}}
\def\O{\mathcal{O}}
\def\Vo{\mathring{V}}
\def\Vb{\overline{V}}
\def\c{\text{\bf c}\,}

\def\gt{\tilde{\gamma}}
\def\go{\mathring{g}}
\def\ks{\text{KS}}

\def\xio{\mathring{\xi}}
\def\phio{\mathring{\phi}}

\def\nh{\hat{n}}

\def\tt{{\bf t}}
\def\gg{{\bf g}}
\def\phifs{\phi_{\text{FS}}}
\def\hphifs{\hat{\phi}_{\text{FS}}}
\def\yo{\mathring{y}}

\def\Db{\bar{D}}
\def\Db{\bar{D}}

\def\chio{\chi_0}
\def\Uo{U_0}

\begin{document}

\title{A quantum kinematics for asymptotically flat spacetimes}
\author{Miguel Campiglia} \email{campi@fisica.edu.uy}
 \affiliation{Raman Research Institute, Bangalore 560080, India}
 \affiliation{Instituto de F\'isica, Facultad de Ciencias,  Montevideo 11400, Uruguay}
\author{Madhavan Varadarajan}\email{madhavan@rri.res.in}
 \affiliation{Raman Research Institute, Bangalore 560080, India}

\begin{abstract}

We construct a quantum kinematics for asymptotically flat 
spacetimes
based on the Koslowski-Sahlmann (KS) representation.
The KS representation is a generalization of the representation underlying Loop Quantum Gravity (LQG) which
supports, in addition to the usual LQG operators,  the action 
of `background exponential operators' which are 
 connection dependent operators labelled by `background' $su(2)$ electric  fields. 
KS states have, in addition to
the LQG state label corresponding to 1 dimensional excitations of the triad, 
a label corresponding to a `background' electric field which describes 3 dimensional excitations of the triad.
Asymptotic behaviour in 
quantum theory is controlled through asymptotic conditions 
on the background electric fields which label the {\em states} and the background electric fields which label
the {\em operators}.
Asymptotic conditions on the 
triad  are imposed as conditions on the background electric field  state label while confining the LQG spin net 
graph labels to compact sets. We show that KS states can be realised as wave functions on a quantum configuration space 
of generalized connections and that the asymptotic behaviour of each such generalized connection is determined
by that of the background electric fields which label the background exponential  operators.
Similar to the spatially compact case,
the Gauss Law and diffeomorphism constraints are then imposed 
through  Group Averaging techniques to obtain a large sector of 
gauge invariant states. It is shown that  this sector  supports
a unitary action of the  group of asymptotic rotations and translations and that,
as anticipated by Friedman and Sorkin, for appropriate spatial  topology,
this sector contains states which display fermionic behaviour under $2\pi$ rotations.

\end{abstract}
\maketitle

\section{Introduction} \label{sec1}

Isolated gravitating systems are modelled by asymptotically flat spacetimes. Key physical results such as the 
positivity of energy and the definition of total angular momentum depend on the delicate asymptotic
behaviour of the gravitational field. Therefore, an
understanding of
quantum gravitational effects in the context of asymptotic flatness requires that these asymptotic conditions be suitably
incorporated in quantum theory. 
The particular quantum gravity approach we are interested
in is  canonical Loop Quantum Gravity (LQG).
LQG is an attempt at canonical quantization of a classical formulation of gravity in terms of $SU(2)$ connections and 
conjugate electric (triad) fields on a Cauchy slice $\Sigma$. 
The basic connection dependent operators are holonomies associated with graphs
and the basic electric field dependent operators are fluxes through surfaces, the graphs and surfaces being 
embedded in the Cauchy slice. The LQG representation can be viewed as a connection 
representation \cite{alm2t}.
A spin network 
basis of quantum states can be built out of holonomies of (generalized) connections 
\cite{leecarlo,baezspinnet}. Consequently each such state `lives' on a graph. Holonomy operators act by multiplication
and electric fluxes through surfaces act, roughly speaking,  by differentiation whenever the surface  
and the spin net graph intersect \cite{leecarlo,aajurekarea} so that the triad field is excited
only along the support of the graph and  vanishes outside this support. 
As a result,
the microsocopic quantum 
spatial geometry can be thought of  as vanishing everywhere except along the spin net graph.

Most work in LQG has been in the context of compact Cauchy slices wherein the spin net graphs
are  chosen to have a finite number of compactly supported edges. In the asymptotically flat case, 
while the connection
vanishes at spatial infinity, the triad approaches a fixed flat triad \cite{abhayaflat}.
Since the spatial geometry is excited only along the support of the graph underlying a spin net state,
and since, in the asymptotically flat case,
 the boundary conditions indicate that the spatial geometry is excited in non-compact regions, it follows
that the graphs underlying spin net states appropriate to asymptotic flatness must be supported in non-compact regions.
Thus these  graphs must have infinitely many (or infinitely long) edges.
Since it seems natural that, as in the spatially compact case,  
such spin net states should be mapped to each other by holonomy operators,
it seems necessary that the compact graph holonomy operators  also admit a generalization to 
ones supported on non-compact graphs.

In this work, instead
of attempting a generalization  based on non-compact versions of the standard  LQG states (namely, spin nets
supported on graphs with non-compact support) and the standard LQG operators (namely holonomies on non-compact graphs), 
we adapt a generalization of LQG itself, on compact spaces, to the asymptotically flat case.
This generalization was introduced by Koslowski  and further developed by Sahlmann 
in their seminal contributions \cite{kos,hanno,koshanno}
and we refer to it as the Koslowski-Sahlmann (KS) representation. The KS representation involves, in addition
 to the standard probes of LQG, an extra set of probes consisting of electric fields. Each such electric field ${\bar E}$
(called a `background electric field' to distinguish it from the dynamical electric field which is a phase space
variable) probes the behaviour of the connection $A_a^i$ through the `background exponential function'
$\beta_{\Eb} (A)= \exp (i\int_{\Sigma} \Eb^a_i A_a^i)$. The KS representation supports the action of 
the corresponding `background exponential operators' \cite{me} and the representation itself can be seen 
as a representation of a `holonomy-background exponential-flux' algebra \cite{mm2}.

In our generalization of the 
KS representation to the asymptotically flat case,
we demand that these extra probes (namely, the background electric fields) 
satisfy appropriate asymptotic conditions.
These  conditions on the background electric fields ensure
%
that the background
exponential functions  are sensitive to the detailed asymptotic behavior of the classical
 connections
and
we are able 
to continue to use compactly  supported graphs to define holonomies.
It then turns out that in the quantum theory  these asymptotic conditions on the background electric fields
control the behaviour of the generalised connections which constitute the quantum configuration space.
This is similar to what happens in standard flat spacetime quantum field theory where the behaviour of the 
elements of the quantum configuration space is determined by that of the `dual probes'; for example, in the standard
Fock representation of the scalar field, the precise distributional nature of the quantum scalar fields is 
determined by the rapid fall off property of the dual smearing functions \cite{glimmjaffe}. As we shall show,
our particular choice of asymptotic behaviour for the background electric fields which label the background exponential
operators implies that elements of the quantum configuration space satisfy a quantum analog  of the classical
asymptotic behaviour. It is in this way that the classical asymptotic conditions on the connection
leave their imprint in quantum theory.

Background electric fields also appear as labels of KS states in addition to the standard LQG spin net
labels. The action of the electric flux operators
 on such states acquires, relative to standard LQG, an
extra contribution corresponding to 
the flux of the background electric field which labels
the state \cite{kos,hanno}. Similar contributions ensue for the Area and Volume operators \cite{hanno}.
It then follows that the classical asymptotic conditions on the electric
field can be incorporated in quantum theory by directly imposing them as conditions on the background electric
field label of the state. 
These conditions are that the electric field asymptotes to the sum of 
a fixed flat triad field and a subleading piece \cite{abhayaflat,ttparity,mcparity}.\footnote{We use  `parity' conditions which are closely related to those of  Reference \cite{mcparity} (see
section \ref{sec2A} for what we mean by the phrase `closely related' and section \ref{sec9} for related issues).
Parity conditions on metric variables
were introduced in \cite{rt} and adapted to the self dual Ashtekar variables in \cite{ttparity}. The 
treatment of  \cite{rt} was improved in \cite{beigom} and the improved treatment adapted to the  Ashtekar-Barbero
variables in \cite{mcparity}. }
The background electric fields which label the background exponential operators
obey the same conditions 
as the background electric fields which label the states except that the `zeroth order' flat triad piece of the latter
is absent in the former.
As we shall see,\footnote{See Footnote \ref{fnconsistent}, section \ref{sec3A}.\label{fn1}}
it is this difference in the  
two sets of conditions (namely on state and operator labels) which allows their consistent imposition.

The resulting quantum kinematics supports a unitary representation of $SU(2)$ rotations and
diffeomorphisms 
subject to appropriate asymptotic behavior. 
We shall be  interested in 
$SU(2)$ rotations and diffeomorphisms which are 
connected to identity.
The diffeomorphisms asymptote to a combination of 
asymptotic rotations, translations and odd supertranslations \cite{mcparity}. 
The $SU(2)$ rotations 
are trivial at infinity  except when they act in combination
with  diffeomorphisms with non-trivial asymptotic rotational action so that 
the triad at infinity is kept fixed under this combined action.
The unitary transformations corresponding to any  combination of $SU(2)$ rotations which are trivial at infinity and
diffeomorphisms with trivial asymptotic rotational and translational parts 
are interpreted as the quantum analog of the finite gauge transformations generated by the 
$SU(2)$ Gauss Law and spatial diffeomorphism constraints \cite{mcparity}.
The unitary transformations corresponding to any  combination of 
diffeomorphisms with non-trivial asymptotic rotational and translational parts together with  appropriate 
$SU(2)$ rotations (so that the combination is an asymptotic symmetry of the fixed flat triad at infinity)
are interpreted as the quantum analog of the finite symmetry  transformations generated by the 
the total angular and linear momenta of the classical spacetime \cite{mcparity}.
We view the ability of the quantum 
kinematics to support these unitary actions as an {\em a posteriori} justification for our treatment of 
asymptotic behaviour in quantum theory.
%
As in the spatially compact case \cite{hanno,mm1}, 
gauge invariant states  can then be constructed by group averaging methods.
We construct large `superselected' sectors \cite{mm1} of such group averaged states and show that 
these sectors support a unitary representation of the finite dimensional group of translations and rotations at infinity.
As in LQG, and as in the
KS representation for the compact case, the Hamiltonian constraint remains a key open issue. This, in turn, precludes
a quantum analysis of the total energy and of the generator of boosts.

While this work seeks to  address the construction of a quantum kinematics appropriate to asymptotic
flatness, it is based on the KS representation rather than the standard LQG one. 
Since we view the KS representation as an effective
description of fundamental LQG states which describe an effectively smooth spatial geometry, we would like to 
base our constructions on 
purely LQG type probes, which, as mentioned above, must now include non-compact graphs.
There have 
been a few extremely interesting and important attempts at 
initiating a study of asymptotically flat LQG \cite{arnsdorf,itp} to accommodate non-compact graphs.
Since these works do not address the full set of asymptotic  conditions adequately, we hope that 
the results of this work will provide a useful supplement to that of References \cite{arnsdorf,itp}
in the putative construction of an LQG quantum kinematics for asymptotically flat spacetimes.

Our work as presented here crucially relies on our recent work \cite{me,mm1,mm2}. 
Specifically, the considerations of sections \ref{sec4}- \ref{sec6} of this paper rest on the analogous results and
 constructions
for the compact case as detailed in Reference \cite{mm2} and those of section 
\ref{sec8} on the 
group averaging analysis of \cite{mm1}.\footnote{The analysis of \cite{mm1} is motivated by Sahlmann's  pioneering work \cite{hanno}.}
The layout of the paper 
is as follows. 

Section \ref{sec2} is devoted to a description of classical structures of relevance to the construction of the KS 
representation in the asymptotically flat case. 
Standard definitions of asymptotic flatness
(see for example \cite{abhayaflat,rt,beigom}) implicitly use the $C^{\infty}$ category.
In anticipation of the formulation of the KS representation based on semianalytic, and hence, 
{\em finitely differentiable} fields,  we tailor the definition of  asymptotically flat 3-manifolds
to the semianalytic category and discuss the differential properties of fields supported on such manifolds.
This is done in section \ref{sec2A0}.
This section also establishes our notation for asymptotic fall-off properties of such fields.
Next, in  section \ref{sec2A} we discuss the 
asymptotic  conditions  on the phase space variables, namely the $SU(2)$ 
connection and its conjugate  electric field,\footnote{
It turns out that the considerations of sections \ref{sec3}- \ref{sec6}
do not depend on the fine details of the asymptotic conditions. The imposition of gauge and diffeomorphism
invariance and the definition of linear  and angular momenta in section \ref{sec7} and \ref{sec8} {\em do} depend on detailed
asymptotic conditions.} 
and display  the elementary phase space functions of interest, namely, the
holonomies, background exponentials and electric fluxes.
The `probes' associated with these functions are, respectively, edges $e$, background electric fields ${\bar E}$ and
$SU(2)$ Lie algebra valued scalars $f$ supported  on surfaces $S$. 
Section \ref{sec2B} discusses the precise  defining properties 
of  probes  
for the connection (namely $e, {\bar E}$) 
and reviews the structures associated with these probes, namely 
the groupoid of compactly supported paths and the Abelian group  (under addition) of these electric
fields subject to appropriate asymptotic conditions.
In section \ref{sec2C}
we detail our choice of electric field probes $(f,S)$. 
In section \ref{sec2D} we discuss the classical Poisson brackets between the 
holonomies, background exponentials
and electric fluxes.

In section \ref{sec3} we define the KS representation (which as we show in this and subsequent sections is) 
appropriate to the asymptotically flat case in terms of the action of the holonomy, background exponential and
electric flux operators on KS spin nets. Each KS spin net is labelled by a standard LQG spin net label 
whose underlying graph is compactly supported and 
a background electric field label which is subject to the same asymptotic conditions as are
satisfied by the asymptotically flat 
electric fields which serve as classical phase space variables. 

Sections \ref{sec4} to \ref{sec6} are devoted to the study of the quantum configuration space 
and to a characterization  of the asymptotic behavior of the generalised connections which constitute this space.
%
We use $C^*$ algebraic 
methods to 
show that the KS representation is unitarily equivalent to one based on square integrable
functions on a quantum configuration space 
 which is a topological completion of the 
classical space of  connections.
Every classical connection $A$ is `smooth' (more precisely, 
finitely differentiable) and satisfies conditions appropriate to asymptotic flatness and thereby 
defines a homomorphism from the space of background electric field labels to $U(1)$
through the background exponential function $\beta_{\Eb} (A)= \exp (i\int_{\Sigma} \Eb^a_i A_a^i)$. Thus any 
classical connection can be thought of as a 
 homomorphism from  the space of all  ${\bar E}$'s to $U(1)$  which is  (a) well defined  and  (b) functionally differentiable with respect to ${\bar E}$.  Property (a) is a 
 result of the fall-off behaviour of $A$ and the  `dual' fall-off behaviour of $\bar E$
\footnote{By this we mean that ${\bar E}$ has a fall off behaviour which 
ensures the well definedness of $\int_\Sigma \Eb^a_i A_a^i$.
}
 and property (b)  is a consequence of the differential properties of $A, {\bar E}$.
We show that any element of the quantum configuration space defines 
a homomorphism which satisfies property (a) but not necessarily property (b). We argue that 
the satisfaction of property (a)  implies that the quantum connections
satisfy a weakened form of  the  classical fall-off conditions on the connection.

The detailed content of sections  \ref{sec4} to \ref{sec6} is as follows. In section \ref{sec4} we prove a key Master Lemma which establishes that 
given a finite set of independent probes and a corresponding set of 
elements in $SU(2)$ (one for each independent edge) and $U(1)$ (one for each rationally independent electric field),
there exists a classical connection satisfying the asymptotic conditions such that the evaluation of the relevant set of holonomies and 
background exponentials on this connection reproduces the given set of group elements to arbitrary accuracy.
The Master Lemma is used extensively in sections \ref{sec5} and \ref{sec6}.
In section \ref{sec5} we complete the Abelian 
Poisson bracket algebra of holonomies and background exponentials, 
$\hba$,
 to the $C^*$ algebra, $\hbabar$ and  identify its spectrum $\spec$ with the quantum configuration space of generalised
connections.
We show that each element of the spectrum corresponds to a pair of homomorphisms,
one homomorphism from the path groupoid to 
$SU(2)$ and
the other  from the Abelian group of electric fields to $U(1)$. The first homomorphism
corresponds to the algebraic structure provided by the holonomies and the second to that 
provided by the background exponentials. We argue that the existence of this 
second homomorphism  and  the satisfaction of asymptotic conditions by every element of the 
Abelian group of electric fields  imply 
%
the imposition of a  
weakened version of  the classical asymptotic boundary conditions 
on elements of the spectrum.
Next, we justify the identification of the spectrum as the quantum configuration space by showing
 that the KS Hilbert space  
is isomorphic to  the space $L^2( \spec, d\muks) $ of square integrable functions on the spectrum with respect 
to a suitably defined measure $d\muks$. We do this by showing that the 
 expectation value function with respect to any KS spin net state of section \ref{sec2} defines a positive linear
function on $\hbabar$. The GNS construction together with  standard Gel'fand theory then provides
an isomorphism between the action of the holonomy and background exponential operators in the KS representation
and their action by multiplication on $L^2( \spec, d\muks) $.
Finally, we show that the electric flux operator is represented as an
operator
on $L^2( \spec, d\muks) $ and that this representation completes the isomorphism between the KS representation 
of section \ref{sec2} and  the $L^2( \spec, d\muks) $ representation. 
In section \ref{sec6} we show that the spectrum $\spec$
is homeomorphic to an appropriate 
projective limit space $\Abar$ 
whose fundamental building blocks are products of 
finite copies of $SU(2)$ and $U(1)$ and that the measure $d\muks$ can be realised as a projective
limit measure of the Haar measure on these building blocks. 

Section \ref{sec7} is devoted to a discussion of gauge transformations and symmetries of the classical theory.
We  describe the group of gauge transformations, $\aut$,  of $SU(2)$ rotations and diffeomorphisms
connected to identity and subject to appropriate asymptotic behaviour. 
Next, we describe the group $\autEo$ of asymptotic symmetries which are connected to identity and show
that $\aut$ is a normal subgroup of $\autEo$. Finally we show that 
the quotient of $\autEo$ by $\aut$ is  the semidirect product of the group of asymptotic  translations with that of
asymptotic rotations so that $\autEo/\aut=\reals^3 \rtimes SU(2)$.
In section \ref{sec8} we show that the groups $\aut$ and $\autEo$ (suitably restricted to the semianalytic category)
are unitarily implemented
on the KS Hilbert space. We construct a large sector of gauge invariant states by 
averaging over the action of the gauge group $\aut$ (suitably restricted to the semianalytic category) and show 
that this sector supports a unitary representation of $\autEo/\aut$. We show that, while for trivial topology,
$\autEo/\aut$ acts effectively as $\reals^3 \rtimes SO(3)$, for the non-trivial topologies considered by Friedman
and Sorkin \cite{fs}, gauge invariant states exist which display fermionic behaviour under $2\pi$ rotations
so that $\autEo/\aut$ is represented non-trivially  as $\reals^3 \rtimes SU(2)$.\footnote{See \cite{arnsdorfspin} for a discussion of Friedman and Sorkin ideas in LQG.} Thus our work yields a fully
rigorous implementation of the beautiful ``Spin $\frac{1}{2}$ from Gravity'' behaviour  predicted by 
Friedman and Sorkin.  We close section \ref{sec8} with a brief  discussion of 
the construction of eigenstates of angular and spatial momenta.
Section \ref{sec9} contains our concluding remarks. 
A number of technicalities are relegated to the Appendices.

Before we proceed to the next section, we  reiterate that the results of 
sections \ref{sec4}- \ref{sec6} represent simple generalizations of results obtained in the spatially 
compact case in Reference \cite{mm2}. Hence, in our presentation, we shall endeavour to employ the notation 
of \cite{mm2} wherever possible and we shall be explicit only
in  aspects of argumentation which differ from those contained in Reference \cite{mm2}. Considerable
parts of our argumentation will be {\em identical} to those in Reference \cite{mm2} and for such parts
we shall simply refer the reader to that work. Similar remarks apply to our exposition in section  \ref{sec8}
of this paper in relation to the contents of Reference \cite{mm1}. 
We 
use units such that $c= 8\pi \gamma G =\hbar =1$ where $\gamma$ is the Barbero-Immirzi parameter. 

\section{Classical Preliminaries} \label{sec2}

\subsection{Differential Structure} \label{sec2A0}
Our classical departure point is a semianalytic version of the  description given in  \cite{mcparity,beigom} for asymptotically flat canonical gravity. 
The Cauchy slice $\Sigma$ is taken to be a $C^k, k >> 1$ semianalytic manifold 
without boundary admitting  certain structure required for the notion of 
asymptotic flatness. This structure is as follows. Let $\Uo \subset \reals^3$ 
be the complement of the unit ball 
in $\reals^3$, $\Uo := \{ (x^1,x^2,x^3) \in \reals^3   :  (x^1)^2+(x^2)^2+(x^3)^2 > 1 \}$.  
We  require the existence of a preferred (semianalytic) chart 
$\chio : \Uo \to \Sigma$ such that $K:=\Sigma \setminus \chio(\Uo)$ is compact. 
We refer to it as the Cartesian chart. In this chart we use 
the notation: $\vec{x}:=(x^1,x^2,x^3)=\{x^\alpha \}$, $r:= \sqrt{(x^1)^2+(x^2)^2+(x^3)^2}$ and $\xh:= \vec{x}/r$. 
Spatial infinity is approached as $r\rightarrow\infty$.

It is important to note that while $\Sigma$ is itself $C^k$ and semianalytic, on $\chio (\Uo)$ we may still define 
semianalytic functions on $\Sigma$ which are of
 {\em arbitrary} degree of differentiability with respect to {\em the preferred
Cartesian coordinates}. More in detail, since $\Uo$ is an open set in $\reals^3$, we can define semianalytic $C^p$ functions,
even for $p >k$, with respect to this preferred $\reals^3$ structure. Such functions  are still $C^k$ semianalytic functions
on $\Sigma$; they just happen to possess higher differentiability with respect to the preferred Cartesian coordinates.
We shall refer to this degree of differentiability in the preferred Cartesian chart as the {\em Cartesian
degree of differentiability}. Next, we detail two examples of such functions which we shall use in this paper.

First note that the discussion above, together with the definition of semianalyticity \cite{lost}, implies
that  a semianalytic $C^k$ function of Cartesian degree of differentiability $p=\infty$
is an {\em analytic} function of the Cartesian coordinates in $\chio (\Uo )$. 
An example of such a function is one which is $C^k$ and semianalytic 
on $\Sigma \setminus \chio (\Uo )$ and which coincides with the function 
$r^{-1}$ in $\chio (\Uo)$.
Since we are away from $r=0$, the function
$r^{-1}$ is   {\em analytic} in the Cartesian chart so that the function in question has
Cartesian degree of differentiability $p=\infty$. A similar conclusion holds for  functions which agree 
asymptotically with the function $r$. 

Another example of the functions which we shall use in this work is 
provided in  Appendix \ref{sphereapp} wherein we show that any $C^{p}$ semianalytic  function
of $\xh$ naturally defines a semianalytic function of Cartesian degree of differentiability $p$ on 
$\chio (\Uo)$. This implies that, for example,
 any semianalytic $C^k$ function on $\Sigma \setminus \chio (\Uo )$ which
agrees with such a function  on $\chio (\Uo )$ for $p\geq k$
is $C^k$ semianalytic
on $\Sigma$. 

To summarise: If we restrict attention to the behaviour of functions in $\chio (\Uo)\subset \Sigma$, then semianalytic 
functions of Cartesian differentiability $p$ are seminalytic $C^k$ functions  on  $\chio (\Uo)\subset \Sigma$ for 
$p\geq k$ else they are seminalytic $C^p$ functions on  $\chio (\Uo)\subset \Sigma$.

Next, we introduce the notion of asymptotic fall offs of functions of Cartesian differentiability $p$.
These fall offs refer to the behaviour of the function in $\chio (\Uo )$ for large enough $r$.
We shall say that a function of Cartesian differentiability $C^p$ is of order 
$O(r^{-\beta})$ if for sufficiently large $r$ and  $m=0,\ldots, p$  the 
$m$-th partial derivatives with respect to the preferred Cartesian coordinates 
are bounded by $c r^{-\beta-m}$ for some constant $c$ independent of $m$.

The notions of Cartesian degree of differentiability together with that of fall-off order can be applied to
tensor fields on $\Sigma$ as follows. The maximum degree of differentiability of tensor fields
on (the $C^k$ semianalytic manifold)  $\Sigma$ is $C^{k-1}$. Components of tensor fields in $\chio (\Uo)\subset \Sigma$
can be evaluated in the preferred Cartesian chart. Each such component can be viewed as a function on 
$\chio (\Uo) \subset \Sigma$. This function in turn can be a linear combination of functions each of different Cartesian
degree of differentiability $p$ and associated order $O(r^{-\beta})$. Indeed, this is the typical situation
we shall encounter in this paper.

Note that $\Sigma$ can of course also be given the less restrictive structure of a $C^k$ manifold 
by completing its maximal semianalytic atlas to a $C^k$ one.
In this setting  $\Sigma$  is still the disjoint union of 
$\Sigma \setminus \chio (U_0)$  and its asymptotic region  $\chio (U_0)$ and the asymptotic region 
is still equipped with the preferred Cartesian chart 
described in the first paragraph of this section.
It is then easy to see that 
the notions of Cartesian degree of differentiability as
well as fall off order continue to be well defined. 

In summary, $\Sigma$ can be thought of as follows. $\Sigma$ is a $C^k$ manifold. On $\Sigma$ there
is a preferred family of $C^k$ charts which endow $\Sigma$ with the structure of a $C^k$ semianalytic manifold, 
with this preferred family comprising  a maximal semianalytic atlas on $\Sigma$. In this family of semianalytic charts, 
one of the charts
is the Cartesian chart on $\chio (U_0)$. From this point of view $\Sigma$ has the ability to support, simultaneously,
both finite differentiability  structures as well as the more restricted semianalytic structures.
This is the point of view we shall employ in the rest of this work. In general, as in the case of compact space
LQG \cite{alrev}, classical structures will be of finite differentiability but not necessarily semianalytic whereas
structures relevant to quantum theory will be chosen to be semianalytic.

\subsection{Phase space variables and functions}\label{sec2A}

In this section we relax the condition of semianalyticity and work with $\Sigma$ as a $C^k$ asymptotically flat manifold

The phase space of interest is coordinatized by pairs $(A^i_a, E^a_i)$ of $su(2)$-valued 1-forms and densitized vector fields satisfying certain fall off conditions at infinity. Here $i=1,2,3$ are internal indices with respect to a fixed $su(2)$ basis $\t_i$ satisfying $[\t_i,\t_j]=\epsilon_{i j k} \t_k$  (we work on a fixed global trivialization of the bundle). $A^i_a,E^a_i$ are taken  to be respectively $C^{k-2},C^{k-1}$ (not necessarily semianalytic) tensors on $\Sigma$.
\footnote{We take  $E^a$ to be of maximal differentiability ($C^{k-1}$).  The reason we take the connection to be $C^{k-2}$ is that in classical theory $A_a$ arises as a combination of extrinsic curvature and spin connection, both of which involve first derivatives of the triad.}    Let $\Eo^a_i$ be a fixed `zeroth order' triad such that in the Cartesian chart $\Eo^\alpha_i=\delta^\alpha_i$. The conditions on 
$(A^i_\alpha, E^\alpha_i), i, \alpha=1,2,3$ (where $\alpha$ refers to  components in the preferred Cartesian
chart $\{x^\alpha\}$)
as $r \to \infty$ are taken to be:
\ba
E^\alpha_i & = &\Eo^\alpha_i+\frac{h^\alpha_i(\xh)}{r}+O(r^{-1-\e}) \label{fallE},\\
A_\alpha^i & = &\frac{g_{\alpha}^i(\xh)}{r^2}+O(r^{-2-\e})\label{fallA},
\ea
where $0<\e<1$,\footnote{The condition $\epsilon<1$ plays no restriction (for $\epsilon_0 >\epsilon$, $f(x)=O(r^{-\epsilon_0}) \implies f(x)=O(r^{-\epsilon})$). Its purpose is to allow for simplifications of the type: $O(r^{-1-\epsilon})+O(r^{-2})=O(r^{-1-\epsilon})$.} and $h^\alpha_i,g_{\alpha}^i$  are functions on the Cartesian sphere satisfying:
\ba
h^\alpha_i(-\xh) &= & h^\alpha_i(\xh), \label{heven} \\
g_{\alpha}^i(-\xh) &= & -g_{\alpha}^i(\xh).  \label{godd}
\ea
For technical reasons (see for instance  footnote \ref{pp1}) we require $h^\alpha_i$, $g_{\alpha}^i$ to be 
respectively $C^{k},C^{k-1}$ 
as functions on the sphere. 
As indicated in the previous section, these specifications are 
consistent with the degree of differentiability  of the tensor fields
$E^a_i$ and $A_a^i$ on $\Sigma$ and correspond to the `$1/r$' and `$1/r^2$'  parts of the triad and connection
fields being of Cartesian degree of differentiability $k$ and $k-1$ respectively. We denote by $\E_{\Eo}$ the space of $su(2)$-valued electric fields satisfying (\ref{fallE}) and   $\A$ the space of $su(2)$-valued one forms satisfying (\ref{fallA}). As in LQG,  $\A$ plays the role of classical configuration space.

Conditions (\ref{fallE}), (\ref{fallA}), (\ref{heven}), (\ref{godd}) are motivated by the parity conditions  of references \cite{mcparity,beigom}
and, indeed, at first glance look identical to them. 
However, whereas the
standard parity conditions
are defined in  the $C^{\infty}$ setting, 
 the parity conditions  (\ref{fallE}), (\ref{fallA})
are defined for finitely differentiable fields.
The conditions ensure the symplectic structure
\be
\Omega(\delta_1,\delta_2):= \int_{\Sigma} (\delta_1 A^i_a \delta_2 E^{a}_i-\delta_2 A^i_a \delta_1 E^{a}_i),  \label{sympfnv}
\ee
is well defined. The elementary phase space functions which admit direct operator correspondents in the KS representation are given by
\begin{eqnarray}
h_{e }(A) &:= & \P e^{\int_{e} A} , \label{hol}\\
\beta_{\Eb}(A) &:= &e^{i \int_\Sigma \tr[\Eb^a A_a]} \label{baexp} \\
F_{S,f}(E) &:=& \int_S dS_a \tr[f E^a]. \label{flux} 
\end{eqnarray}

Here 
$h_{e }(A)$ is the $SU(2)$ matrix valued holonomy of the connection along the compactly supported edge $e$,
$\beta_{\Eb}(A)$ is the background exponential labelled by the 
background electric field $\Eb$ subject to appropriate asymptotic conditions, 
and $F_{S,f}(E)$ is the electric flux 
smeared with the $su(2)$-valued function $f$ through the surface
$S\subset \Sigma$.
``$\tr$'' stands for -2 times the standard   matrix trace,   $\tr[M]:= -2 ( M_{1 1}+ M_{2 2})$, so that   $\tr[\t_i \t_j]=\delta_{ij}$. Anticipating the key role that the  `probes' $e,\Eb, S,f$  play in quantum theory, we shall choose
them to be semianalytic. 
The detailed definition of $e$ and the asymptotic conditions on $\Eb$ are displayed in section \ref{sec2B}.
The detailed choice of surfaces $S$ and associated smearing functions $f$
is displayed in section \ref{sec2C}.


\subsection{Probes of the connection}\label{sec2B}
\subsubsection{Holonomy related structures}
An oriented piecewise semianalytic curve $c$ is defined as a piecewise semianalytic map
$c: [0,1] \to \Sigma$
(see Definition 6.2.1 of \cite{ttbook}). 
The compact semianalytic parameter range $[0,1]$ in the definition of $c$ implies that 
curves are compactly supported in $\Sigma$ and, hence,  do not extend to spatial infinity.
Paths  are defined by identifying curves 
differing in orientation preserving reparametrizations and retracings.  They form a groupoid $\P$ with composition law 
given by concatenation. The compact support of  the curves $c$ and the definition of paths in terms of 
equivalence classes of {\em finite}
compositions of curves implies that any path $p$ is compactly supported in $\Sigma$.
An edge $e$ is a path $p$ that has a representative curve such that  the  
image $\et$ of this representative curve is a connected 1 dimensional semianalytic submanifold with 2 point boundary. 
Thus, edges are 
elementary paths  which generate $\P$ by composition. 
We denote by $\hom(\P,SU(2))$ the set 
of all  homomorphisms from $\P$ to $SU(2)$. The set of holonomies $\{ h_p[A],  p \in \P\}$ for a  given connection
 $A \in \A$ define an element in $\hom(\P,SU(2))$ by virtue of the composition law:  
$h_{p' p}[A]_{C}^{\; D}=h_{p'}[A]_{C}^{\; C'} h_{p}[A]_{C'}^{\; D}$, whenever the endpoint 
$f(p)$ of $p$ coincides with the beginning point $b(p')$ of $p'$.

A set of edges $e_1,\ldots,e_n$ is said to be independent if their intersections can only occur at their endpoints, i.e. if $\et_i \cap \et_j \subset \{b(e_i),b(e_j),f(e_i),f(e_j) \}$. We denote by $\gamma:=(e_1,\ldots,e_n)$ an ordered set of independent edges and by $\Lh$ the set of all such ordered sets of independent edges.  Given $\gamma,\gamma' \in \Lh$, we say that  $\gamma' \geq \gamma$ iff all edges of $\gamma$ can we written as composition of edges (or their inverses) of $\gamma'$.  With this relation $(\Lh,\geq)$ becomes a directed set \cite{mm2}.

To summarise: The compactness of the curves $c$ 
ensures that no subtelities ensue relative to the treatment of the compact 
case in \cite{mm2}.

\subsubsection{Background exponential related structures} 
The probes associated to the  background exponentials are electric fields.
Each such electric field ${\Eb}$ 
satisfies the fall offs:
\be
\Eb^\alpha_i =\frac{\hb^\alpha_i(\xh)}{r}+O(r^{-1-\e}) \label{fallEb} ,
\ee
with
\be
\hb^\alpha_i(-\xh)=\hb^\alpha_i(\xh),\label{evenEb}
\ee
where, as before,  the expressions refer to components in the (semianalytic) 
Cartesian chart $\{ x^\alpha \}$, $\Eb^a$ is $C^{k-1}$ semianalytic and $\hb^\alpha_i$ is $C^{k}$ semianalytic as a function on the sphere. 
Note that these conditions, modulo semianalyticity, correspond to those associated   with {\em variations} of 
asymptotically flat triads (\ref{fallE})
and that these conditions  ensure convergence of  the 3-dimensional integral in (\ref{baexp}).

We denote by $\E$ the set of all such electric fields $\Eb^a_i$. 
 $\E$ is an Abelian group with composition law given by addition. We denote by $\hom(\E,U(1))$  the set of homomorphism from $\E$ to $U(1)$.  The set of background exponentials $\{ \beta_{\Eb}[A], \Eb \in \E \}$  for a  given connection $A \in \A$ define an element in $\hom(\E,U(1))$ by virtue of  $\beta_{\Eb'}[A] \beta_{\Eb}[A]=\beta_{\Eb'+\Eb}[A]$.

A set of electric fields $\Eb_1, \ldots, \Eb_N$ is said to be independent, if they are algebraically independent, i.e. if they are independent under linear combinations with integer coefficients:
\be
\sum_{I=1}^N q_I \Eb_{I}=0, \; q_I \in \integers \iff q_I =0 , I=1,\ldots,N .\label{algind}
\ee
We denote by $\Upsilon=(\Eb_1, \ldots, \Eb_N)$ an ordered set of independent electric fields. The set of all ordered  sets of independent electric fields is denoted by $\Lb$. Given $\Upsilon,\Upsilon' \in \Lb$ we say that  $\Upsilon' \geq \Upsilon$  iff all the electric fields in $\Upsilon$ can be written as algebraic combinations of  those in $\Upsilon'$. One can 
verify that $(\Lb,\geq)$ is a directed set exactly as done in \cite{mm2} (the proof is purely algebraic and insensitive 
to the  detailed properties of $\E$ such as the particular fall-off (\ref{fallEb})). 

\subsubsection{Combined Holonomy and Background exponential structures} \label{holbelabels}

The combined set of labels associated to holonomies and background exponentials is given by pairs $l=(\gamma,\Upsilon) \in \Lh \times \Lb =: \L$ with preorder relation given by $(\gamma',\Upsilon') \geq (\gamma,\Upsilon)$ iff $\gamma' \geq \gamma$ and $\Upsilon' \geq \Upsilon$. It immediately follows that $\L$ is a directed set. 
Given $l=(e_1,\ldots,e_n,\Eb_1,\ldots,\Eb_N) \in \L$ we define the group
\be
G_l := SU(2)^n \times U(1)^N \label{gl}
\ee
and the map
\ba
\pi_l: \A & \to &  G_l \\
A & \mapsto & \pi_l[A] := (h_{e_1}[A],\ldots,h_{e_n}[A], \beta_{\Eb_1}[A],\ldots ,\beta_{\Eb_N}[A]). \label{pil} 
\ea
It will be useful for later purposes to describe  the relation between $\pi_{l}$ and $\pi_{l'}$ whenever $l' \geq l$.

Let  $l' \geq l$. Edges $e_{i} \in l$ can then be written as compositions of edges in $l'$. Let us denote this relation by: $e_i= \tilde{p}_i(e'_1,\ldots )$, where $\tilde{p}_i$ denotes a particular composition of edges (and their inverses) in $l'$. This corresponds to a relation on the holonomies of the form:
\be
h_{e_i}[A] = h_{\p_i(e'_1,\ldots)}[A] = p_i(h_{e'_1}[A], \ldots ), \label{hepp}
\ee
where  $p_i: SU(2)^{n'} \to SU(2)$  is the map determined by interpreting the compositions rules of $\tilde{p}_i$ as matrix multiplications. For example, if $e_1=e'_2 \circ  (e'_1)^{-1}$ then $p_1(g'_1,\ldots,g'_{n'})=g'_2 (g'_1)^{-1}$. Similarly,  electric fields $\Eb_I \in l$ can be written as integer linear combinations of electric fields in  $l'$:
\be
\Eb_I = \tilde{P}_I (\Eb'_1 ,\ldots) :=  \sum_{J=1}^{N'} q_{I}^{J} \Eb'_{J}, \quad q_{I}^{J} \in \integers, \quad I=1,\ldots,N. \label{Ptilde}
\ee
Associated to (\ref{Ptilde}) there is the map $P_I: U(1)^{N'} \to U(1)$ given by $P_I(u'_1,\ldots,u'_{N'})= \Pi_{J=1}^{N'}(u'_J)^{q_{I}^{J}}$ so that
\be
\beta_{\Eb_I}[A] = \beta_{ \tilde{P}_I (\Eb'_1 ,\ldots)}[A] = P_I(\beta_{\Eb'_1}[A], \ldots ). \label{bepp}
\ee 
The above maps combine in a map
\be
p_{l , l'}:=(p_1,\ldots,p_n,P_1,\ldots,P_N): G_{l'} \to G_l \label{projmap}
\ee
in terms of which equations (\ref{hepp}) and (\ref{bepp}) are summarized as:
\be
\pi_l[A] =p_{l , l'}( \pi_{l'}[A]), \label{pip}
\ee
expressing $\pi_l$ in terms of $\pi_{l'}$ and $p_{l , l'}$.



\subsection{Probes of the Electric field}\label{sec2C}
 The probes of the electric field are the pairs $(S,f)$ which go into the construction of the electric flux
$F_{S,f}$ of equation (\ref{flux}). We choose the surfaces $S$ to be oriented compact 2 dimensional 
semianalytic submanifolds of $\Sigma$ with or without boundary. For $S$ without boundary, we require the $su(2)$ valued
function $f$ to be semianalytic as a function on $S$. For $S$ with boundary, we require that $f$ be semianalytic
and compactly supported  as a function on the interior, Int$S$, of  $S$.
The choice of surfaces ensures that any  such surface $S$ intersects any edge (as defined in section \ref{sec2B})  in at 
most  a
{\em finite} number of connected semianalytic submanifolds (i.e. in a finite number of isolated points  and/or a 
finite number  semianalytic edges tangential to $S$).
 This in turns ensures that the Poisson bracket between the holonomies and fluxes,
is, as in the standard LQG case, well defined. The restriction of the support of $f$ within a compact set of 
Int$S$ is to avoid any further  technicalties associated with semianalyticity in the presence of boundaries.

\subsection{Poisson brackets between Elementary functions}\label{sec2D}

The classical Poisson brackets of the above functions is as follows. Holonomies and background exponentials Poisson commute among themselves and each other. The Abelian Poisson bracket algebra they generate plays a crucial role in the construction of the quantum configuration space to be described in later sections.  Poisson brackets between  fluxes and holonomies are exactly as in LQG, since our choice of edges and surfaces is such that they do not `reach' infinity and, as in the semianalytic, compact $\Sigma$ case,  they  can only intersect each other finitely many times. The Poisson bracket between fluxes and background exponentials is 
\be
 \{ \beta_{\Eb}, F_{S,f} \} = i F_{S,f}(\Eb)
\label{pbbef}
\ee
where $F_{S,f}(\Eb) = \int_S dS_a \tr[f \Eb^a]$. 
In order to obtain 
a classical algebra involving brackets between fluxes, it would be necessary to repeat the analysis of Reference 
\cite{mm2} which, following References \cite{acz,lost} identifies these Poisson brackets with 
commutators of derivations on the space of connection dependent functions 
generated by sums and products of holonomies and background
exponentials. While we anticipate no obstruction to doing so, we leave this for future work.

\section{The KS representation for the asymptotically flat case}\label{sec3} 
In section \ref{sec3A} we display the KS Hilbert space representation for the asymptotically flat case. 
Section \ref{sec3B} discusses the implementation of the classical
asymptotic boundary conditions  on the electric field in this representation.

\subsection{The KS Hilbert space and operator actions thereon.}\label{sec3A}
The KS Hilbert space, $\Hks$,  is spanned by states of the form $|s, E \rangle$, where $s$ is an LQG spin network with edges as described in the previous section,  and $E^a$ a {\em semianalytic}
 asymptotically flat background electric field satisfying 
(\ref{fallE}). The  inner product is given by
\begin{equation}
\langle s^{\prime},{E}^{\prime }_{}|s,{E}\rangle = \bra s|s^{\prime} \ket_\lqg
\delta_{{E}^{\prime}, {E}}, \label{ksip}
\end{equation}
where $\bra s|s^{\prime} \ket_\lqg$ is the LQG spin network inner product and $\delta_{{E}^{\prime}, {E}}$ the Kronecker delta.

Holonomies (\ref{hol}) and background exponentials (\ref{baexp}) act by
\begin{eqnarray}
{\hat h}^{\phantom{e\;}C}_{e\;D}|s,E \rangle &=& \vert {\hat h}^{\lqg \,A}_{e\; \quad B} s,E_{}\rangle ,
\label{holhat}
\\
{\hat \beta}_{\Eb} |s,E \ket &=&  |s, E+\Eb \ket \label{betahat}.
\end{eqnarray}
Above, we have used the notation of \cite{me} wherein 
given an LQG operator ${\hat O}$ with action 
$\hat O|s\rangle = \sum_IO^{(s)}_I|s_I\rangle$  in standard LQG, we have defined the state $|{\hat O}s, { E}\rangle$
in the KS representation through
\begin{equation} 
|{\hat O}s ,{E}_{}\rangle:= \sum_IO^{(s)}_I|s_I,{ E}_{} \rangle .
\label{oks}
\end{equation}
Note that the action (\ref{betahat}) is well defined because the new background field state label $E+\Eb$
satsifies the boundary conditions (\ref{fallE}) by virtue of the boundary conditions (\ref{fallEb}) on $\Eb$.\footnote{\label{fnconsistent}It is in this sense that the imposition of the asymptotic conditions (\ref{fallE}) 
on the background electric field labels 
of  {\em states} is  consistent with the imposition of 
the 
asymptotic conditions (\ref{fallEb}) on the background electric field labels of {\em operators} (see the remarks in the
main text of section \ref{sec1} preceding Footnote \ref{fn1}).}

The action of fluxes is given by 
\be
{\hat F}_{S,f}|s,E \rangle = |{\hat F}^\lqg_{S,f} s,E \rangle + F_{S,f}(E)|s,E \rangle ,
\label{fluxhat}
\ee
where $f=f^i \t_i$ is the $su(2)$-valued smearing scalar on the surface $S$ and $F_{S,f}(E) = \int_S dS_a f^i E^a_i$ the  
flux associated to the background electric field $E^a_i$.

It is easily verified that with these definitions the background exponentials act as unitary operators with 
$({\hat \beta}_{\Eb})^{\dagger}= {\hat \beta}_{-\Eb}$, the fluxes as self adjoint (or, more precisely, symmetric) operators, and that adjointness relations of holonomy operators  reproduce the  relations of classical holonomies under complex conjugation.

The (unitary) action of spatial diffeomorphisms and $SU(2)$ gauge transformations will be discussed in section \ref{sec8}.

\subsection{Implementation of the electric field boundary conditions.}\label{sec3B}
From a phase space perspective, the classical boundary conditions described in section \ref{sec2A} can be cast in the following form: Given a phase space point $(A,E)$, there exists a radius $r_{(A,E)}$ in the asymptotic region $\Sigma \setminus K$ such that Eqns. (\ref{fallE}) , (\ref{fallA}) hold for $r>r_{(A,E)}$. By $r$ greater than a given radius $r_0$ we  mean points in $\Sigma$ lying in:
\be
 \Sigma \setminus \Sigma_{r_0} :=\{ \vec{x} \in \reals^3 : (x^1)^2+(x^2)^2+(x^3)^2>r_0^2 \} \label{sr0}
 \ee
where (\ref{sr0}) is described in the Cartesian chart. $\Sigma_{r_0}$, defined as the complement of the set (\ref{sr0}), represents the points of $\Sigma$ `inside' the 2-sphere of radius $r_0$.

An analogue statement in quantum theory (regarding the electric field boundary conditions) is:  Given a KS spinnet $|\psi \ket \equiv |s,E \ket$,  there exists a radius $r_{\psi}$ in the asymptotic region  such that (i) the spin network $s$ lies in  $\Sigma_{r_{\psi}}$ (as defined above), 
(ii)  $E$ satisfies (\ref{fallE}) for $r> r_{\psi}$.   Property (i),  ensured by the compactness of $s$, implies that the spinnet $|\psi \ket$ can be thought of as  an eigenvector of an electric field operator $\hat{E}^a(x)$ with $x$ outside  $\Sigma_{r_{\psi}}$:
\be
\hat{E}^a(x) |\psi \ket = E^a(x)| \psi \ket , \quad x \in \Sigma \setminus \Sigma_{r_{\psi}}. \label{hatE}
\ee
From property (ii), the eigenvalue in (\ref{hatE}) satisfies (\ref{fallE}). Thus Eq. (\ref{hatE}) represents a quantum version of the electric field boundary condition (\ref{fallE}).

 An alternative strategy more attuned to standard LQG would be to describe the electric field boundary conditions purely in terms of fluxes.  This is however  more involved as it requires the use of surfaces approaching infinity.  In appendix \ref{fluxapp} we present the first steps towards the implementation of this idea.

\section{The Master Lemma} \label{sec4}

\subsection{Statement of the Lemma}\label{sec4A}
Let $(e_1,\ldots,e_n)$ be a set of $n$ independent edges. Let $(\Eb_1,\ldots,\Eb_N)$ 
be a set of $N$ algebraically independent electric fields, 
each with asymptotic behaviour (\ref{fallEb}). Define the group
\be
G := SU(2)^n \times U(1)^N \label{g}
\ee
and the map
\ba
\pi:  \A & \to &  G \\
A & \mapsto & \pi [A] := (h_{e_1}[A],\ldots,h_{e_n}[A], 
\beta_{\Eb_1}[A],\ldots ,\beta_{\Eb_N}[A]). \label{pi} 
\ea
Then the  map $\pi$ has dense range in $G$.\\

To prove the lemma, it clearly suffices to construct $A^{{ g},\delta}\in \A$ such that, 
given $g:=(g_1, \ldots, g_n, u_1, \ldots, u_N) \in G_l=SU(2)^n \times U(1)^N$ and any $\delta>0$, $A^{{ g},\delta}$
has the property that
\ba
|h_{e_{\alpha}}[A^{{ g},\delta}]_{\;C}^{\;\;D}- g_{\alpha C}^{\;\;D}| &\leq & C_1 \delta\; \forall \;\;\alpha=1,..,n\;  {\rm and}\; 
C,D=1,2.
\label{hgdelta}
\\
|\beta_{\Eb_I}[A^{{ g},\delta}] - e^{i \theta_I}| & \leq & C_2 \delta \;\forall\; \;I=1,..,N.
\label{budelta}
\ea
Here $C_1,C_2$ are $\delta$- independent constants, $C,D$ are $SU(2)$ matrix indices and 
$u_I =: e^{i \theta_I}\;, \theta_I\in \reals$. 

The proof parallels the proof of the Master Lemma  for the case of  compact $\Sigma$ in \cite{mm2}.  It consists of five steps, of which 
only the first one requires a minor adaptation to the present, non-compact $\Sigma$ case.  
We outline the steps in section \ref{sec4B} below referring to \cite{mm2} for technical proofs. 
The end result is a connection $A^{{ g},\delta} \in \A$ which implements equations (\ref{hgdelta}) and (\ref{budelta})
and which falls off faster than  $r^{-2}$ near spatial infinity thus satisfying  the asymptotic conditions (\ref{fallA})
with no leading order $r^{-2}$ term. 

We remark here that 
connections with  only subleading  behaviour at infinity  cannot describe spacetime solutions with non-vanishing linear
momenta since the surface terms corresponding to the evaluation of the linear momenta 
only involve the leading order $r^{-2}$ part of the connection.
Note however that we have only shown that a {\em finite} number of arbitrarily prescribed 
$SU(2)$ and $U(1)$ elements can be approximated to arbitrary accuracy using such connections with no leading order 
asymptotic behaviour. As we show in section \ref{sec5A} the set of {\em all} edge holonomies and 
{\em all} background exponentials
seperate points in $\A$. This together with the Master Lemma implies
that one needs infinitely many 
evaluations of holonomies and background exponentials to distinguish between connections with vanishing and non-vanishing
leading order fall offs.

As a final remark, note that by setting $l=(e_1,\ldots,e_n,\Eb_1,\ldots,\Eb_N)$, the Master Lemma implies 
that the map $\pi_l$ of equation (\ref{pil}) is dense in $G_l$  for any $l$.

\subsection{Steps in the proof of the Lemma}\label{sec4B}

\noindent (i) Construct a connection ${\bar A}^{B, \delta}$ which satisfies (\ref{budelta}):\\
Assume without loss of generality that 
the electric fields $\{\Eb_1,\ldots,\Eb_N\}$  are ordered so that the first $M$  are linearly independent and 
that the remaining  $P:=N-M$ electric fields can be written as linear combinations of the first $M$ ones so that:
\be
\Eb_{M+j}= \sum_{\mu=1}^{m} k^\mu_j \Eb_\mu , \quad j=1,\ldots,P , \label{Emj}
\ee
for some real constants $k^\mu_j$. Next, we construct    $M$ connections  $A^\nu, \nu=1,\ldots,M$  satisfying:
\be
\int_\Sigma \tr[\Eb_\mu^a  A^{\nu}_a]= \delta_\mu^\nu ,\quad \mu,\nu = 1,\ldots,M . \label{dualE}
\ee
An adapted version of the argument given in \cite{mm2} for the existence of such connections is as follows. Let $q_{ab}$ 
be an asymptotically flat metric in $\Sigma$ so that $q_{ab}= \qo_{ab} + O(1/r)$ with $\qo_{ab}$ being the 
fixed flat metric at infinity.  Let $\Omega_p$ be a strictly positive function on $\Sigma$ with an  $O(r^{p})$ 
behavior at infinity. Define the metric  $q^{(p)}_{ab}:= \Omega_p^2 q_{ab}$ so that we have 
\be
q^{(p)}_{ab}:= \Omega_p^2 q_{ab}, \;\;  q^{(p)}_{ab} = O(r^{2p}) \;\;{\rm as}\;\;r\rightarrow \infty.
\label{qpab}
\ee
Define the following bilinear form on $\E$:
\be
\bra \Eb, \Eb' \ket := \int_\Sigma q^{-1/2} q_{ab}\tr[\Eb^a \Eb'^b] .\label{ipE}
\ee
Let us now set $p>1$. Then 
the asymptotic behavior of $\Omega_p$ implies the integral (\ref{ipE}) is convergent.  
Nondegeneracy of $q_{ab}$  and positivity of $\Omega_p$ implies  $\bra, \ket$ is positive definite. 
It then follows that the $M \times M$ matrix defined by:  $\bra \Eb_\mu , \Eb_\nu \ket , \mu, \nu =1,\ldots,M$ is invertible. Denote its inverse by $c_{\mu \nu}$ . Then the one-forms $A^{\nu}_a :=  q^{-1/2} \sum_{\rho=1}^{M}  c_{\rho \nu} q_{ab} \Eb^b_\rho$  satisfy (\ref{dualE}) and, from equation (\ref{qpab}), fall off as $r^{-p-1}$ at infinity so that 
$A^{\nu}_a \in \A$. By exactly the same steps as in \cite{mm2}, one can then find $M$ real numbers $t^{(\delta)}_\mu, 1 \leq \mu \leq M $ such that
\be
{\bar A}^{B, \delta}_a:= \sum_{\mu=1}^{M} t^{(\delta)}_\mu A^\mu_a
\ee
 satisfies  (\ref{budelta}) with $C_2=1$. \\

\noindent (ii)  For sufficiently small $\epsilon$  and appropriately chosen $\epsilon$- independent charts,
construct balls $B_{\alpha}(2\epsilon ), \alpha =1,..,n$ of coordinate size  $2\epsilon$ such that 
\ba
B_{\alpha}(2\epsilon ) \cap B_{\beta}(2\epsilon ) = \emptyset \;\; {\rm iff} \;\; \alpha \neq \beta && \\
B_{\alpha}(2\epsilon ) \cap \et_{\beta} = \emptyset \;\; {\rm iff} \;\; \alpha \neq \beta &&\\
\bar{B}_{\alpha}(2\epsilon ) \cap \et_{\alpha} \;\; {\rm is}\;\; {\rm a}\;\; {\rm  semianalytic}\;\; {\rm  edge}&&
\label{2epsilonedge}
\ea

Since the edges are compactly supported, the construction of (ii) is identical to the one given in \cite{mm2}.  
\\
\noindent (iii) Construct a real semianalytic function $f_{\epsilon}$ such that $|f_{\epsilon}|\leq 1$ on $\Sigma$
with 
\ba 
f_{\epsilon}&=& 1\;\; {\rm on}\;  \Sigma - \cup_{\alpha} B_{\alpha}( 2\epsilon ) \\
            &=& 0 \;\; {\rm on}\;  \cup_{\alpha} B_{\alpha}( \epsilon ),
\ea
where $B_{\alpha}(\epsilon )$  denotes the $\epsilon$ size ball with the same centre as $B_{\alpha}(2\epsilon )$.

The
construction of $f_\epsilon$ is identical to that in \cite{mm2}.  \\

\noindent (iv) 
From (\ref{2epsilonedge}) it follows that each $e_{\alpha}$ can be written as the composition of three edges
$ s^1_{\alpha},s^2_{\alpha},s_{\alpha}$,
\be
e_{\alpha} = s^1_{\alpha} \circ s_{\alpha} \circ s^2_{\alpha},
\ee
with 
\ba
\tilde{s}_{\alpha} &:= & \et_{\alpha} \cap {\bar B}_{\alpha}(\epsilon )\\
 \tilde{s}^1_{\alpha} \cup \tilde{s}^2_{\alpha} & = & \et_{\alpha} \cap (\Sigma -{B}_{\alpha}(\epsilon )) .
\ea
Define:
\be
h_{s^i_{\alpha}}[A^{B,f}] =: g^i_{\alpha}, \;\;i=1,2
\ee
where $A^{B,f}:= f_{\epsilon} {\bar A}^{B,\delta}$.
and construct a connection $A^{\epsilon}$ supported in $\cup_{\alpha} B_{\alpha}(\epsilon )$ such that 
\be
h_{s_{\alpha}}[A^{\epsilon}] = (g^1_{\alpha})^{-1}g_{\alpha}(g^2_{\alpha})^{-1} .\label{step4}
\ee
The construction of $A^{\epsilon}$ is identical to that in \cite{mm2}. \\

\noindent (v)  Set $A^{g,\delta}:= A^{B,f} + A^{\epsilon}$.
It is then possible to show that  for small enough $\epsilon$, $A^{g,\delta}$ satisfies
equations 
(\ref{hgdelta}) and (\ref{budelta})  with $C_1=0$ and $C_2=2$.
The proof of this assertion is identical to that in \cite{mm2}.

\section{The quantum configuration space}\label{sec5} 
In section \ref{sec5A} we introduce the (Abelian) holonomy-background exponential algebra $\hba$ and its associated $C^*$ algebra $\hbabar$. The spectrum $\spec$ of $\hbabar$ is shown to be a topological completion of the space of connections $\A$. In section \ref{sec5B} we give a characterization of $\spec$ in terms of certain  homomorphisms, to be later used in section \ref{sec5E}.   In section \ref{sec5C} we realize $\Hks$ as an $L^2$ space over $\spec$ (with an integration measure 
$d \mu_{\ks}$) and show that the 
holonomies and background exponentials act by multiplication in this realization.
Section \ref{sec5D} is concerned with the action of flux operators on $L^2(\spec, d \mu_{\ks})$.
We note that 
while the general arguementation of sections \ref{sec5C} and \ref{sec5D} follows that for the spatially compact case  in 
\cite{mm2}, a key difference from the compact case stems from the unavailability of the \\
``$|s=0,E=0\ket$''
cyclic state of \cite{mm2} in the asymptotically flat context due to the incompatibility of the $E=0$ label with 
asymptotic flatness. Notwithstanding this the considerations of sections \ref{sec5C} and \ref{sec5D} establish 
the unitary equivalence of the standard KS representation of the holonomy, background exponential and flux operators
with their representations on $L^2( \spec ,d  \mu_{\ks})$. 
In section \ref{sec5E} we use the characterization of section \ref{sec5B} to show that elements of $\spec$ satisfy a 
weakened version of
the asymptotic conditions satisfied by elements of  $\A$.
\subsection{The algebras $\hba$ and $\hbabar$} \label{sec5A}

We denote by $\hba$ the $*$-algebra of functions of $\A$ generated by the elementary functions (\ref{hol}) and (\ref{baexp}), with * relation given by complex conjugation. By the same arguments as those given in \cite{mm2}, any element $a[A] \in \hba$ can be written as
\be
a[A] = a_l(\pi_l[A]), \label{Ff}
\ee
for some $l=(e_1,\ldots,e_n,\Eb_1, \ldots, \Eb_N) \in \L$, $\pi_l[A]$ as in (\ref{pil}), and $a_l \in \pol(G_l)$, where $\pol(G_l)$ is the set of functions on  $G_l$ that depend polynomially on 
the $SU(2)$ and $U(1)$ entries and the complex conjugates of these entries. 
Given  $a,l,a_l$ for which (\ref{Ff}) holds, $a$  is  said to be {\em compatible} with $l$, $l$ is said to be compatible with 
$a$ and  $a,l$ are said to be mutually compatible.
That a given mutually compatible pair $a,l$ uniquely specifies $a_l \in \pol(G_l)$ follows from the 
continuity of $a_l$ together with the denseness of $\pi_l[\A]$ in $G_l$.
If $l' \geq l$ and $l$ is  compatible with $a$, then $l'$ is compatible with $a$ and
\be
 a_{l'} = a_{l} \circ p_{l , l'} \label{rellabels},
 \ee
with $p_{l , l'} : G_{l'} \to G_l$ the map  induced by the way probes in $l$ are written in terms of probes in $l'$ as described in section \ref{holbelabels} (see  Reference \cite{mm2} for a proof of this claim).

On $\hba$ we define the norm:
\be
|| a || := \sup_{A \in \A} |a[A]|,  \quad a \in \hba .\label{algnorm}
\ee
When $a$ is written as in (\ref{Ff}), the lemma of section \ref{sec4} implies:
\be 
|| a ||= \sup_{A \in \A} |a_l(\pi_l[A]) | = \sup_{g \in G_l} |a_l(g)|.\label{normF}
\ee
Being a sup norm, (\ref{algnorm}) is compatible with the product and complex-conjugation star relations on $\hba$. 
The completion $\hbabar$ of $\hba$ in the norm (\ref{algnorm}) is then  a unital $C^*$ algebra. We denote the Gel'fand spectrum of $\hbabar$  by $\spec$. 
$\spec$ is a compact, Hausdorff space and $\hbabar \simeq C(\spec)$ where $C(\spec)$ is the $C^*$ algebra of continuous 
functions on $\spec$. We denote by $\cyl(\spec) \subset C(\spec)$ the subalgebra of continuous 
functions corresponding to $\hba$ in the Gel'fand identification. 

From \cite{rendall}, the fact that $\hba$ separates points in $\A$, implies  that $\A$ is topologically  
dense in $\spec$ or, equivalently, that 
$\spec$ is the completion of $\A$ in the Gel'fand topology. To see that  $\hba$ separates points in $\A$, we proceed
as follows.
Given $A'_a \neq A''_a \in \A$ we want to find $a \in \hba$ such that $a[A'] \neq a[A'']$. It is  enough to consider 
elements of the form $a=\beta_{\Eb}$. Setting $A_a := A_a' -A_a''$ the condition translates to show that for any $A_a \neq 0$ there exists  $\Eb^a \in \E$ such that $\beta_{\Eb}[A] \neq 1$. Since $A_a \neq 0$ we can find a local chart where at least one of its components, say $A_1^{1}$, is nonzero. Further, we can find a small enough open ball $U$ where this component is of definite sign, say $A_1^{1}|_{U}>0$. By using  semianalytic bump functions similar to those used in step (iii) of the master Lemma (see Eq. (A7) of \cite{mm2}) one can construct an electric field $\Eb^a \in \E$ with support in $U$ such that $\Eb_1^1|_{U}>0$ and with all remaining components being zero. Then $s:=\int_\Sigma \Eb^a_i A^i_a = \int_U \Eb^1_1 A^1_1 >0$. One can further take $\Eb^a$ such that $s \notin 2 \pi \integers$ so that  $\beta_{\Eb}[A]= e^{i s} \neq 1$.

\subsection{Characterization of the spectrum $\spec$}\label{sec5B}

Recall from  \ref{sec2B}1 and \ref{sec2B}2, that  $\hom(\P,SU(2))$ and $\hom(\E,U(1))$ denote the space of homomorphisms
from the groupoid of paths $\P$ to the group $SU(2)$  and from the Abelian group of electric fields $\E$ to the group
$U(1)$. 
As in the case of compact $\Sigma$, the spectrum admits a characterization in terms of these homomorphisms.
More in detail, each element of the spectrum is in unique correspondence with a pair of homomorphisms, one 
member of the pair in $\hom(\P,SU(2))$ and the other in $\hom(\E,U(1))$ i.e. there is 
a bijection between $\spec$ and $\hom(\P,SU(2)) \times \hom(\E,U(1))$.
As in the compact $\Sigma$ case \cite{mm2}, we construct this bijection as follows.

First recall from standard Gel'fand theory that elements of the spectrum are in correspondence with $C^*$ algebraic
homomorphisms from $\hbabar$ to $\complex$. Accordingly, given such a homomorphism 
$\phi \in \spec$ define:
\ba
s_\phi : & \P& \to SU(2) \\
& p & \mapsto s_\phi(p)_{C}^{\;D} := \phi(h_{p\; C}^{\; \;\;D})
\ea
and
\ba
u_\phi : & \E& \to U(1) \\
& \Eb & \mapsto u_\phi(\Eb):= \phi(\beta_{\Eb}).
\ea
As in \cite{mm2}, one can verify that the $*$-homomorphism properties of $\phi$ imply that $s_\phi \in \hom(\P,SU(2))$ and $u_\phi \in \hom(\E,U(1))$.  Conversely, given $s \in \hom(\P,SU(2))$ and $u \in \hom(\E,U(1))$ one can find   $\phi \in \spec$ such that $u_\phi=u$ and $s_\phi=s$ as follows. $\phi$ is first defined as a $*$-homomorphism from $\hba$ to $\complex$ by:
\be
\phi(a)
:= a_l(s(e_1),\ldots,s(e_n),u(\Eb_1),\ldots,u(\Eb_N)) \label{defhom},
\ee
where $l=(e_1,\ldots,e_n,\Eb_1, \ldots, \Eb_N)$ is compatible with $a$. From the homomorphism properties of $s$ and $u$, and the considerations of section \ref{sec5A} one can verify that (\ref{defhom}) is independent of the choice of $l$ and satisfies the $*$-homomorphism properties \cite{mm2}. Finally one can show boundedness of $\phi$  by means of  Eq. (\ref{normF}), so that $\phi$ admits a unique extension to $\hbabar$ \cite{mm2}. Thus $\phi \in \spec$, and by construction $u_\phi=u$ and $s_\phi=s$.

\subsection{Realization of the KS Hilbert space as the space $L^2(\spec,d \muks)$} \label{sec5C}
 
Fix a KS state $|0, E \ket$ where $E^a$ is an asymptotically flat electric field. We wish to define a positive linear functional (PLF) on $\hba$ associated to this state.\footnote{Since we have not  shown that the operators (\ref{holhat}), (\ref{betahat}) induce a representation of $\hba$ on the KS Hilbert space, we  here verify explicitly that expectation values yield a well defined  PLF on $\hba$.}  Given $a \in \hba$ and $l \in \L$ compatible with $a$, let $\hat{a}_l$ be the operator on the KS Hilbert space associated to $a_l$. Define:
\be
\w(a):= \bra 0,E | \hat{a}_l | 0,E  \ket =  \int_{G_l}  a_l(g) d \mu_l  , \label{defplf}
\ee
where $d \mu_l$ is the Haar measure on the group $G_l$ normalized so that $\int_{G_l} d \mu_l =1$. The second equality in (\ref{defplf}) is described in appendix B.2 of \cite{mm2}.  That this definition is independent of the choice of compatible label $l \in \L$ follows from the `cylindrical consistency' of the measures  $\{ \mu_l, l \in \L\}$, as shown in appendix C.3 of \cite{mm2}. Positivity is also easily verified from the integral representation in (\ref{defplf}). 

From the master lemma and Eq. (\ref{Ff}) one finds $\w$ is bounded:
\be
| \w(a)| =\left| \int_{G_l} a_{l}(g) d \mu_l \right| 
 \leq   \sup_{g \in G_l}  |a_l(g)| = ||a|| . \label{wbounded}
 \ee
Eq. (\ref{wbounded}) serves two purposes. On the one hand it tell us that $a[A]=0 \implies \w(a)=0$, so that $\w$ is well defined on the abstract algebra $\hba$ (i.e. it respects all quotienting relations like $\beta_{\Eb_1} \beta_{\Eb_2} - \beta_{\Eb_1+ \Eb_2}=0$).  On the other hand, the boundedness property implies the existence of a unique extension of $\w$ to  $\hbabar \simeq C(\spec)$ \cite{ttbook}.  It then follows by  Riesz-Markov theorem that $\w$ defines a regular measure $\muks$ on $\spec$ satisfying:
\be
\w(a)=\int_{\spec} a d \muks, \label{riesz}
\ee
where in the RHS of (\ref{riesz}) $a$ is seen as an element of  $C(\spec)$ via the standard Gel'fand identification.  

We now show that $\Hks \simeq L^2(\spec,d\muks)$ as Hilbert space representations of $\hbabar$.
Note that $| 0,E  \ket \in \Hks$ is a cyclic state i.e. the action of $\hba$ on  $| 0,E  \ket$
yields a dense subspace ${\cal D}_{\ks}$ in  $\Hks$.\footnote{The action of a background exponential changes the label $E$ of $|0,E\ket$ to any desired
background field state label and appropriate 
finite sums and products of holonomy operators change the label $0$ of $|0,E\ket$ to any desired LQG spin net 
label. It follows the dense set of the finite span of KS spinnets may be obtained by the action of operators in $\hba$.}
Next, choose an element $\Psi_0 \in C(\spec)$ such that 
\be
|\Psi_0|^2=1 \;\;{\rm on}\;\;\spec \label{psio2=1}.
\ee
Define the vector space ${\cal D}_{C(\spec)}:=\{a\Psi_0 \in C(\spec ), \; a\in \hba \subset C(\spec)\}$ and equip
${\cal D}_{C(\spec )}$ with the inner product
\be
(a\Psi_0, b\Psi_0) = \int_{\spec} (a\Psi_0)^* (b\Psi_0) d \muks , \;\;a,b\in \hba .
\label{psioip}
\ee
Define $V_{E,\Psi_0}:{\cal D}_{\ks}\rightarrow {\cal D}_{C(\spec)}$ by $V_{E,\Psi_0}({\hat a} | 0,E  \ket ):= a\Psi_0$
where ${\hat a}$ denotes the KS operator correspondent of the element $a\in \hba$ and $a\Psi_0 \in  C(\spec)$ as in
equation (\ref{psioip}) above.
Then it is straightforward to check using equations (\ref{psio2=1}), (\ref{psioip}) and (\ref{riesz}) that 
 $V_{E,\Psi_0}$  is a one to one map from ${\cal D}_{\ks}$ to ${\cal D}_{C(\spec)}$ which preserves the inner product.
It follows  that $V_{E,\Psi_0}$
 admits a unique extension from $\Hks$ to $L^2(\spec,d\muks)$. From the fact that $\hba$ is commutative,
it follows that $V_{E,\Psi_0}$ unitarily maps the KS representation of $\hba$ to a representation by multiplication
on $L^2(\spec,d\muks)$ by the continuous functions in $\hba\subset C(\spec)$. Since the holonomies and 
background exponentials are represented as bounded operators in the KS representation, and since elements of 
$\hba$ are finite polynomials in these basic operators, it follows that $\hba$ is represented as an algebra of 
bounded operators on $\Hks$. This boundedness is readily seen in the representation on  $L^2(\spec,d\muks)$
as a consequence of the boundedness of continuous functions on the compact Haussdorff space $\spec$. 

Finally, note that $L^2(\spec,d\muks)$ supports not only the operator algebra $\hba$ 
but also its $C^*$ completion, $\hbabar$.
Once again, since continuous functions on compact Haussdorff space are bounded, it follows that elements of 
$\hbabar$ are represented as bounded operators on $L^2(\spec,d \muks)$.
It follows by using the inverse map
$V^{-1}_{E,\Psi_0}$ that we may define the action of elements of $\hbabar$ on $\Hks$ also as bounded operators
on $\Hks$. This completes the demonstration that $\Hks\equiv L^2(\spec,d\muks)$ as Hilbert space representations
of $\hbabar$.

Next we note  that the identification of $\Hks$ with $L^2(\spec,d\muks)$ is not unique. The non-uniqueness
is two fold. First, as seen above, the map $V_{E,\Psi_0}$  may be defined for any $\Psi_0 \in C(\spec)$
subject to equation (\ref{psio2=1}). Second, it is easy to check that the  PLF $\omega_E$ of equation (\ref{defplf})
is independent of the choice of $E$ i.e. for any $a\in \hba$ and any $E_1,E_2$ subject to the asymptotic
conditions (\ref{fallE}) $\omega_{E_1} (a)= \omega_{E_2}(a)$. Thus the unitary map
$V_{E,\Psi_0}$ may be defined for any $E$ subject to (\ref{fallE}) and any $\Psi_0 \in C(\spec)$.
In the compact case, there is a natural choice, namely $E=0$ and $\Psi_0 =1$ and this together with the 
natural action of flux operators as derivations yields the unitary equivalence of $L^2(\spec, d\muks)$
and $\Hks$ as Hilbert space representations of the holonomy, background exponential {\em and} flux operators
\cite{mm2}.

In contrast, in the asymptotically flat case, there is no such natural choice because $E=0$ is not consistent with 
the requirements of asymptotic flatness (\ref{fallE}). In order to proceed further, we make {\em some} choice, $E$,
consistent with (\ref{fallE}) and, henceforth, set $\Psi_0=1$ and use the notation $V_E:= V_{E,\Psi_0=1}$.
As we shall see in the next section, the unitary equivalence  of the KS representation of 
holonomies and background exponentials with the one on $L^2(\spec, d\muks)$ 
defined by $V_E$ can be extended to include the action of  flux operators. While, in contrast to the 
action of operators in $\hbabar$, the explicit action of the flux operators on (a dense subspace of) $L^2(\spec,d\muks)$
{\em will} involve the choice of $E$, the extension will be seen to hold for any choice of $E$ subject to (\ref{fallE}).


\subsection{Action of Fluxes on $L^2(\spec,d \muks)$} \label{sec5D}

The  discussion of section \ref{sec5C} 
indicates that  elements $a \in \hbabar \simeq C(\spec)$ have a dual interpretation: 
When seen as elements of $C(\spec)$ 
they  are `wavefunctions' in the $L^2$ representation, i.e. \emph{vectors} in the Hilbert space; 
when seen as elements of $\hbabar$, they are naturally associated with   \emph{operators} $\hat{a}$ on the Hilbert space 
$\Hks$. Accordingly, when we use the former interpretation we shall refer to $a$ as a wavefunction
and when we use the latter we shall denote $a$ by ${\hat a}$ and refer to it as an operator.

Let us fix a `reference' KS spinet $|0,E \ket$ as before. The wavefunction associated to a KS spinnet $|s,\tilde{E} \ket$ 
via the map $V_E$ is then given by: $T_s \beta_{\tilde{E}-E} \in \cyl(\spec)$, 
where $T_s[A] \in \hba$ is the spin network function associated to  $s$ \cite{ttbook}, 
$\beta_{\tilde{E}-E}[A] \in  \hba$ the background exponential function (\ref{baexp}), and $T_s,\beta_{\tilde{E}-E}$ 
the respective elements in 
$\cyl(\spec)$ under the Gel'fand identification $\hba \simeq \cyl(\spec)$.
\footnote{Notice that there are no elements in $\cyl(\spec)$ that could correspond to  ``$\beta_{\tilde{E}}$''
 or ``$\beta_{E}$'' since these labels do not satisfy the required fall off (\ref{fallEb}). 
However $\tilde{E}-E \in \E$ is a valid label for the background exponentials.}   

From the action of the flux operator ${\hat F}_{S,f}$ on KS spinnets (\ref{fluxhat}), and the above, $E$-dependent identification of KS spinnets with elements of $\cyl(\spec)$, we obtain the following action of the fluxes in the latter space:
\be
{\hat F}^{(E)}_{S,f}( T_s \beta_{\Eb})= 
({\hat F}^{\lqg}_{S,f}T_s) \beta_{\Eb}+ F_{S,f}(E+\Eb) T_s \beta_{\Eb}.  \label{fluxwf}
\ee

As indicated in section \ref{sec5C},
even though  the natural choice $E=0$ is not available if we wish to describe asymptotically flat geometries, 
all different choices of asymptotically flat `reference' $E$ yield unitarily equivalent Hilbert spaces. 
The unitary transformation that maps the wave functions in the ``$E'$ representation'' associated with $V_{E'}$
 to those in the  ``$E$ representation'' associated with $V_{E}$ is given by the multiplication operator:
\ba
\hat{U}_{E, E'}: & L^2(\spec,d\muks) & \to  L^2(\spec,d\muks) \\
& \Psi & \mapsto \beta_{E'-E} \Psi.
\ea
The multiplicative operators from $\hba$ remain unaltered by this action, whereas  flux operators (\ref{fluxwf}) 
in the  $E'$ representation are  mapped into the corresponding flux operators in the $E$ representation:
\be
\hat{U}_{E, E'} {\hat F}^{(E')}_{S,f} \hat{U}_{E', E} =  {\hat F}^{(E)}_{S,f}
\ee
as can be verified by straightforward computation. 


\subsection{Asymptotic behaviour of elements of $\spec$.}\label{sec5E} 
From section \ref{sec5A}, the quantum configuration space $\spec$ is a topological  completion of the 
classical configuration space of connections $\A$. Since elements of $\A$ are subject to the asymptotic conditions
(\ref{fallA}), one may interpret this feature of $\spec$ as evidence for the incorporation of 
(\ref{fallA}) in the quantum theory. However, as we note in  section \ref{sec6}, while {\em topologically} dense, 
$\A$ is {\em measure theoretically} sparse in $\spec$ with respect to the measure $\mu_{KS}$. Since expectation value
computations in quantum theory do not obtain contributions from zero measure sets, 
if  there exists
a measure theoretically thick
set in $\spec$ whose elements can be thought of as  blatantly violating the conditions (\ref{fallA}), 
we would conclude that 
these conditions do not leave an imprint in quantum theory.
That this is not so follows from the fact that 
{\em every} element of $\spec$ satisfies a quantum analog these conditions. This quantum analog can be thought of 
as weaker than the classical conditions (\ref{fallA}).
To see this we proceed as follows.

From section \ref{sec5B}, every element of the quantum configuration space $\spec$ is identified with 
an element of $\hom (\E, U(1))$. In particular since $\A \subset \spec$, every $A\in \A$ defines
$h_A\in \hom (\E, U(1))$ through $h_A({\bar E})= \beta_{\bar E}(A)=  e^{i\int_{\Sigma} A_a^i\Eb^a_i}\in U(1)$. 
The asymptotic fall offs
on $A$ (\ref{fallA}) ensure that the integral $\int_{\Sigma} A_a^i\Eb^a_i$ is well defined for every $\Eb \in \E$
by virtue of the boundary conditions (\ref{fallEb}) on elements of $\E$.
This in turn 
ensures the well- definedness of  $h_A$. However, given the conditions (\ref{fallEb}), the conditions
(\ref{fallA}) are only sufficient (rather than necessary) for the well definedness of this integral.
For example, consider a  connection ${\bf A}$ with asymptotic behaviour
\be
{\bf A}_{\alpha}^i = \frac{df}{dr}\frac{g^i_{\alpha}({\hat x})}{r^2} \;+\; \frac{C^i_\alpha}{(r  \log r)^2}
\ee
with $f(r)= \cos (e^r)$ and $C^i_\alpha$ constant. A by parts integration shows that the contribution of the first term to 
the integral $\int_{\Sigma} {\bf A}_{a}^k{\Eb}^a_k$ (with $\Eb \in \E$) is finite and it is easy to check the finiteness of the contribution of
the second term.
Clearly ${\bf A}\notin \A$.
Nevertheless, 
$h_{{\bf A}} \in   \hom (\E, U(1))$ with  
$h_{{\bf A}}({\bar E})= \beta_{\bar E}({\bf A})=  e^{i\int_{\Sigma} {\bf A}_{ a}^i\Eb^a_i}\in U(1)$.
Since elements of $\spec$ include {\em all} homomorphisms in $\hom (\E, U(1))$ such connections are also 
a part of the quantum configuration space. Thus, instead of the detailed fall offs (\ref{fallA}), such
quantum connections $A^{\prime}$  satisfy the weaker condition that the integral:
\be
\int_{\Sigma} A_{b}^{\prime \, i}\Eb_i^b <\infty \quad \forall \Eb \in \E .
\label{weakercond}
\ee

Of course not every homomorphism arises from a connection field; just as in the spatially compact case, there
may be {\em no} 
connection field associated with a given homomorphism.
To see this, note that the homomorphisms defined by elements of $\A$ have a `smooth' dependence
with respect to the background field which allows the computation of the functional derivative of any such 
homomorphism with respect to its background field dependence.
More in detail, $h_A \in \hom (\E, U(1))$
is defined through $h_A({\bar E}\in \E )= \exp( i\int A_a^i{\bar E}^a_i)$ so that $\frac{\delta h_A}{\delta {\bar E}(x)}
= i A(x) h_A$. However, it is possible to define elements of 
$\hom (\E, U(1))$ which are not functionally differentiable.\footnote{As an example consider $A \in \A$ and $\phi \in \hom(\reals,U(1))$ such that $\phi(t)$ is nowhere differentiable in $\reals$. Then $\Eb \to \phi(\int A_a^i{\bar E}^a_i)$ is a non-functionally differentiable element of $\hom(\E,U(1))$.}

For such `distributional' elements of $\spec$ which cannot be realised as a connection field,
a quantum analogue of condition (\ref{weakercond}) is provided by the defining property of elements of $\spec$:
namely that every such element yields a well defined
element of $\hom (\E, U(1))$. It is in this sense that elements of the quantum configuration space
satisfy a weakened version of the classical fall offs (\ref{fallA}).

Note that the quantum configuration space is an enlargement of the classical one in two 
different ways. First, as in the compact case, the elements of $\spec$ may be `distributional' in the sense that the homomorphisms they define are not functionally differentiable with respect to their electric field argument. Second, elements of $\spec$ satisfy the (distributional) analog of the weaker boundary condition (\ref{weakercond}) than (\ref{fallA}). The weakening of the functional differentiability property may be thought of as indicative of  a certain `blindness' of (distributional) elements of   $\spec$  to the differential structure  of the manifold (the differential structure is, after all, what allows the definition of infinitesimal variations of $\bar E$ and the consequent computation of the functional derivative. We speculate  in closing that the weakened asymptotic  behaviour may similarly be thought of as an insensitivity to the function space  structure underlying the infinite dimensional symplectic structure of the classical phase space.  
More in detail, one expects that pairs of elements, one in the space of variations of the classical triad field  
and one in (the tangent space to) $\A$  lie in the tangent space to the infinite dimensional classical phase space.  
From this point of view the condition  (\ref{fallA}) and the (variation of the) condition (\ref{fallE}) 
could, in a more rigorous treatment, possibly be seen as linked to the function 
space topology of the phase space. The  weakened condition (\ref{weakercond}) 
is too coarse to define typical function space
topologies and in this sense the  weakened 
asymptotic behaviour of elements of $\spec$ could be seen to arise from their 
insensitivity to the detailed function space structure.

\section{The quantum configuration space as a projective limit}\label{sec6}
In section \ref{holbelabels} we defined the label set $\L$, the spaces $\{G_l, l \in \L \}$, and the maps $\{ p_{l,l'} : G_{l'} \to G_l, l' \geq l \in \L \}$. It is easy to verify that if  $l'' \geq l' \geq l$ then $p_{l,l''}=p_{l,l'} \circ p_{l',l''}$. Further, by the same argument as given in \cite{mm2}, the maps $p_{l,l'}$ can be shown to be surjective.  The projective limit space $\Abar$ can then be constructed from the family $( \L,\{ G_l \} ,\{ p_{l l'} \})$   as in \cite{aajurekproj} as follows. 

One first considers the space $\A_\infty := \times_{l \in \L} G_l$.  A point in $\A_\infty$ is then given by a collection of points $\{x_l \in G_l \}$ for every $l \in \L$. $\A_\infty$ is given the so-called  Tychonov topology \cite{ttbook}, i.e. the weakest topology  such that the canonical projections
\ba
p^\infty_l : & \A_\infty & \to  G_l  \label{plinf} \\
& \{ x_{l'} \} & \mapsto  x_l 
\ea
are continuous. Under this topology,  $\A_\infty$  is compact and Hausdorff \cite{ttbook}. Next, one considers the subset of $\A_\infty$ given by:
\be
\Abar: = \{ \{ x_l \} \in \A_\infty \; | \; p_{l,l'}(x_{l'})=x_l, \;  \forall l' \geq l \}  \label{defAbar}
\ee
with the topology induced by $\A_\infty$. One can then show that $\Abar$ is also compact and Hausdorff \cite{aajurekproj}. Finally, the projections  $p_l: \Abar \to G_l$ defined by $p_l:=p^\infty_l |_{\Abar} $ can be shown to be continuous and surjective \cite{aajurekproj}.  

The proof that $\Abar$ is homeomorphic to $\spec$ is the same as the one given in \cite{mm2} following \cite{velhinho} and consists in showing that  $C(\Abar)$ and $C(\spec) \equiv \hbabar$ are isomorphic as $C^*$ algebras. This is done by constructing a norm preserving $*$-isomorphism $T$, between $\cyl(\spec) \equiv \hba$ and 
\be
\cyl(\Abar) := \cup_{l \in L} \; p^*_l \; \pol(G_l) \subset C(\Abar),
\ee
given by:
\ba
T : \cyl(\Abar) & \to & \hba \\
 f = f_l \circ p_l& \mapsto & T(f):= f_l \circ \pi_l.\label{T}
\ea
Finally, one makes use of Stone-Weierstrass theorem to show that the completion of $\cyl(\Abar)$ coincides with  $C(\Abar)$. Since $C(\spec)$ can be seen as the completion of $\cyl(\spec)$, it follows that $C(\Abar) = C(\spec)$ as $C^*$ algebras. By Gel'fand theory it then follows that $\Abar$ and $\spec$ are homeomorphic

As in \cite{mm2}, one can consider the two separate projective limit spaces $\Ah$ and $\Ab$ associated to $( \Lh,\{ G_\gamma \} ,\{ p_{\gamma \gamma'} \})$ and $( \Lb,\{ G_\Upsilon \} ,\{ p_{\Upsilon \Upsilon'} \})$, and show that $\Abar = \Ah \times \Ab$. From this perspective, it may look as if one got a `too big' space, somehow involving two copies of space of connections. Let us clarify this point. $\Abar$ is understood as a topological completion of $\A$. $\A$ embeds in $\Abar$ via the map $\pi_l$ of section \ref{holbelabels}:
\ba
& \A & \to \Abar = \Ah \times \Ab \\
& A & \mapsto \{ \pi_l[A] \} = (\{ \pi_\gamma[A] \}, \{ \pi_\Upsilon[A] \}) \label{AinAbar}
\ea 
since by virtue of (\ref{pip}) $\{ \pi_l[A] \} \in \A_\infty$ lies in $\Abar$ (\ref{defAbar}).  Above we also displayed the embedding in the $\Ah \times \Ab$ picture of $\Abar$, where $\pi_\gamma[A]$ and $\pi_\Upsilon[A]$ are respectively given by the first $n$ and  last $N$ components of (\ref{pil}). 
This product structure  allows for other, inequivalent embeddings, for instance:
\ba
& \A & \to \Abar \\
& A & \mapsto \{ (\pi_\gamma[A] , \id_{G_\Upsilon}) \},
\ea 
and in this sense one could perhaps argue that $\Abar$ is `too big' .
However the interpretation of $\Abar$ as a completion of $\A$ is only with respect to the `correct' embedding (\ref{AinAbar}).

The description of the measure $\muks$ on $\spec \approx \Abar$ can also be given in terms of cylindrically consistent measures $\{ \mu_l \}$ on $\{G_l \}$ exactly as in \cite{mm2}, and the same argument given to  show that $\A \subset \Abar$ is of measure zero  applies  here  as well.

\section{Gauge transformations and Asymptotic Symmetries: Classical considerations}\label{sec7}

We describe the group of gauge transformations, $\aut$,  in section \ref{sec7A}
 and the group of 
asymptotic symmetries, $\autEo$, in section \ref{sec7B}. 
The asymptotic behaviour of elements of $\aut$ and $\autEo$ displayed in sections \ref{sec7A} and \ref{sec7B}
is derived as follows.
We study the space of $C^{k-1}$ $SU(2)$ rotations and $C^k$ diffeomorphisms subject to the restriction that their
combined action
preserves the space of classical triad fields with fall offs (\ref{fallE}).
In appendix \ref{symgroup} we show that
this  restriction on $SU(2)$ rotations and diffeomorphisms completely specifies their asymptotic behaviour.
The restriction is {\em a priori} weaker than the  restriction that such transformations
preserve both the asymptotic fall offs on the triad as well those on the connection (\ref{fallA}).
However, as can be easily verified, the additional restriction of preservation of (\ref{fallA}) is automatically
satisfied by the transformations subject to the behaviour specified by the considerations of Appendix \ref{symgroup}.

The conditions derived in Appendix \ref{symgroup} are as follows.
The diffeomorphisms asymptote to a combination of rotations, translations and `odd supertranslations' 
(odd supertranslations are defined in 
\cite{mcparity}).
Diffeomorphisms with trivial rotational part are accompanied by
 $SU(2)$ transformations which asymptote to identity and diffeomorphisms with non-trivial rotational part are accompanied
by $SU(2)$ transformations which compensate for this rotation so as to preserve the flat triad at infinity.
Comparison with the asymptotic fall offs of the finite transformations generated by 
the $su(2)$ multipliers and shift vectors
which serve as smearing functions for the $SU(2)$ Gauss Law, the spatial
diffeomorphism constraint 
and the total angular and linear momenta in the $C^{\infty}$ setting of Reference \cite{mcparity} allow us to 
identify which of the transformations
of the previous paragraph are to be interpreted as  gauge transformations and which as asymptotic symmetries.\footnote{Reference \cite{mcparity} specifies the asymptotic fall offs of the multipliers. 
In appendix  \ref{infintesimalapp} we use these 
conditions on the multipliers to 
deduce  corresponding conditions on the finite $SU(2)$ transformations and diffeomorphisms 
generated by these multipliers.}
This comparison indicates we identify   gauge transformations
with   combinations of 
diffeomorphisms with trivial rotational and translational parts together with  $SU(2)$ transformations which asymptote to identity, and that  we identify
asymptotic symmetries with combinations of  diffeomorphisms with non-trivial rotational  and
translational parts together with $SU(2)$ transformations which compensate for the rotational diffeomorphism
so as to leave the fixed flat triad invariant at infinity.
%


In section \ref{sec7C} we show that $\aut$ is a normal
subgroup of $\autEo$ and that the quotient group $\autEo/\aut$ 
is the finite dimensional group $\reals^3 \rtimes SU(2)$.
In section \ref{sec7D} we display the action of semianalytic 
elements of $\aut$ and $\autEo$  on the holonomy, flux and background
exponential functions. The reason for the additional restriction of semianalyticity is that these functions
are defined using semianalytic probes and the semianalyticity of these probes as well as the 
 Poisson bracket algebra of these functions is preserved by semianalytic elements of $\aut, \autEo$.

\subsection{Kinematical gauge group $\aut$}\label{sec7A}
A kinematic gauge transformation $a$ consists of  
a pair $(g,\phi)$ where $g$ is a $C^{k-1},\; SU(2)$ internal rotation, $\phi$ is a 
$C^k$ diffeomorphism 
and $a=(g,\phi)$ is connected to identity.\footnote{\label{fnconn}We shall say that $a$ is connected to identity if there exists a  one parameter family of automorphisms 
$a (s) \in \aut, s\in [0,1]$ such that
(i) $a(0)=\id$, $a(1)=a$  (ii) for every point $p$ in $\Sigma$, $\phi (s)(p)=: p(s)$ is continuous with respect to the
topology of  $\Sigma$ and (iii) for every point $p\in \Sigma $, $g(s,p)$ is continuous in $SU(2)$.}

The transformation $a\equiv (g,\phi )$ acts on  phase space as: 
\ba
(g, \phi ) \cdot A_a & := & g \cdot (\phi \cdot A_a) =  g \phi_* A_a g^{-1} - (\partial_a g) g^{-1}  ,\\
(g, \phi ) \cdot E^a & := & g \cdot (\phi \cdot E^a) =  g  \phi_* E^a g^{-1} . \label{autAE}
\ea
$\phi_*$ denotes push-forward so for instance:  $(\phi_* g)(x) \equiv g(\phi^{-1}(x))$. 

As explained in the beginning of  section \ref{sec7}
the considerations of Appendix \ref{symgroup} together with 
the classical phase space analysis \cite{ttparity,mcparity} and the considerations of Appendix \ref{infintesimalapp} 
imply 
the following  asymptotic conditions as $r \to \infty$ on each such gauge transformation $a$:
\ba
g(x) & = &\idtwo+ \frac{\lambda(\xh)}{r}+O(r^{-1-\e}), \label{fallg0} \\
\phi^\alpha(x) & = & x^\alpha+ s^\alpha(\xh)+O(r^{-\e}) ,\label{fallphi0}
\ea
where
\ba
\lambda(-\xh) & = &\lambda(\xh) ,\label{evenlam} \\
s^\alpha(-\xh) & = & -s^\alpha(\xh), \label{odds}
\ea
and $\lambda$ and $s^\alpha$ are respectively  $C^{k},C^{k+1}$ 
functions on the sphere.
It follows from  appendices \ref{symgroup} and \ref{aAapp} that the set $\aut$  of all such gauge transformations preserves the asymptotic
conditions (\ref{fallE}) and (\ref{fallA}).  From  appendix \ref{autgroupapp} it follows that $\aut$  is a group.  From (\ref{autAE}) one finds  the 
composition rule on $\aut$ has the following `semi-direct product' form:
\be
(g, \phi)  (g', \phi') =( g \phi_* g', \phi \circ \phi') \label{aap}.
\ee


\subsection{Group $\autEo$}\label{sec7B}
 In the Cartesian chart $\{x^\alpha \}$ there is a preferred  flat metric $\mathring{q}_{\alpha \beta} = \delta_{\alpha \beta}$ with isometries generated by (asymptotic)  rotations $R^\alpha_\beta \in SO(3)$ and translations $t^\alpha \in \reals^3$:
 \be
 x^\alpha \to R^\alpha_\beta x^\beta +t^\alpha. 
 \ee
From the analysis of References \cite{ttparity,mcparity} we expect that the group of these asymptotic isometries is represented
on the phase space variables $(A,E)$. We refer to this putative `symmetry' group as $\autEo$.

 Recall that the  fall-off conditions on the electric field $E$ are such that the  zeroth order value of $E$ 
at infinity is fixed  by the triad $\Eo^\alpha$  (\ref{fallE}).  
Whereas under asymptotic translations $\Eo^\alpha$ remains invariant, under asymptotic rotations it changes as
\be
\Eo^\alpha \to R^\alpha_\beta \Eo^\beta,
\ee
thus violating the fall-offs (\ref{fallE}).  If however the asymptotic rotation is accompanied by an internal $SU(2)$ rotation $\go$ satisfying
\be
R^\alpha_\beta \go \Eo^\beta \go^{-1} = \Eo^\alpha, \label{Rgg}
\ee
 then the  boundary condition (\ref{fallE}) is preserved.  
This  discussion  suggests that elements of $\autEo$ may be obtained by augmenting  those in $\aut$
with appropriate 
rotational and translational diffeomorphisms, together with gauge transformations of the type (\ref{Rgg}).
Indeed as  explained in the beginning of  section \ref{sec7} this is exactly what happens. 

We proceed as follows.
We define every element $a \in \autEo$ to be in correspondence with a pair $(g,\phi)$ 
where $g$ is a 
$C^{k-1}$ $SU(2)$
gauge transformation, $\phi$ a 
$C^k$ diffeomorphism, and $a= (g,\phi)$ is connected to identity (where
by connected to identity we mean that there exists a path in $\autEo$ subject to the conditions (i)- (iii)
of Footnote \ref{fnconn}).
Each such element has phase space action given by (\ref{autAE}).
As explained in the beginning of section \ref{sec7}, 
the considerations of Appendix \ref{symgroup} together with 
the classical phase space analysis \cite{ttparity,mcparity} and the considerations of Appendix \ref{infintesimalapp}
then imply the following  conditions 
on  $(g,\phi)$ as $r \rightarrow \infty$:
\ba
g(x) & = & \go+\frac{\lambda(\xh)}{r}+O(r^{-1-\e}) \label{fallg} ,\\
\phi(x)^{\alpha} & = & R^\alpha_\beta(\go)x^\beta+ t^\alpha +  s^\alpha(\xh)+O(r^{-\e}) ,\label{fallphi}
\ea
where $\lambda$ and $s^\alpha$ are as in (\ref{evenlam}), (\ref{odds}) and $R^\alpha_\beta(\go) \in SO(3)$ is given by the adjoint action of $\go$ where we identify  $su(2)$ with the $\reals^3$ of the Cartesian chart. In other words,  
it is the matrix satisfying,
\be
\go (x^\alpha \t_\alpha) \, \go^{-1} =R^\alpha_\beta(\go) x^\beta \t_\alpha \label{Rgo}.
\ee
Condition (\ref{Rgo}) can be seen to be equivalent to  (\ref{Rgg}) by noting that $\Eo^\alpha=\t_\alpha$ and $(R^{-1})^\alpha_\beta=R^\beta_\alpha$.

It follows from  Appendix \ref{autapp} that the set $\autEo$ of all such  transformations preserve the asymptotic
conditions (\ref{fallE}) and (\ref{fallA}) and that  $\autEo$ is a group.
From (\ref{autAE}) it follows that 
the group composition rule on $\aut$ has the same `semi-direct product' form (\ref{aap}) as in the $\aut$ case.
For later purposes we display the asymptotic form of a composed element $(g'',\phi''):= (g,\phi)(g',\phi')= (g \phi_* g', \phi \phi')$ (see appendix \ref{autgroupapp}):
 \ba
 g'' & = & \go \go' + (\even) r^{-1}+ O(r^{-1-\epsilon}) \label{gpp} \\
 \phi''^\alpha &= & R^\alpha_\beta(\go \go') x^\beta + R^\alpha_\beta(\go) t'^\beta +t^\alpha +(\odd)+O(r^{-\epsilon}). \label{phipp} 
 \ea

\subsection{$\autEo/\aut$} \label{sec7C}
From the above discussion it is clear that $\aut \subset \autEo$. We now show it is actually a normal subgroup. 
Let $a \in \autEo$ and $b \in \aut$. Since both groups are defined to be connected to identity, there exist paths  $a(s) \in \autEo$ and  $b(s) \in \aut$ with parameter $s \in [0,1]$ such that $a(0)=b(0)=\id$ and $a(1)=a, b(1) = b$. Using (\ref{gpp}), (\ref{phipp}) it is straightforward to verify that 
\be
c(s) :=a(s) b(s) a(s)^{-1} 
\ee
satisfies the fall-offs associated to $\aut$. Since $c(0) = \id$, $c(s)$ represents a path in $\aut$ so that $c(1)=a b a^{-1} \in \aut$. This shows normality of $\aut \subset \autEo$. We now describe the resulting quotient group $\autEo/\aut$.

Elements of  $\autEo/\aut$ are equivalence classes $[a], a\in \autEo$ where two elements $a,a'\in \autEo$ are equivalent if  $a a'^{-1} \in \aut$.  Let  us denote by
\be
\tt: \autEo \to \reals^3, \quad \gg:\autEo \to SU(2) ,
\ee
the maps that assign to $a \in \autEo$ the value of its asymptotic translation and $SU(2)$ rotation respectively. Thus for $a=(g,\phi)$ as in (\ref{fallg}) and (\ref{fallphi}), $\tt(a)=\vec{t}$ and $\gg(a)=\go$. We now show that 
\be
[a]=[a'] \iff \tt(a)=\tt(a'), \; \gg(a)= \gg(a'), \; \label{quotient}
\ee 
so that the space of equivalence classes may be identified with $ \reals^3 \times SU(2)$. We will later describe the product rule between equivalence classes.

From Eqs. (\ref{gpp}), (\ref{phipp}) one has that for any $c \in \aut, a \in \autEo$:  $\tt(a c)=\tt(a)$ and $\gg(a c) =\gg(a)$. This establishes the $\implies$ implication in (\ref{quotient}). To prove  the converse we will construct a particular family  of $\autEo$ elements with prescribed asymptotic values and show they exhaust all classes.

Any element $\go \in SU(2)$ can always be written as
\be
\go=e^{\theta \nh} ,
\ee
where $\nh \equiv \nh^i \t_i  \in S^2 \subset su(2)$ and  $\theta \in [0,2 \pi]$ (values greater than $2 \pi$ can be reached by flipping $\nh$, $e^{(2 \pi+\alpha) \nh}=e^{-(2 \pi-\alpha) \nh}$).  This parametrization is one-to-one except at the `poles' $\go=\pm \idtwo$, where  $\theta=0,2 \pi$ and $\nh \in S^2$ is undetermined.   Given $(\nh,\theta) \in S^2 \times [0,2 \pi]$ and $\vec{t} \in \reals^3$ define the diffeomorphisms:
\ba
\phi_{\vec{t}}^\alpha (x)& := & x^\alpha + t^\alpha f(r) ,\\
\phi_{(\nh,\theta)}^\alpha (x) &:=& R^\alpha_\beta(e^{\nh f(r) \theta}) x^\beta. \label{phinhth}
\ea
Here $f$ is a {\em semianalytic} $C^k$ interpolating function such that $f(r)=1$ for $r>r_2$ and $f(r)=0$ for $r<r_1$ (see  Eq. (A7) of \cite{mm2} for explicit example of such function). In addition we require that  $f'(r) \ll |\vec{t}|^{-1}$ so as to ensure invertibility of $\phi_{\vec{t}}^\alpha$.\footnote{ Let $\tilde{\phi}_{\vec{t}}: \reals^3 \to \reals^3$ be the map  induced by $\phi_{\vec{t}}$ on $\reals^3$ by setting  $\tilde{\phi}_{\vec{t}}|_{B_{r_1}}=\id$ where $B_{r_1}$ is the solid ball of radius $r_1$ in $\reals^3$.  Let $D_\alpha^\beta(x):= \partial_\beta \tilde{\phi}_{\vec{t}}^\alpha (x) = \delta_\beta^\alpha + t^\alpha \xh_\beta f'(r)$ be the differential of such map. Condition $f'(r) \ll |\vec{t}|^{-1}$ implies that $(D^{-1})^\alpha_\beta(x)$ exists and its coefficients are bounded as functions on $\reals^3$. A version of Hadamard theorem then tells  $\tilde{\phi}_{\vec{t}}$ is invertible in $\reals^3$ (see theorem 6.2.3 of \cite{kp2}). It then follows that $\tilde{\phi}^{-1}_{\vec{t}}|_{B_{r_1}}=\id$  so that it can be extended to $\Sigma$ to define an inverse for $\phi_{\vec{t}}$.}   $R^\alpha_\beta(e^{\nh f(r) \theta})$ is the ($r$-dependent) $SO(3)$ rotation associated to the ($r$-dependent) $SU(2)$ element $e^{\nh f(r) \theta}$ as in Eq. (\ref{Rgo}). The inverse of (\ref{phinhth}) is given by $\phi_{(-\nh,\theta)}$.  By construction $\phi_{\vec{t}}\;$ and $\phi_{(\nh,\theta)}$ are semianalytic diffeomorphisms that asymptote to a given translation and rotation respectively. Finally  define\footnote{To see that (\ref{defb}) is connected to identity consider the path $b(s):= (e^{s \theta \nh} , \phi_{s \vec{t}} \circ \phi_{(\nh, s \theta)} )$. Then $b(s)=\id$ and $b(1)=  b_{(\vec{t},\nh,\theta)}$.}
  \be
  b_{(\vec{t},\nh,\theta)} := (e^{\theta \nh} , \phi_{\vec{t}}  \circ \phi_{(\nh,\theta)} ) \in \autEo. \label{defb}
  \ee
  By construction we have  $\tt(b_{(\vec{t},\nh,\theta)})=\vec{t}$ and $\gg(b_{(\vec{t},\nh,\theta)})=e^{\theta \nh}$.   
  We now take care of a subtlety  due to the $SU(2)$ parametrization being used. At $\theta=0$,  $b_{(\vec{t},\nh,0)} =(\idtwo , \phi_{\vec{t}})$ is independent of $\nh$ and no issue arises. At $\theta=2 \pi$,   $b_{(\vec{t},\nh, 2 \pi)} = (-\idtwo,  \phi_{\vec{t}}  \circ \phi_{(\nh, 2 \pi)} )$ depends on $\nh$.  We thus need to show that all such elements are in the same class, that is:
\be 
b^{-1}_{(\vec{t},\nh', 2 \pi)} b_{(\vec{t},\nh, 2 \pi)} =( \idtwo, \phi^{-1}_{(\nh', 2 \pi)} \circ \phi_{(\nh, 2 \pi)})\in \aut, \quad \forall \nh,\nh' \in S^2. \label{eqnnp}
\ee
To show (\ref{eqnnp}) we  need to find a one parameter family of diffeomorphisms $\phi_s \in \aut$  such that $\phi_0=\id$ and  $\phi_1=\phi^{-1}_{(\nh', 2 \pi)} \circ \phi_{(\nh, 2 \pi)}$.  Consider a path  $\nh(s) \in S^2$ such that $\nh(0)=\nh$ and $\nh(1)=\nh'$. Then
\be
\phi_s:=\phi^{-1}_{(\nh(s), 2 \pi)} \circ \phi_{(\nh, 2 \pi)} \label{defphis}
\ee 
provides such path.   
  
We finally show that any element in $\autEo$ is equivalent to one of the elements (\ref{defb}). Let $a \in \autEo$ with asymptotic values $\vec{t}$ and $\go=e^{\theta \nh}$. We want to show that 
\be
c:=b_{(\vec{t},\nh,\theta)} a^{-1} \label{defc}
\ee
is in $\aut$. By construction $c$ has $\aut$-type fall-offs, so that all we  need to show is that it is connected to identity.   Let  $a(s) \in \autEo$ be  a path connecting identity at $s=0$ to $a$ at $s=1$. Set  $\vec{t}(s):=\tt(a(s))$ and $\go(s):=\gg(a(s))$. 
For reasons that will become clear in a moment, we choose a parametrization $s \mapsto a(s)$ such that each time $\go(s)$ goes through $-\idtwo$, it stays there for some closed interval $s\in [s_{i_1},s_{i_2}] \subset [0,1]$  (this can always be achieved by an appropriate reparametrization $s \to s'(s)$). A path $(\nh(s),\theta(s))$ that is continuous in $S^2 \times [0,2 \pi]$ and satisfies   $e^{\theta(s) \nh(s) }=\go(s)$ can then be constructed as follows:  For $\go(s) \neq -\idtwo$, the values of $(\nh(s),\theta(s))$ are uniquely defined and the path is continuous.  When $\go(s)$ goes through $-\idtwo$, $\nh$ may undergo a `flip' $\nh \to -\nh$.  For  $s \in [s_{i_1},s_{i_2}]$ where $\go(s)=-\idtwo$, we   specify $\nh(s)$ so as to implement this flip in a continuous fashion i.e. we hold $\theta$ fixed at $2\pi$ and continuously deform 
${\hat n }$ to $-{\hat n}$.    The corresponding path $b_{(\vec{t}(s),\nh(s),\theta(s))}$ is thus continuous in $\autEo$ (in the sense of footnote \ref{fnconn}) and
\be
c(s):=b_{(\vec{t}(s),\nh(s),\theta(s))} a^{-1}(s), \label{defcs}
\ee
provides the continuous path in $\aut$ connecting identity with $c$ (\ref{defc}). 

The above discussion establishes the one to one correspondence (\ref{quotient}) which allow us to  parametrize elements in the quotient space by pairs $(\vec{t},\go) \in  \reals^3 \times SU(2)$. We now discuss the product rule.  

Recall the product on the quotient space is defined by $[a] [a']:=[a a']$. Given $a,a' \in \autEo$ with asymptotic values $\vec{t},\go$ and $\vec{t'},\go'$, the asymptotic values of the product $a'':=a a'$ can be read-off from (\ref{gpp}), (\ref{phipp}) (see appendix \ref{autgroupapp}). From this we conclude the product rule on the quotient space is:
\be
(\vec{t},\go) (\vec{t'},\go') =  (R(\go) \vec{t'} +\vec{t}, \go \go'), \label{prodquot}
\ee
which corresponds to the semidirect product of translations with $SU(2)$ rotations, $\autEo/\aut=\reals^3 \rtimes SU(2)$. 

We conclude the section with some comments. An analogous analysis in metric variables (where the relevant groups are the diffeomorphism parts of $\aut$, $\autEo$) shows that there can be  two  possible quotient groups: The standard isometry group of Euclidean space $\reals^3 \rtimes SO(3)$, or its cover $\reals^3 \rtimes SU(2)$. Which one arises depends on the topology of the manifold \cite{fs}. For simple topologies, for instance $\Sigma=\reals^3$, the quotient group has only the $SO(3)$ factor. In \cite{fs} it is described how certain topologies, associated with an $SU(2)$ factor in the quotient group,  support odd-spin states in quantum theory. We will recover this result when studying asymptotic symmetries in quantum theory. In particular we will find that there are no `new' odd-spin states arising from the use of triad rather than metric variables. Eventhough in the triad formulation the quotient group has always an  $SU(2)$ rather than $SO(3)$ factor, the condition for existence of nontrivial odd-spin sates is still  the same as the one described in \cite{fs} for metric variables.

\subsection{Action of $\aut, \autEo$ on elementary phase space functions} \label{sec7D}

Since gauge transformations are a subset of asymptotic symmetry  transformations, the action of $\aut$ on elementary
phase space functions can be obtained by suitable restriction of that of $\autEo$ on these functions. Accordingly,
we focus on the action of the larger group $\autEo$. 
Given $(g,\phi)\equiv a \in \autEo$, we define its action on  phase space functions $F[A,E]$  by:   $(a \cdot F)[A,E]:=F[a^{-1}\cdot A, a^{-1}\cdot E]$ so that $(a b) \cdot F= a \cdot (b \cdot F)$.  
The elementary  phase space functions (\ref{hol}), (\ref{flux}), (\ref{baexp}) then transform as follows:
\ba
a \cdot h_e(A) &= & g^{-1}(\phi(f(e)))h_{\phi(e)}(A)g(\phi(b(e))) , \label{ahol}\\
a \cdot F_{S,f}(E)&  = &  F_{\phi(S),g \phi_* f g^{-1}}(E) , \label{aflux} \\
a \cdot \, \bE [A] & =& e^{i \alpha(a,\Eb) }\beta_{a \cdot \Eb}[A], \label{abaexp}
\ea
where $b(e),f(e)$ denote the beginning and end points of the edge $e$ and 
\be
\alpha(a,\Eb):= \int_\Sigma \tr[\phi_* (\Eb^a) g^{-1}\partial_a g] = \int_\Sigma \tr[(a \cdot \Eb^a) \partial_a g g^{-1}]. \label{alpha}
\ee
Note that the set of phase space functions above is not preserved by action of arbitrary elements of $\autEo$
because such elements do not, in general, preserve the semianalyticity property of the probes.
In order that the semianalyticity is preserved we further restrict attention to those elements of $\autEo$ which are
semianalytic. It is then straightforward to check that such elements form a group. Let us call this group
$\autEo({\rm sa})$.

Next, we show that the transformation laws (\ref{ahol})-(\ref{abaexp}) hold.
Since edges and surfaces are of compact support, Eqns. (\ref{ahol}) and (\ref{aflux}) follow by the same argument as in the case of compact $\Sigma$ \cite{mm1}.  To show (\ref{abaexp}) it will be convenient to study  separately the actions $g \cdot \bE \equiv (g,\id) \cdot \bE$ and  $\phi \cdot \bE \equiv (\id, \phi) \cdot \bE$, and later combine them by $(g,\phi) = (g,\id) \circ (\id, \phi)$.  Even though   asymptotic rotations do not admit such `splitting' in phase space, one can make sense of such splitting in the space of connections $\A$ and  of `barred' electric fields $\E$. 

First we notice that if $\rho$ is a scalar density such that $I:=\int_\Sigma \rho < \infty$ and $\phi$ a diffeomorphism of $\Sigma$ then $I= \int_\Sigma \phi_* \rho$. In particular for  $\rho= \tr[\Eb^a \phi^{-1}_*A_a]$ and $\phi\in \diff_\infty$ this implies
\be
\phi \cdot \bE = \beta_{\phi \cdot \Eb}. \label{phibeta}
\ee 
On the other hand, $g \cdot \bE$ is  given by the exponential of:
\be
\int_\Sigma \tr[\Eb^a( g^{-1}A_a g+ g^{-1} \partial_a g)] = \int_\Sigma \tr[ g  \Eb^a g^{-1} A_a] +\int_\Sigma \tr[\Eb^a g^{-1} \partial_a g].
\ee
Both integrals on the RHS are convergent since $g = \go+(\even) r^{-1}+O(r^{-1-\epsilon})$ and  $\partial_a g = (\odd) r^{-2}+O(r^{-2-\epsilon})$. We thus conclude that $g \cdot \bE = e^{i \alpha(g,\Eb)} \beta_{g \cdot \Eb}$.  Applying this result to $g \cdot \beta_{\phi \cdot \Eb}$ and using (\ref{phibeta}) we obtain (\ref{abaexp}) with $a=(g, \phi)$.

By construction, (\ref{ahol}), (\ref{aflux}) and (\ref{abaexp}) provide a representation of $\autEo({\rm sa})$ on holonomies, fluxes and background exponentials.

\section{Gauge transformations and Asymptotic Symmetries: Quantum Implementation.}\label{sec8}
As in compact space LQG, we restrict attention to  convenient subgroups of the automorphism groups $\autEo , \aut$.
The restricted subgroups are chosen so as to incorporate the property of semianalyticity in such a way that 
the ensuing quantum theory is tractable, elegant and structurally rich.
\footnote{See section \ref{sec9C} for further discussion on the role of semianalyticity.} 
As seen in the previous section, the  set of  holonomy-background exponential- flux functions is  invariant
under the action of the subgroup $\autEo({\rm sa})$ of semianalytic elements of $\autEo$.
Here we further restrict the set of these transformations by a  requirement of connection to identity.
More in detail, we define the quantum symmetry group  $\autqEo$ to be 
the set of elements of $\autEo({\rm sa})$ which are connected to identity by paths  in $\autEo({\rm sa})$.
\footnote{Note that any element in  $\autEo({\rm sa})$ {\em is}  connected to identity by paths in  $\autEo$. However
it is not clear if any of these paths lie completely in $\autEo({\rm sa})$. 
This is why we explicitly define $\autqEo$. 
Similar comments apply to the necessity of defining $\autq$.}
It is easy to see that $\autqEo$ is a subgroup of  $\autEo({\rm sa})$.
We also define $\aut({\rm sa})$ to be the set of semianalytic elements of $\aut$. It is easy to check that 
$\aut({\rm sa})$ is a subgroup of $\autEo({\rm sa})$. Finally we define the quantum gauge group $\autq$
 to be the set of elements
of $\aut({\rm sa})$ which are connected to identity by paths in $\aut({\rm sa})$.
It is straightforward to check that $\autq$ is a subgroup of  $\aut({\rm sa})$.
The groups $\autqEo$, $\autq$ are to be thought of as the quantum counterparts of the classical groups
$\autEo$,  $\aut$. Clearly we have that  $\autq$ is a subgroup of $\autqEo$. 
It is then straightforward to see that the arguments of section \ref{sec7C} apply unchanged 
if we replace $\autEo$,  $\aut$ by $\autqEo$, $\autq$. It follows that $\autq$ is a normal subgroup of 
$\autqEo$ and that the quotient group $\autqEo / \autq$ is, once again, 
$\reals^3 \rtimes SU(2)$. 

In section \ref{sec8A} we present a unitary action of $\autqEo$ on $\Hks$. 
This unitary action, restricted to $\autq$, is used to construct an $\autq$-invariant Hilbert space $\Haut$ in 
section \ref{sec8B}. Section \ref{sec8C} discusses the unitary action of asymptotic rotations and translations on 
$\Haut$. Supplementary material to this section is given in appendix \ref{appsec8}.
\subsection{Unitary representation of $\autqEo$ on the KS Hilbert space} \label{sec8A}
We wish to construct unitary operators $U(a) : \Hks \to \Hks, \; a \in \autqEo$ satisfying $U(a b)=U(a) U( b), \; a,b \in \autqEo$ and such that they implement the analogue of the transformations (\ref{ahol}), (\ref{aflux}), (\ref{abaexp}) for the corresponding operators (\ref{holhat}), (\ref{betahat}), (\ref{fluxhat}).  The natural candidate is:
\be
U(a) |s, E \ket := e^{i \alpha(a,E)} |U^{\lqg}(a) s, a \cdot E \ket ,\label{UasE}
\ee
with $U^{\lqg}(a) s$  the usual action of $\autq$ on spin networks and  $a\cdot E$ as in  (\ref{autAE}). In the case of compact  $\Sigma$ the phase $\alpha(a,E)$ in (\ref{UasE}) is given by  (\ref{alpha}). This strategy does not work in the present case since  the integral (\ref{alpha}) is not guaranteed to be convergent if we replace $\Eb^a$ by $E^a$. Fortunately the  divergent term is a total derivative that can be removed without affecting the composition property of the phases (Eq. (\ref{phasesaap}) below) required for the satisfaction of  $U(a b)=U(a) U( b)$. Given $a=(g,\phi)$, let
\be
\go:= \lim_{r \to \infty} g ,\label{defg0}
\ee
be the zeroth order term of $g$. Thus $\go=\idtwo$ for $a \in \autq$ and for asymptotic translations, whereas for asymptotic rotations $\go$ is an  internal rotation that ensures $\Eo$ is fixed under the action of $a$. We define the phase $\alpha(a,E)$ in (\ref{UasE})  by
\be
\alpha(a,E):=\int_\Sigma \rho(a,E) , \label{alpha2}
\ee
with
\be
\rho(a,E):= \tr[a \cdot E^a \, \partial_a g g^{-1}] - \partial_a\tr[E^a g \go^{-1}]. \label{rho}
\ee
In appendix \ref{finitealphaapp} we show that the integral (\ref{alpha2}) is finite. We  now verify   $U(a)$ is  a representation of $\autqEo$:
\be
U(a) U(a') = U(a a')   \; \forall a,a' \in \autqEo. \label{Uaap}
\ee
The action on a KS spinnet $|s,E \ket$ of the operators on each side of (\ref{Uaap}) is:
\ba
U(a) U(a') |s,E \ket &=& e^{i\alpha(a,a' \cdot E)}e^{i \alpha(a',E)} |U^{\lqg}(a a') s, a a' \cdot E \ket   \label{lhsaa}\\
U(a a') |s,E \ket &=& e^{i \alpha(a a',E)} |U^{\lqg}(a a') s, a a' \cdot E \ket\label{rhsaa},
\ea
where we used that $U^{\lqg}(a a') =U^{\lqg}(a)U^{\lqg}(a') $ and $a  \cdot (a' \cdot E)=(a a') \cdot E$.  In appendix \ref{phasecompapp} we show that
\be
\alpha(a a',E) =\alpha(a,a'\cdot{E})+\alpha(a',{E}) \label{phasesaap}
\ee
from which (\ref{Uaap}) follows.

It is straightforward to check that $U(a)$ preserves the inner product between KS spinnets. Finally, one can verify that $U(a)$ implements the transformations (\ref{ahol}), (\ref{aflux}), (\ref{abaexp}) for the operators (\ref{holhat}), (\ref{betahat}), (\ref{fluxhat}):
\ba
U(a) {\hat h}_e U(a)^\dagger &=& g^{-1}(\phi(f(e))){\hat h}_{\phi(e)}g(\phi(b(e))) \label{uahol} \\
U(a) {\hat F}_{S,f} U(a)^\dagger & = & {\hat F}_{\phi(S),g \phi_* f g^{-1}}  \label{uaflux} \\ 
U(a) {\hat \beta}_{\Eb} U(a)^\dagger &=&e^{i \alpha(a, \Eb)} {\hat \beta}_{a \cdot \Eb} , \label{uabe}
\ea
where $a=(g,\phi)$. Again, because of the compact support property of $e$ and $S$,  (\ref{uahol}) and (\ref{uaflux}) follow from the  same arguments given in the case of compact $\Sigma$ \cite{mm1}. Eq. (\ref{uabe}) can be shown by the same steps given in Eq. (C5) of  \cite{mm1} thanks to (\ref{phasesaap}) and the fact that (\ref{alpha2}) satisfies (see appendix \ref{finitealphaapp}):
\be
\alpha(a,E+\Eb)=\alpha(a,E)+\alpha(a,\Eb), \label{alphaEEb}
\ee
 with $\alpha(a,\Eb)$ as in (\ref{alpha}).

\subsection{Group averaging and $\autq$ invariant Hilbert space $\Haut$} \label{sec8B}
\subsubsection{Setup}
In this section we denote KS spinnets  labels as $\psi = (s,E)$, and the action of $\autq,\autqEo$ on these labels as $a \cdot \psi$.  Let $\D$ be the dense subspace of $\Hks$  given by the finite linear span of KS spinnets.  Let   
\be
\Ph_\psi  :=  \{ a \in \autq  :   U(a) | \psi \ket \propto | \psi \ket  \} , \quad \sym_\psi  :=  \{ a \in \autq  :   U(a) | \psi \ket=| \psi \ket  \} \subset \Ph_\psi .
\ee
We start with the candidate for a group averaging map as in \cite{mm1}:
\be
\eta(|\psi \ket)= \left\{
\begin{array}{lll}
\quad \quad 0 & \text{if }  & \sym_{\psi} \subsetneq \Ph_{\psi} \\
& &\\
\eta_{[\psi]}\displaystyle\sum_{\c \in \autq/\sym_\psi} (U(a_{\c}) |\psi \ket )^{\dagger}  &\text{if } &\sym_{\psi} = \Ph_{\psi}  ,
\end{array} \right. 
\label{etaks}
\ee
where the sum is over the set of right  cosets of  $\autq$ by $\sym_\psi$, that is, 
 the set of distinct $\sym_\psi$-orbits $a \,  \sym_\psi \subset \autq$ for all $a \in \autq$.  $a_{\c}$ is a choice of 
representative on each orbit $\c$ and $\eta_{[\psi]}$ are as yet unspecified  positive constants that depend only on the $\autq$-orbit $[\psi]$ of $\psi$, that is $\eta_{[a \cdot \psi]}=\eta_{[\psi]} \; \forall a \in \autq$. From the same arguments as in \cite{mm1}, (\ref{etaks}) gives a well defined antilinear map   $\eta : \D \to \D'$   ($\D'$ the algebraic dual of $\D$), satisfying 
\be
\eta( U(a) |\psi \ket)  =\eta(|\psi \ket) \; \forall a \in \autq ,
\ee
\be
\eta(|\psi_1 \ket)[|\psi_2 \ket] = \overline{\eta(|\psi_2 \ket)[|\psi_1 \ket]}  \; ,  \quad \eta(|\psi_1 \ket)[|\psi_1 \ket] \geq  0 \;  \quad \forall \, |\psi_1 \ket, |\psi_2 \ket \in \D .
\ee
The last property allows one to define the inner product $\bra \eta(|\psi_1 \ket)|\eta(|\psi_2 \ket) \ket_{\eta}:= \eta(|\psi_1 \ket)[|\psi_2 \ket]$ that  is used in the definition of the  $\autq$-invariant Hilbert space $\Haut$ \cite{alm2t,mm1}.

There remains to be imposed a third requirement on $\eta$ that allows to define `observables' in $\Haut$ from operators in $\Hks$. Let  $\O$ be the set of operators on $\Hks$ such that $O \in \O$ satisfies (i)  both $O$ and $O^\dagger$ are defined on $\D$ and their action preserves $\D$; (ii) $U(a)O=O U(a) \; \forall a \in \aut$.  Following \cite{alm2t} we  require:
\be
\eta(|\psi_1 \ket)[O |\psi_2 \ket]=\eta(O^\dagger |\psi_1 \ket)[|\psi_2 \ket]  \quad \forall \, \psi_1,\psi_2 \in \D, \; \forall \,  O \in \O. \label{prop3}
\ee
This condition ensures  that all such $O$ obey, as  operators in  $\Haut$, the same  adjointness relations as operators in $\Hks$. In particular, if $O$ is unitary as operator in $\Hks$, it  guarantees $O$ is unitary as an operator in  $\Haut$.  As in \cite{mm1}, we shall see that  (\ref{prop3}) determines the coefficients $\eta_{[\psi]}$ within a `superselection sector' (see below).

It is however not obvious that condition (\ref{prop3})  will guarantee a unitary action of asymptotic rotations and translations, $U(b), b \in \autqEo$, since these operators do not belong  to $\O$ for the following reason: whereas  $\O$ defined above consists of operators that commute `strongly' with the gauge group, asymptotic rotations and translations commute `weakly' with the gauge group. That is, instead of satisfying  (ii) in the definition of $\O$, they satisfy: (ii)': Given $a \in \autq$,   $O U(a)=U(a')O$ for some $a' \in \autq$. Whereas (i) and (ii)'  are sufficient conditions to  be able to define  $O$ as an operator on $\Haut$, it is a priori not clear that such  operator will satisfy the correct adjointess properties. In subsection 
\ref{sec8C} we will verify that rotations and translation do act unitarily in $\Haut$.

\subsubsection{Superselection sectors}

By `superselection sector' we will mean a subspace of $\Haut$ such that any two elements in the subspace can be mapped into each other by an operator in $\O$ or by some $U(b), b \in \autqEo$.\footnote{The inclusion of $U(b), b \in \autqEo$ in this definition may be redundant. 
It may be the case that any  superselection sector defined only with respect to $\O$ 
is automatically invariant under the action of  $U(b), b \in \autqEo$.}
If $\eta(|\psi \ket)$ and $\eta(|\psi' \ket)$ lie in the same (different) superselection sector, we shall  say that $|\psi \ket$ and $|\psi' \ket$ lie in the same (different) superselection sector. 

Given an asymptotically flat $E^a$, we define its rank sets as in \cite{mm1} by
\ba
V^{E}_0 & := &\{ x \in \Sigma \; : \; \rk(E)=0\} ,\label{V0}\\
V^{E}_1 & := &\{ x \in \Sigma \; : \; \rk(E)=1\}, \label{V1}\\
V^{E}_2 & := &\{ x \in \Sigma \; : \; \rk(E) \geq 2 \}. \label{V2} 
\ea
In appendix \ref{rsapp} we show the rank sets can be decomposed into finite union of semianalytic submanifolds. $\diff$-classes of these sets can be used as partial labels for superselection sectors as follows.

By the same arguments given in \cite{mm1}, it follows that
\ba
\bra s, E | O |s',  E' \ket \neq 0 \implies & i)  &\; \Vo^E_0=\Vo^{E'}_0 ,   \quad \Vb^{E}_2=\Vb^{E'}_2   \label{cond1}\\
&  ii) &\; \dim(V^E_{n} \cap (\cup_{n' \neq n}  V^{E'}_{n'})) <3 , \; n=0,1,2   \label{cond2}  \\
&  iii) & \gt(s) \cap \Vo^E_0= \gt(s') \cap \Vo^E_0, \label{cond3}
\ea
where  $\gt(s), \gt(s')$ are the graphs of the spin networks, and dim denotes the dimension of the sets in (ii) (which are also composed of finite union of semianalytic submanifolds, see appendix \ref{rsapp}). As in \cite{mm1},  these conditions imply that  for any pair of  KS spinnets $| s, E \ket$ and $|  E' , s' \ket$ lying in a same superselection sector there  exists a diffeomorphism $\phi$ such that (\ref{cond1}), (\ref{cond2}) and (\ref{cond3}) hold with $\phi(V^{E'}_n)$ in place of $V^{E'}_n$ and $\phi(s')$ in place of $s'$.

\subsubsection{Group averaging in the absence of rank 1 backgrounds} \label{sec8B3}
We  now focus on a particular class of superselection sectors, and discuss how the coefficients $\eta_{[\psi]}$ are determined. The type of states we consider are the analogue of those described in section 7 of \cite{mm1}:  $|\psi \ket = |s,E \ket$ such that: (1) $V^{E}_1$ is of zero measure, (2) the only infinitesimal  $\autq$-symmetries of $E^a$ are those associated to $V^{E}_0$, (3) $s$ is an $SU(2)$ gauge invariant spin network, (4) $\Ph_\psi = \sym_\psi$.  Conditions (1) to (4) give a set of consistent restrictions in the sense described in \cite{mm1}.\footnote{Condition (3) should actually be improved upon to allow for gauge variant spin networks, see discussion of section 9 B in \cite{mm1}.}  Next, we define 
\be
\sym^0_\psi \equiv \sym^{0}_{(s,E)}:= \{ a \in \autq : a|_{\Vb^E_2}= \id , \; a(\et)=\et \quad \forall \; \et \in s \}, 
\ee
where $\Vb^E_2$ is the closure of $V^E_2$, $\et \in s$ denotes the edges in $s$ (as 1-dimensional manifolds) and  $a(\et) \equiv (g,\phi)(\et):=\phi(\et)$.  It is easy to verify that $\sym^0_\psi$  is a normal subgroup of $\sym_\psi$, and as in \cite{mm1} we  assume the corresponding quotient group
\be
D_\psi := \sym_\psi/ \sym^0_\psi \label{Dpsi}
\ee
is finite, this finiteness being expected from that of the group of allowed edge permutations of $\gamma(s)$ and of the discrete symmetries of the background. By the same arguments given in \cite{mm1}, condition (\ref{prop3}) is then satisfied if and only if 
\be
\eta_{[\psi]}= C |D_{\psi}| \label{etaD}
\ee
for some constant $C>0$. Thus, the ambiguity in the coefficients $\eta_{[\psi]}$ is reduced to one constant per each superselection sector.

\subsection{Asymptotic rotations and translations on $\Haut$}  \label{sec8C}
\subsubsection{Unitary action}
We now discuss the analogue of condition (\ref{prop3}) for asymptotic rotations and translations:\footnote{If (\ref{prop3p}) holds for some $b$, it automatically holds for  $b'=b a$ and $b''=a b$, $\forall a \in \autq$, so that the set of independent conditions in (\ref{prop3p}) are parametrized by $\autqEo/\autq$. Similarly,  one has that  $U(b)$, $U(a b)$ and  $U(b a)$, define the same operator on $\Haut \; \forall a \in \autq$.}
\be
\eta(|\psi_1 \ket)[U(b) |\psi_2 \ket]=\eta(U^\dagger(b) |\psi_1 \ket)[|\psi_2 \ket] \quad \forall \, |\psi_1 \ket ,|\psi_2\ket  \in \D, \; \forall \,  b \in \autqEo  .\label{prop3p}
\ee
We first note that the coefficients $\eta_{[\psi]}$ determined in the last section by Eq. (\ref{etaD}) satisfy
\be
\eta_{[b \cdot \psi]}= \eta_{[\psi]}, \; \forall b \in \autqEo.  \label{etabpsi}
\ee
The reason is the same as to why (\ref{etaD}) is $\autq$ invariant:  There is a one to one correspondence between elements of  $\sym_\psi/ \sym^0_\psi$ and elements of $\sym_{b\cdot \psi}/ \sym^0_{b\cdot \psi}= b\, \sym_\psi b^{-1} /b \sym^0_\psi b^{-1}$ given by: $[s] = s \sym^0_\psi \in \sym_\psi/ \sym^0_\psi \iff b [s] b^{-1} \in \sym_{b\cdot \psi}/ \sym^0_{b\cdot \psi}$, so that $|D_\psi|= |D_{b \cdot \psi}|$.

We now show that Eq. (\ref{etabpsi}) implies Eq. (\ref{prop3p}).  Consider the following equivalent form of Eq. (\ref{prop3p}):
\be
\eta(U(b) |\psi \ket) ( U(b) |\phi \ket) = \eta(|\psi \ket)(|\phi \ket) \quad \forall \, |\psi \ket, |\phi \ket  \in \D, \; \forall \,  b \in \autqEo , \label{prop3pp}
\ee
Using (\ref{etaks}), the left hand side of (\ref{prop3pp}) can be written as:
\be
\eta(U(b) |\psi \ket) (U(b) |\phi \ket)= \eta_{[b \cdot \psi]}\displaystyle\sum_{\c \in \autq/\sym_{b \cdot \psi}} \bra \psi| U^\dagger(b^{-1} a_{\c} b) | \phi \ket. \label{Ubeta}
\ee
Next, we note that $\sym_{b \cdot \psi}= b\, \sym_\psi b^{-1} $ and that there is a one to one correspondence between $\autq/\sym_\psi$ and $\autq/(b\, \sym_\psi b^{-1})$ given by
\be
\c \in \autq/(b\, \sym_\psi b^{-1}) \iff b^{-1} \c b \in  \autq/\sym_\psi .
\ee
Further, $b^{-1} a_{\c} b \in b^{-1} \c b $. Thus, the sum on the right hand side of (\ref{Ubeta}) takes the same form as the sum in the definition of  $\eta[|\psi \ket](|\phi\ket)$. Using (\ref{etabpsi}), equation (\ref{prop3pp}) follows.

Recall from section \ref{sec7C} that  $\autqEo/\autq =\reals^3 \rtimes SU(2)$.  Let $b_{\vec{t}},b_{\go} \in \autqEo$  denote representatives  of a translation $\vec{t} \in \reals^3$ and a rotation $\go \in SU(2)$. The unitary action of $\vec{t},\vec{\go}$ on  $ |\psi ) := \eta( |\psi \ket) \in \Haut$ can then be written as:
\be
U(\vec{t})  |\psi ):=  \eta( U(b_{\vec{t}}) | \psi \ket), \quad U(\go) |\psi )  :=  \eta( U(b_{\go}) | \psi \ket).
\ee
  
\subsubsection{Friedman-Sorkin `Spin 1/2 from Gravity' States}
In \cite{fs}, Friedman and Sorkin (FS) provide a general argument for the appearance of odd-spin states in a quantum theory of the gravitational field.  The argument may be summarized as follows. Consider a theory of quantum gravity with states given by wavefunctions of the 3-metric on the spatial slice. The diffeomorphism constraint implies the physical wavefunctions are invariant under the action of diffeomorphisms that are trivial at infinity and connected to identity. Let us denote by  `gauge' such diffeomorphisms.
Let $\phifs$ be a diffeomorphism  that interpolates between a $2 \pi$ rotation at infinity with identity in the interior (for instance take $\phifs := \phi_{(\t_3,2\pi)}$ as in Eq. (\ref{phinhth})).
Doubly connectedness of $SO(3)$ implies that $\phifs \circ \phifs $ is guage (an argument for this follows from the  discussion leading to Eq. (\ref{defphis})). Thus on the physical space: 
\be
\hphifs \, \hphifs=\id. \label{phifs2}
\ee
We now ask if  $\phifs$ is gauge, i.e. if it can be continuosly deformed to identity in such a way that it stays trivial at infinity.  It turns out that the answer depends on the topology of the manifold $\Sigma$. The  characterization of all possible topologies for which $\phifs$ is not gauge is available in the mathematics literature and described in \cite{fs}. Manifolds $\Sigma$ with such topologies are termed  `spinorial'.  For spinorial manifolds, there are states $\psi$ such that 
\be
\psi':=\hphifs \psi \neq \psi .
\ee
It then follows that the state $\psi'-\psi$ exhibits odd spin behaviour under the asymptotic $2 \pi$ rotation:
\be
\hphifs (\psi'-\psi)=-(\psi'-\psi).
\ee

The adaptation of these ideas to our model are straightforward. We only need to keep track of additional $SU(2)$ rotations.\footnote{Another difference which does not alter the argument is the presence of odd supertranslations which are non-trivial at infinity but still  `gauge'.} In our language, odd spin states can arise if $-\idtwo \in SU(2) \subset \autqEo/\autq$ acts non-trivially in $\Haut$. Let
\be
b_{-\idtwo} := (-\idtwo, \phifs), \label{bmo}
\ee
be a representative of the type given in section \ref{sec7C}.  If $\Sigma$ is not spinorial, we can deform $b_{-\idtwo}$ through $\aut$ by deforming the diffeomorphisms while keeping the $-\idtwo$ factor fixed. In this way one then obtains a new representative  $b'_{-\idtwo}:=(-\idtwo,\id)$. It is easy to see that $U(b'_{-\idtwo})| s,E \ket= | s,E \ket$ for all KS spinnets. This implies $U(-\idtwo)= \id$ on  $\Haut$ so that no odd spin states arise.  

For spinorial $\Sigma$ the argument above does not apply and  odd spin states may be constructed as in the above FS argument, for instance by considering states of the form $|(\phifs)_*\Eb \ket -| \Eb \ket$.  Note  that in this case $(-\idtwo,\id) \notin \autqEo$ as there is no path in $\autqEo$ connecting this element to identity. 
In particular  $(-\idtwo,\id)$ is not a representative of $-\idtwo \in SU(2) \subset \autqEo/\autq$ as it was in  the non-spinorial manifold case.

\subsubsection{Comments on eigenstates of linear and angular momentum}

We conclude the section with some comments regarding 
the possibility of  finding states with definite linear or angular momentum.   
A first example is given by the case where $\Sigma=\reals^3$ and $|\psi_0 \ket: =|0,\Eo \ket$ where $0$ denotes the trivial spin network and $\Eo^a$ the flat triad on $\reals^3$.  The corresponding $\aut$-invariant state $\eta(|\psi_0 \ket)$ is left invariant under $U(\go), U(\vec{t}), \; \forall \, \go \in SU(2), \vec{t} \in \reals^3$ and hence  $\eta(|\psi_0 \ket)$  is a state with zero linear and angular momenta.  A similar state with $\Eo^a$ replaced by a spherically symmetric triad (but not translation invariant) yields a state with zero angular moment and indefinite linear momentum.
We have not succeeded in finding examples of normalizable states in $\Haut$ with  definite nonzero momenta. 
On the other hand, while we do not do so here, {\em distributional} states 
with definite linear/angular momenta can  be constructed 
by group averaging over the appropriately weighted action of asymptotic translations/rotations  
on suitable states in $\Haut$.

\section{Conclusions}\label{sec9}
\subsection{Summary of results}
In this work we developed a quantum kinematics for asymptotically flat gravity which suitably addresses the 
classical asymptotic conditions on the triad and connection variables of the theory.
The quantum  kinematics supports a unitary representation of the gauge group $\autq$ 
(suitably restricted by the requirement
of semianalyticity  and connectedness to identity) 
of internal rotations which asymptote to 
identity  and  spatial diffeomorphisms which asymptote to odd supertranslations.
The kinematics also supports a unitary representation of the larger group $\autqEo$  of asymptotic 
symmetries (also subject to semianalyticity and connectedness to identity restrictions) 
which, modulo gauge,  asymptote to non-trivial rotations and translations 
at spatial infinity.
Similar to the case of  compact space LQG, 
solutions to the Gauss Law and spatial diffeomorphism constraints are identified with 
states which are invariant under the action of $\autq$.
We constructed a  large sector of the Hilbert space of such gauge invariant states using group averaging methods.
As in the case of compact space LQG,  this sector is  superselected with respect 
to the action of a  certain set of gauge  invariant  
observables.\footnote{We do not expect that these  `kinematical' superselection sectors will be preserved by the Hamiltonian constraint.}
We showed that the 
group of asymptotic rotations and translations 
(obtained as the quotient of 
the symmetry group $\autqEo$  by the gauge group $\autq$)  is implemented unitarily on this sector. 
Finally, following Friedman and Sorkin \cite{fs}
we showed that for $\Sigma$ with  appropriate topology, 
the gauge invariant Hilbert space contains states that change by a sign under a $2 \pi $ rotation. 

\subsection{Remarks and caveats of a technical nature} \label{sec9B}
While, in  sections 
\ref{sec3}-\ref{sec6}, we used  asymptotic boundary conditions 
corresponding to finite differentiability  versions of the    
so called `parity' conditions \cite{ttparity,mcparity,beigom},  our considerations 
in these  sections 
are expected to 
go through for any choice of asymptotic conditions which implement asymptotic flatness in such a way that the 
phase space is realized as a cotangent bundle over the configuration space of connections. The subsequent developments in sections \ref{sec7} and \ref{sec8}
do, however, depend on the detailed choice of these parity conditions. 
More in detail, the demonstration of the finiteness of 
the phase factors (\ref{alpha}) and (\ref{alpha2}) as well as the  composition property (\ref{phasesaap}) 
depend crucially on the parity conditions and would have to be checked explicitly for any other choice  of 
asymptotic behaviour. 

Regarding the restriction to semianalytic fields: The KS background electric field labels have the double condition of satisfying (\ref{fallE}) \emph{and} being semianalytic. In appendix \ref{sphereapp} we show there exist a rich family of functions satisfying both the required fall-offs and semianaliticity conditions. Similarly, the group elements of $\autq$ and $\autqEo$  are required to be semianalytic in addition to possess the fall-offs described in sections \ref{sec7A} and \ref{sec7B}. Explicit examples of such  (connected to identity) elements with nontrivial asymptotic translations and rotations appear in section \ref{sec7C}. It is possible to construct by the same methods examples of   $\autq$ elements with non-trivial asymptotic supertranslations. 

 Our final remark is concerned with a technicality in our treatment of group averaging
in  section \ref{sec8B3}: Similar to Reference \cite{mm1} we assumed finiteness of the group $D_{\psi}$ defined in Eq. (\ref{Dpsi}).

\subsection{The role of semianalytic structures}\label{sec9C}

Let us first discuss the role of  semianalytic structures in compact space LQG.
One may attempt to view compact space LQG  (on a $C^k$ Cauchy slice admitting a semianalytic
structure) as  a quantization of a Hamiltonian
formulation of gravity with $C^{k-1}$ triads and $C^{k-2}$ connections  defined on this slice.

In this view, the classical holonomy--flux algebra may be obtained by smearing electric fields along semianalytic $C^k$ surfaces $S$ and semianalytic $C^{k-1}$ functions $f$ thereon; and by smearing the  connections along  semianalytic $C^k$ edges. This algebra is invariant under semianalytic $C^{k-1}$ $SU(2)$ rotations and semianalytic $C^k$ diffeomorphisms.  Therefore the kinematical  gauge group of LQG  is taken to be  the semidirect product of semianalytic $SU(2)$ rotations with semianalytic  diffeomorphisms. 
In the process, we see that the generic finite differentiability  gauge transformations of the classical theory have been restricted to be semianalytic. Nevertheless, since the quantum theory is so elegantly  
formulated for the semianalytic category and since the quantum excitations which are invariant under both the
kinematical gauge transformations as well as the (putative) action of the Hamiltonian constraint are expected to
be very different from the kinematic ones, it is a good strategy to explore where this relatively mild restriction
leads us to.  Indeed, our view is that (i) this strategy is likely to be {\em too conservative} to construct the final
quantum theory but  (ii) that nevertheless pursuing this strategy will provide enough information
to motivate more radical but educated departures therefrom so as to construct fundamental quantum spacetime 
structures.  Indeed, in  the final picture one would perhaps expect that the spacetime manifold only emerges at scales
much larger than the Planck scale. Similar remarks apply to the KS representation for compact spaces. 
Therefore our view is that one should not worry
too much about the exact class of differential structures one uses in quantum theory relative to classical theory.

At this point it is necessary to make the following caveat regarding classical Hamiltonian theory in the $C^k$ setting.
The infinitesimal transformations generated by the diffeomorphism and Hamiltonian constraints involve
single spatial derivatives of the fields so that the infinitesimal variations of the fields suffer a drop 
of degree of differentiability. Consequently, these transformations do not 
leave the phase space of finite differentiability  fields invariant. While it may well be the case that the consideration of {\em finite}
transformations remedies this defect for flows corresponding to a combination of 
arbitrary shifts and {\em non-vanishing} lapses,\footnote{If the lapse vanishes, standard results \cite{lang} indicate that the diffeomorphism constraint
smeared with a $C^{k-1}$ shift 
generates
$C^{k-1}$ diffeomorphisms. Such diffeomorphisms in general map a $C^{k-1}$ tensor field to a $C^{k-2}$ one, so that 
phase space is not preserved.} 
we are not aware of a demonstration of this property. However, with our view
on classical structures emerging  from quantum ones at large distance scales, we are not unduly concerned 
if such a property holds; 
it is enough for us that the quantum theory is based
on structures which are similar but not necessarily identical to the classical case. Since
the problems discussed in this paragraph disappear for $C^{\infty}$ fields, in our view, it is an acceptable
strategy 
to (a) restrict the $C^{\infty}$  category to the semianalytic one 
so as to construct an algebra which admits
an elegant quantization for which  a large class of diffeomorphisms 
(namely the semianalytic ones) act unitarily, 
(b) to define kinematically gauge invariant states as those invariant under suitable
semianalytic transformations and (c) to then attempt to impose the Hamiltonian constraint on this space.

Relative to the compact case, in the asymptotically flat case we need to specify not only
the differential structure of the fields considered but also their asymptotic behaviour.
Once again for the reasons discussed above, the simplest consistent Hamiltonian formulations
are defined in the $C^{\infty}$ setting. The fall offs we would like to use are those
provided by the parity conditions of References \cite{beigom,mcparity}. As indicated in section \ref{sec2A}
the subleading terms in these fall offs are only defineable in the $C^{\infty}$ setting.
Nevertheless, guided by the role of semianalytic structures in quantum theory in the compact space setting, we would
like to base our considerations on the $C^k$ semianalytic differential structure while still making 
contact with the parity conditions in some way. For the phase space variables, 
this leads to the conditions (\ref{fallE}), (\ref{fallA})
which are closely related to the parity conditions in that while the leading order terms look 
identical to those for the $C^{\infty}$ case, 
the fall offs of the subleading terms are restricted by their finite differentiability.
Similarly, the asymptotic behaviour of the gauge and symmetry transformations 
(as described in sections \ref{sec7A} and \ref{sec7B})
is closely related to their $C^{\infty}$ counterparts in appendix \ref{infintesimalapp}.

\subsection{Future work}
Our considerations are based on the KS representation which is a generalization of standard compact space LQG
\cite{kos,hanno,me,mm1,mm2} to account for non-trivial  `background' spatial geometries, the LQG
representation being associated with a `zero background spatial geometry' state corresponding to a vanishing 
(and hence, degenerate) triad field. Two distinct research directions suggest themselves depending on  
whether one views the KS representation as fundamental or whether one views the KS representation as 
an effective description of fundamental LQG excitations. If one views the KS representation as fundamental,
the next step would be to develop an understanding of the Hamiltonian constraint operator in this representation.
If one views the LQG representation as fundamental then it would be of great interest 
to use the KS kinematics developed here as a template for the development of asymptotically flat LQG, building on 
the earlier work of references \cite{arnsdorf,itp}.

Irrespective of which line of thought one pursues, key physical issues can only be addressed once one has
an adequate understanding of the Hamiltonian constraint (in the LQG context see
\cite{ttqsd,aloklqg,aloku11,aloku12,meu11,meu12}; in the KS context see \cite{sandipan1,sandipan2}). 
In our view, the two most important and crucial issues
are the existence of (A)  a quantum positive energy theorem 
\footnote{See Reference \cite{leepositive} for first steps, albeit in the context of self dual rather than 
real connection variables.}
and (B) a unitary 
representation of asymptotic Poincare transformations (including time translations and boosts, both of which
require an understanding of the Hamiltonian constraint operator). 
Indeed, any understanding of these issues, even without a full understanding of the Hamiltonian constraint 
itself,  would be very exciting and would go a long way in showing the physical  viability of the 
theory.\\

\textbf{Acknowledgements:} 
We are indebted to Fernando Barbero and Hanno Sahlmann for their comments on  a draft version of this work. MC thanks Alok Laddha for helpful discussions. 

\appendix

\section{Semianalyticity of functions on $\chi_0 (U_0)$ from that of functions on $S^2$.} \label{sphereapp}

$\chi_0 (U_0)$ can be given an analytic structure compatible with the preferred Cartesian chart 
$x^{\alpha}$. In (1)- (3a) below we shall treat $\chi_0 (U_0)$ as an analytic manifold with this analytic 
structure. \\

\noindent (1) Let $S^2$ denote the topological 2 sphere. 
We endow $S^2$ with an analytic atlas and denote the resulting analytic manifold by $S^2_w$ 
so that $S^2$ and $S^2_w$ are identical as point sets and topologies.
Let $(u,v)_{N,S}$ be the (analytic) stereographic
coordinate charts associated with  $S^2-N, S^2-S$ where $N,S$ are the North and South Poles of $S^2$.
Define the natural analytic map $g:\chi_0 (U_0)\rightarrow S^2_w$ in the charts $x^{\alpha}$ and $(u,v)_{N, S}$
as:
\be 
(u,v)_N = \left(\frac{x^1}{1-x^3},\frac{x^2}{1-x^3} \right),  \quad
(u,v)_S = \left(\frac{x^1}{1+x^3},-\frac{x^2}{1+x^3} \right).
\ee
This implies that given any analytic chart $x^A$ on $S^2_w$, we have that:\\
(i) $g^A(x^{\alpha})\equiv x^A(x^{\alpha})$ are analytic functions of $x^{\alpha}$.\\
(ii) $\left(r(x^{\alpha}) = \sqrt{x^{\alpha}x_{\alpha}}, x^A(x^{\alpha})\right)$ are analytic coordinates on $\chi_0(U_0)$.
\\

\noindent (2) Consider a maximal $C^n$ semianalytic atlas on $S^2$ which contains the analytic charts
of (1). This atlas endows $S^2$ with the structure of a $C^n$ semianalytic manifold. We refer to it as $S^2_n$
so that as point sets and topologies $S^2=S^2_w=S^2_n$.
It is easy to check that the function $g$ is a semianalytic function from $\chi_0 (U_0)$ to $S^2_n$.
\\

\noindent (3)(a) Let $f:S^2_n \rightarrow \reals$ be a $C^n$  semianalytic  function. Define 
$h=f\circ g : \chi_0(U_0)\rightarrow \reals$. It is easy to check that $h$ is a semianalytic function using the 
fact that 
$f$ and $g$ are semianalytic functions.
(This can be checked by using semianalytic charts on $S^2_n$ , analytic ones on $\chi_0(U_0)$ 
and then applying Proposition A.4  and Definition A.11 of \cite{lost}).

In terms of the preferred Cartesian chart on $\chi_0 (U_0)$ and any analytic
chart $x^A$ on $S^2_w$ (which, by (2) is also a semianalytic chart on $S^2_n$),  we may write
$h\equiv f (x^A(x^{\alpha}))$.
Using the expression of Cartesian derivatives in terms of spherical ones in (4)  below, it follows that $h$ is $C^n$.
Hence $h$ is a semianalytic function from  $\chi_0 (U_0)$ as an analytic manifold to $S^2_n$.
\\
\noindent (3)(b) Let us restore the semianalytic $C^k$ structure which $\chi_0 (U_0)$ inherits from $\Sigma$.
It immediately follows from (3a) that for $n\leq k$,  $h$ is $C^n$ semianalytic function from 
$\chi_0 (U_0) \subset \Sigma$ to $S^2_n$ and that for $n>k$, $h$ is a $C^k$ semianalyic function from 
$\chi_0 (U_0) \subset \Sigma$.
\\

\noindent (4)
Start with:
\be
\left(\frac{\partial}{\partial x^\alpha} \right)^a = \qo^{ab}\partial_b(x_\alpha) \label{0}.
\ee
In spherical $(r,x^A)$ coordinates the flat metric $\qo_{ab}$ takes the form:
\be
ds^2=dr^2 + r^2 q_{AB} d x^A dx^B
\ee
where $q_{AB}$ is the unit sphere metric. Then in spherical coordinates (\ref{0}) becomes:
\ba
\frac{\partial}{\partial x^\alpha}  & = & \partial_r(x_\alpha) \partial_r +   r^{-2} q^{AB} \partial_B(x_\alpha) \partial_A. \label{1} \\
& = & \xh_\alpha \partial_r + r^{-1} q^{AB}  \partial_B(\xh_\alpha) \partial_A. \label{partialsph0}
\ea

\section{Electric field boundary conditions in terms of fluxes} \label{fluxapp}
In this appendix we present the first steps in coding the classical boundary conditions (\ref{fallE})  in terms of  limiting behaviour of fluxes 
associated with families of surfaces obtained by translations `towards infinity'  of a suitably chosen surface.
The surfaces considered are all in the asymptotic region $\Sigma \setminus K$ and in the rest of this section we shall
work in the preferred Cartesian chart $(x_1,x_2,x_3)\equiv {\vec x}$ in this region. Unit vectors are normalized
with respect to the coordinate metric. We denote the unit vector along the $I$th direction ($I=1,2,3$) 
by ${\hat I}$ and the unit vector along ${\vec x}$ by ${\hat x}$ so that 
${\hat x}= \frac{{\vec x}}{|{\vec x}|}$ with $|{\vec x}| = \sqrt{x_1^2+ x_2^2 +x_3^2}$.

For each $I=1,2,3$, let $S({\vec x_0}, {I})$ be the planar disc of radius $r_0$ centred at ${\vec x_0}$  
with unit normal 
${\hat I}$, with $|{\vec x_0}|, r_0$ chosen so that the disc is contained entirely in $\Sigma \setminus K$. Thus we have
that 
\be 
S({\vec x_0}, {I})= \{ {\vec x}:{\vec x}= {\vec x_0} + {\vec r}, |{\vec r}|\leq r_0, {\vec r}\cdot{\hat I}=0\}.
\ee
Let $S^+_{ R}({\vec x_0}, {I})$ be the rigid translation of $S({\vec x_0}, {I})$ by the vector  
${\vec R}= R{\hat x_0}, R>0$. Let
$S^-_{ R}({\vec x_0}, {I})$ be the rigid translation of $S({\vec x_0}, {I})$ by the vector  $-2\vec{x_0} -{\vec R}$
so that $S^-_{R}({\vec x_0}, {I})$ is obtained by reflecting
$S^+_{ R}({\vec x_0}, {I})$ through the origin while assigning its orientation such that its unit normal is 
also ${\hat I}$.
Let $f^i_I({\vec x}), {\vec x} \in S({\vec x_0}, {I})$ be the $su(2)$ valued smearing function associated with 
$S({\vec x_0}, {I})$. Define the smearing functions $f^i_{\pm, I}$
associated with $S^{\pm}_{ R}({\vec x_0}, {I})$ by appropriate translations of $f^i_I$: 
\ba
f^i_{+,I} ({\vec x}) &= &f^i_I( {\vec x} -{\vec R})\\
f^i_{-,I} ({\vec x})  &= &f^i_I( {\vec x} +2\vec{x_0}+ {\vec R}).
\ea

Then it is easy to check that the fall-off
 conditions (\ref{fallE}) on the triad field induce the following limiting behaviour on the 
fluxes on the above  families of probes for each $I=1,2,3$:
\be
\lim_{R\rightarrow \infty}F_{ S^{\pm}_{R}({\vec x_0}, {I}), f_{\pm, I}}
= \int_{S({\vec x_0}, {I})}d^2S_If^{i=I}_{ I}({\vec x})  ,
\label{flux1}
\ee
\be 
\lim_{R\rightarrow\infty}R \big(F_{S^{\pm}_{R}({\vec x_0}, {I}), f_{\pm, I}}- 
\int_{S({\vec x_0}, {I})}d^2S_If^{i=I}_{\hat I}({\vec x})\big)
=h^I_i ({\hat x_0})\int_{S({\vec x_0}, {I})}d^2S_If^{i}_{ I}({\vec x}),
\label{flux2}
\ee
\be
\lim_{R\rightarrow\infty} R^{1+\epsilon}\big( F_{S^{\pm}_{R}({\vec x_0}, {I}), f^i_{\pm, I}} - 
\int_{S({\vec x_0}, {I})}d^2S_If^{i=I}_{\hat I}({\vec x})
- R^{-1} h^I_i ({\hat x_0})\int_{S({\vec x_0}, {I})}d^2S_If^{i}_{I}({\vec x})\big) =0,
\label{flux3}
\ee
for some $\epsilon >0$. Here 
$d^2S_I:= dx_Jdx_K, I\neq J \neq K$ and 
$h^I_i ({\hat x_0})$ is the evaluation of the (even) function $h^I_i$
which parameterizes the 
next to leading order part of the triad (see equation (\ref{fallE})) at the point  ${\hat x_0}$
on the unit 2-sphere. We have used the summation convention over repeated occurrences of the $su(2)$ index $i$ but not 
over the Cartesian coordinate index $I$.

It is easy to verify that the existence of  $su(2)$  valued functions  $h_I$ on the unit 2-sphere for which 
equations (\ref{flux1}), (\ref{flux2}), (\ref{flux3}) are satisfied for all choices of  $f^i_I, {\hat x_0}$ imply an asymptotic expansion of $E^a$ of the form:
 \be
E^\alpha_i  = \Eo^\alpha_i+\frac{h^\alpha_i(\xh)}{r}+g^\alpha_i , \label{fallEapp}
\ee
with $\lim_{r \to \infty} r^{1+\epsilon}g^\alpha_i =0 $.  This is  however not the full set of  conditions on $E^a$. For example, there is no control over the fall-offs of $\partial_\beta g^\alpha_i$. This requires further specifications,  involving pair of surfaces approaching to each other at appropriate rates. 
Whereas we do not envisage obstacles in doing so, we leave for future work the determination of a full set of conditions on fluxes capturing (\ref{fallE}).

We conclude by showing that conditions (\ref{flux1}), (\ref{flux2}), (\ref{flux3})  hold in the quantum theory defined by the KS 
representation in the following sense. Consider any KS spinnet $| s, E\ket$. From section \ref{sec3A},
each such background triad state label $E^a_i$ satisfies the asymptotic conditions (\ref{fallE}) for some
appropriate $h^a_i({\hat x})$. Since the spinnet graphs are confined to some compact region, it follows
that for large enough $R$ the surfaces $S^{\pm}_{R}({\vec x_0}, {I})$ do not intersect the 
graph underlying the spinnet label $s$. It follows that the standard LQG contributions to the action of 
flux operators  associated with these surfaces 
vanish for sufficiently large $R$. This implies that 
\ba
{\hat F}_{S^{\pm}_{R}({\vec x_0}, {I}), f_{\pm, I}}| s,E\ket
&= & \lambda_{E,S^{\pm}_{R}({\vec x_0}, {I}),f_{\pm, I}} \; |s, E\ket \nonumber \\
\lambda_{E,S^{\pm}_{R}({\vec x_0}, {I}),f_{\pm, I}} &=&
\int_{S^{\pm}_{R}({\vec x_0}, {I})}d^2S_af^i_{\pm, I}({\vec x}) E^a_i({\vec x}) 
\label{fluxevalue}
\ea
so that, for sufficiently large $R$ the KS spinnet $| s, E\ket$ is an {\em eigenvector} of the 
operators ${\hat F}_{S^{\pm}_{R}({\vec x_0}, {I}), f_{\pm, I}}$.
It is then easy to check that if we replace the classical fluxes  in 
the conditions (\ref{flux1}), (\ref{flux2})  by their quantum 
 eigenvalues (\ref{fluxevalue}), these conditions hold. 

\section{Supplementary material for section \ref{sec7}} \label{autapp}
\subsection{Preservation of $\A$ and $\E_{\Eo}$ by $\autEo$} \label{aAapp}

Let $\phi$ and $g$ be a $C^k$ diffeomorphism and  $C^{k-1}$ $SU(2)$ rotation as in section \ref{sec7B},
\ba
\phi(x)^{\alpha} & = & R^\alpha_\beta x^\beta+ t^\alpha +  s^\alpha(\xh)+O(r^{-\e}) , \label{phapp} \\
g(x) & = & \go+\frac{\lambda(\xh)}{r}+O(r^{-1-\e}) \label{gapp} ,
\ea
where $s^\alpha$ is a $C^{k+1}$ odd function on the sphere and $\lambda$ a $C^{k}$ even function on the sphere. In this section we show that for any asymptotically flat electric field $E^a \in \E_{\Eo}$ as in (\ref{fallE}) and any connection $A_a \in \A$ as in (\ref{fallA}) one has:
\ba
\phi_* E^\alpha (x) &= & R^\alpha_\beta \Eo^\beta+ (\even)/r+O(r^{-1-\e})  \label{phE} \\
\phi_* A_\alpha (x) &= & (\odd)/r^2+O(r^{-2-\e}),  \label{phA} \\ 
 g(x) E^\alpha(x) g^{-1}(x) &= &  \go \Eo^\alpha \go^{-1}+ (\even)/r+O(r^{-1-\e}), \label{gE} \\
 g(x)  A_\alpha (x) g^{-1}(x) -(\partial_\alpha g(x)) g^{-1}(x)&= & (\odd)/r^2+O(r^{-2-\e}),\label{gA}
\ea
where ``$(\even)$'' denote  $C^k$ even functions on the sphere and ``$(\odd)$'' denote $C^{k-1}$ odd functions on the sphere (from here onwards, all appearances of ``$(\even)$'' and   ``$(\odd)$''  will refer to functions with these degrees of differentiability). Our main tool will be given by general properties of  remainders described in appendix \ref{Oapp}.

We first note  that from appendix \ref{phinvapp} $\phi^{-1}$ has the asymptotic expansion:
\be
\phi^{-1 \alpha}(x)  =  (R^{-1})^\alpha_\beta x^\beta +t'^\alpha +s'^\alpha(\xh)+O(r^{-\e}) , \label{phinv} \\
\ee
with $t'^\alpha=-(R^{-1})^\alpha_\beta t^\beta $ and $s'^\alpha(x)=-(R^{-1})^\alpha_\beta s^\beta(R^{-1}(x))$. Next we note that:
\ba
\partial_\beta \phi^\alpha(x) & = & R^\alpha_\beta +(\even)/r +O(r^{-1-\e}) ,\label{Dph} \\
\partial_\beta \phi^{-1 \alpha}(x) & = & (R^{-1})^\alpha_\beta +(\even)/r +O(r^{-1-\e}).  \label{Dphinv} \\
\ea

The Cartesian  coordinate expression of (\ref{phE}) consists of a product of three terms:
\be
\phi_* E^\alpha (x) = \left[ \det(\partial \phi^{-1}(x))\right] \left[ \partial_\beta \phi^{\alpha}(\phi^{-1}(x)) \right] \left[E^\beta(\phi^{-1}(x))\right]. \label{phE1}
\ee
From (\ref{Dphinv}) and (\ref{prodf}) one finds: $\det(\partial \phi^{-1}(x))= 1 +(\even)/r +O(r^{-1-\e})$. From (\ref{Dph}),  (\ref{phinv}) and Eq. (\ref{fphiR}) one has: $\partial_\beta \phi^{\alpha}(\phi^{-1}(x)) = (R^{-1})^{\alpha}_\beta +(\even)/r +O(r^{-1-\e})$. Finally, from (\ref{fallE}) and (\ref{fphiR}) one finds: $E^\beta(\phi^{-1}(x))= \Eo^\beta+(\even)/r+ O(r^{-1-\e})$. Multiplying these three expansions one obtains (\ref{phE}).

The Cartesian  coordinate expression of (\ref{phA}) consists of a product of two terms:
\be
\phi_* A_\alpha (x) = \left[ \partial_\alpha \phi^{-1 \beta}(x) \right] \left[ A_\beta(\phi^{-1}(x) \right].
\ee
The first term is given by (\ref{Dphinv}). The second term is $A_\beta(\phi^{-1}(x)) =(\odd) r^{-2}+O(r^{-2-\e})$ by use of Eq. (\ref{grala2}). Multiplying the two terms and using Eq. (\ref{prodfa}) one recovers (\ref{phA}).

Next, taking the complex conjugate and transpose of (\ref{gapp}) one gets:
\be
g^{-1}(x)  =  \go^{-1}+(\even)/r+O(r^{-1-\e}) \label{ginv}.
\ee
Eq. (\ref{gE}) follows straightforwardly by multiplying the expansions of $g$ (\ref{gapp}), $E^a$ (\ref{fallE}) , and $g^{-1}$ (\ref{ginv}) and use of Eq. (\ref{prodfa}). Similarly the linear in $A_a$ term in (\ref{gA}) directly follows by use of Eq. (\ref{prodfa}). The inhomogeneous term in (\ref{gA}) follows by multiplying $\partial_\alpha g(x) = (\odd)/r^{2}+O(r^{-2-\e})$ with (\ref{ginv}) and using Eq. (\ref{prodfa}).


\subsection{Characterization of $\autEo$ as symmetry group of $\E_{\Eo}$} \label{symgroup}
In this section we  refer to $\autEo$ as the set of pairs  $(g',\phi')$ 
leaving the space of asymptotically flat electric fields $\E_{\Eo}$ invariant: 
\be
\autEo:= \{ (g',\phi')  :  g' \phi'_* E^a g'^{-1} \in  \E_{\Eo},  \quad \forall E^a \in \E_{\Eo} \}. \label{EEoinv}
\ee
Here $g'$ are $C^{k-1}$ $SU(2)$ rotations and $\phi'$  $C^k$ diffeomorphisms of $\Sigma$. 
The purpose of the section is to show that the (component connected to identity of the) set (\ref{EEoinv}) coincides with what is referred to as $\autEo$ in section \ref{sec7B}.  
Let $\phi:= \phi'^{-1}$ and $g:=\phi'^{-1}_* g'^{-1}$ so that $(g,\phi) (g',\phi') =(\idtwo,\id)$ and condition (\ref{EEoinv}) takes the form $(\phi^{-1})_*( g^{-1}   E^a g) \in  \E_{\Eo}$. For the particular case of an electric field  $E^a$ that exactly  coincides  with the flat $\Eo^a$ in the asymptotic region (for sufficiently large $r$), condition (\ref{EEoinv}) reads:\footnote{Such an 
electric field can  be obtained by interpolating $\Eo^a$ for $r>r_2$ with a  zero electric field for $r<r_1<r_2$ through a  function $f(r)$ of the type given in  Eq. (A7) of \cite{mm2} by $E^a(x)=f(r) \Eo^a$.} 
 \be
E'^\alpha:=(\phi^{-1})_*( g^{-1}   \Eo^\alpha g)(x) = \Eo^\alpha + (\even)/r + O(r^{-1-\e}).\label{gphEo}
 \ee
It will be convenient to work with undensitized quantities.  Taking the determinant of $E'^\alpha$ on both sides of (\ref{gphEo}) and using Eq. (\ref{prodf}) one finds $\det(E')=1+(\even)/r + O(r^{-1-\e})$. From Eq. (\ref{flam}) any power of the determinant obeys similar fall-offs. Since $\det(E')=\phi^{-1}_* \det(\Eo)$ (for large enough $r$) one can then obtain an `undensitized' version of (\ref{gphEo}):
 \be
e'^\alpha=(\phi^{-1})_*( g^{-1}   \eo^\alpha g)(x) = \eo^\alpha + (\even)/r + O(r^{-1-\e}),\label{gpheo}
 \ee
   where $e'^\alpha :=\det^{-1/2}(E') E'^\alpha$ and $\eo^\alpha$ is the flat undensitized triad associated to $\Eo^\alpha$. Setting $q'^{\alpha \beta}:=e'^\alpha_i e'^\beta_i$ one finds $q'^{\alpha \beta}=\qo^{\alpha \beta}+ (\even)/r + O(r^{-1-\e})$ where $\qo^{\alpha\beta}$ is the contravariant flat metric in the asymptotic  region.  Writing $q'_{\alpha \beta}$ in terms of $q'^{\alpha \beta}$ by the standard inverse matrix formula and using (\ref{prodf}), (\ref{finv}) one finds:
\be
q'_{\alpha \beta}(x) =  \qo_{\alpha \beta} + (\even)/r + O(r^{-1-\e}),
\ee
where $\qo_{\alpha\beta}= \delta_{\alpha \beta}$  is the flat metric in the asymptotic region.   We then conclude  $\phi$ satisfies:
\be
\phi^{-1}_*  \qo_{\alpha \beta}(x)    =  \qo_{\alpha \beta} + (\even)/r + O(r^{-1-\e})  =: C_{\alpha \beta}(x). \label{phqo}
\ee
We  now use  (\ref{phqo}) to determine the asymptotic form of $\phi$.  In what  follows we use $\qo_{\alpha \beta}$ to rise and lower indices. 
Let
\be
D^{\; \; \beta}_\alpha(x)  :=  \partial_\alpha \phi^{ \, \beta}(x) ,  \label{defD} 
\ee
so that from the definition of push-forward in (\ref{phqo}) we have:
\be
C_{\alpha \beta}(x) = D^{\; \; \mu}_\alpha(x) D_{\beta \mu}(x) . \label{defC}
\ee
Eq. (\ref{phqo}) and its first derivative then read: 
\ba
C_{\alpha \beta}(x) & =& \delta_{\alpha \beta} +  r^{-1} (\even) +  O_{k-1}(r^{-1-\e}) ,  \label{phqo2} \\
\partial_\gamma C_{\alpha \beta}(x)& = &r^{-2} (\odd) + O_{k-2}(r^{-2-\epsilon}) \label{partialC}
\ea
(subscripts in remainders denote number of derivatives with known fall-offs, see  appendix \ref{Oapp}). For later purposes, we note that the $r^{-2} (\odd)$ term in (\ref{partialC}) is of Cartesian degree of differentiability $k-1$ (this follows as direct consequence of the $r^{-1}(\even)$ term in (\ref{gphEo}) being of Cartesian degree of differentiability $k$).
From the  `integrability condition'  $\partial_\gamma \partial_\beta \phi^\alpha= \partial_\beta \partial_\gamma \phi^\alpha$:
\be
\partial_\gamma D_{\alpha \beta}(x)= \partial_\alpha D_{\gamma \beta}(x), \label{integrability}
\ee
one can verify the identity:
\be
\partial_\gamma D_{\alpha \beta} = \frac{1}{2}(D^{-1})_\beta^{\;\; \mu}(\partial_\alpha C_{\gamma \mu} +\partial_\gamma C_{\alpha \mu}- \partial_\mu C_{\alpha \gamma}), \label{partialD}
\ee
where $D^{-1}$ is the inverse matrix of $D$. Contracting indices in  (\ref{phqo2}) we find  $D^{\alpha \beta}(x) D_{\alpha \beta}(x) = 3+O_0(r^{-1})$ which implies $D^{\; \; \beta}_\alpha$ is bounded. Equations similar to (\ref{defC})
(\ref{phqo2}) hold for an analysis of the push forward of the {\em contravariant} metric with $D$ replaced by $D^{-1}$.
As a result, we  have that $D^{-1\; \; \beta}_\alpha$ is also bounded.
This, together with Eqns. (\ref{partialD}) and (\ref{partialC}) imply:
\be
\partial_\gamma D^{\; \; \beta}_\alpha(x) = O_0(r^{-2}). \label{fallpartialD1}
\ee
Let $r,\theta, \phi$ denote polar coordinates. 
Integration of (\ref{fallpartialD1})  along the radial direction implies that 
$\lim_{r\rightarrow \infty} D(r,\theta ,\phi)=: D(\theta , \phi)$. 
Angular integration of (\ref{fallpartialD1}) at fixed $r$ followed by taking the $r\rightarrow \infty$ then implies
that $D$ asymptotes to a  constant matrix:
\be
R_\alpha^\beta  := \lim_{r \to \infty} D^{\; \; \beta}_\alpha(x). \label{defR}
\ee
From (\ref{phqo2}) and (\ref{defC}) this matrix is orthogonal. Furthermore,  since $\phi$ is connected to identity, it has to be a rotation.\footnote{Even if we drop the connected to identity condition on $\phi$, the preservation of the triad at infinity in (\ref{gphEo}) implies $\phi$ is orientation preserving so that $D$ can only asymptote to a rotation.}  It will be convenient to factor out this rotation and write $\phi$ and  $D$  as:
\be
\phi^\alpha(x) = R^\alpha_\beta (x^\beta+y^\beta(x)) ,  \quad D_\alpha^{\;\; \beta}(x)  =   R^\beta_\gamma  \Db_\alpha^{\;\; \gamma}(x) ,
\ee
where 
\be
\Db_\alpha^{\;\; \beta}(x) = \delta^\beta_\alpha +Y_\alpha^{\;\; \beta}(x), \quad Y_\alpha^{\;\; \beta}(x)  = \partial_\alpha y^\beta(x) , \label{defDb}
\ee
with
\be
\lim_{r \to \infty} Y_\alpha^{\;\; \beta}(x)=0 , \quad \partial_\gamma Y^{\; \; \beta}_\alpha(x) = O_0(r^{-2}). \label{condY}
\ee
The overall rotation does not affect relations (\ref{defC}) and (\ref{partialC}) and one has:
\ba
C_{\alpha \beta}   & =  & \Db^{\; \; \mu}_\alpha \Db_{\beta \mu} , \label{Cbar} \\
\partial_\gamma Y_{\alpha \beta} \equiv \partial_\gamma \Db_{\alpha \beta}  & = & \frac{1}{2}(\Db^{-1})_\beta^{\;\; \mu}(\partial_\alpha C_{\gamma \mu} +\partial_\gamma C_{\alpha \mu}- \partial_\mu C_{\alpha \gamma}). \label{partialDb}
\ea

We now  determine the form  of $y^\alpha(x)$.  Contracting the second equation in (\ref{condY}) with  $\xh^\gamma$  and using $\partial_r = \xh^\gamma \partial_\gamma$ one finds $\partial_r  Y_\alpha^{\;\; \beta}=O_0(r^{-2})$. Integrating with respect to $r$ and using the first equation in (\ref{condY}) we find:
\be
 Y_\alpha^{\;\; \beta}(x) =O_0(r^{-1}). \label{fallY1}
\ee
From the first equation in (\ref{defDb}) and (\ref{fallY1}) it is easy to verify that:
\be
(\Db^{-1})_\alpha^{\;\; \beta}(x)= \delta^\beta_\alpha+O_0(r^{-1}). \label{Dbinv}
\ee
Using (\ref{Dbinv}) and (\ref{partialC}) in (\ref{partialDb}) one finds:
\be
\partial_\gamma Y_\beta^{\;\; \alpha}(x)= r^{-2}(\odd)+O_0(r^{-2-\epsilon}). \label{ddy}
\ee
Contracting (\ref{ddy}) with $\xh^\gamma$ one obtains $\partial_r (Y_\beta^{\;\; \alpha})= r^{-2}(\even)+O_0(r^{-2-\epsilon})$ so that
\be
Y_\beta^{\;\; \alpha}(x)=r^{-1}f^\alpha_\beta(\xh) + O_1(r^{-1-\epsilon}), \label{partialy}
\ee
where $f^\alpha_\beta$ is even. 
That the remainder in (\ref{partialy}) is $O_1$ rather than $O_0$ can be seen by applying $\partial_\gamma$ to (\ref{partialy}) and comparing the result with (\ref{ddy}).
Contracting (\ref{partialy}) with $\xh^\beta$   we obtain an equation for $\partial_r y^\alpha$ which upon integration over $r$ gives:
\be
y^\alpha(x)=\ln r f^\alpha(\xh) + \yo^\alpha(\xh)+ O_2(r^{-\epsilon}), \label{y2}
\ee
where 
\be
 f^\alpha(\xh):= \xh^\beta f^\alpha_\beta(\xh) =(\odd), \label{oddf}
 \ee
   $ \yo^\alpha(\xh)$  an integration `constant' and the fact that the remainder is $O_2$ follows from the same logic as in (\ref{partialy}).  In order to further determine the form of the terms in (\ref{y2}), we apply $\partial_\beta$ to (\ref{y2}) and compare the result with (\ref{partialy}). In spherical coordinates $(r, x^A)$ we have:
\be
\partial_\beta = r^{-1} D^A \xh_\beta D_A + \xh_\beta \partial_r  , \label{partialsph}
\ee
where sphere indices $A,B,\ldots$ are raised with the unit sphere metric and $D_A$ is the derivative on the unit sphere (see Eq. (\ref{partialsph0})).  Acting on (\ref{y2}) with  (\ref{partialsph})  we get:
\be
 \partial_\beta y^\alpha(x) = r^{-1} \ln r D^A \xh_\beta D_A f^\alpha(\xh)  + r^{-1} \xh_\beta  f^\alpha(\xh) + r^{-1} D^A \xh_\beta D_A\yo^\alpha(\xh)+ O_1(r^{-1-\epsilon}) . \label{py3}
\ee
In order for (\ref{py3}) to be compatible with (\ref{partialy}) it must be the case that $D^A \xh_\beta D_A f^\alpha(\xh)=0$ which is only satisfied if  $f^\alpha(\xh) =c^\alpha=$ constant. From the  parity condition (\ref{oddf})  this constant must be zero and we conclude that $f^\alpha=0$.\footnote{In the absence of parity conditions the resulting diffeomorphisms $x^\alpha \mapsto x^\alpha+ \ln r c^\alpha$ are a spatial version of the  so-called `logarithmic translations'  \cite{aalog}.}  Finally, for the  third term in (\ref{py3}) to be  compatible with (\ref{partialy}) we must have:
\be
\yo^\alpha(\xh) = t^\alpha + s^\alpha(\xh),\label{yots}
\ee
with $t^\alpha$ constant and $s^\alpha$ of odd parity.  Note that from Eq. (\ref{ddy}) we have that $\partial_\gamma \partial_\beta s^\alpha$ is of Cartesian degree of differentiability $k-1$ (this differentiability coming from that in the leading order term of (\ref{partialC})). Thus $s^\alpha(x)$ is of  Cartesian degree of differentiability $k+1$ as a function on $\Uo$. This in turn implies $s^\alpha(\xh)$ is $C^{k+1}$ as a function on the sphere.

To summarize, we have thus far shown that $\phi$ has the asymptotic form:
\be
\phi^\alpha(x)=R^\alpha_\beta(x^\beta +t^\beta + s^\beta(\xh)+ y_1^\beta(x)) , \label{fallphib}
\ee
with $y_1^\alpha(x)=O_2(r^{-\epsilon})$ and $s^\alpha(\xh)$ a $C^{k+1}$ odd function on the sphere.  We finally show that $y_1^\alpha = O_k(r^{-\epsilon})$.
 
If one computes $\phi^{-1}_*  \qo_{\alpha \beta}(x) \equiv C_{\alpha \beta}(x)$ from  (\ref{defC}) and (\ref{fallphib}) one gets: 
\be
C_{\alpha \beta}(x)=\delta_{\alpha \beta}+ 2\partial_{(\alpha} s_{\beta)}(x) +O_1(r^{-1-\epsilon}) \label{Cb2}.
\ee
Comparing (\ref{Cb2}) with (\ref{phqo2}) we conclude that the $r^{-1} (\even)$ part of $C_{\alpha \beta}$ in (\ref{phqo2}) is given by $2\partial_{(\alpha} s_{\beta)}$.  Eq. (\ref{partialC}) can then be written as:
\be
 \partial_\gamma C_{\alpha \beta}(x) = 2 \partial_\gamma \partial_{(\alpha} s_{\beta)}(x) + O_{k-2}(r^{-2-\epsilon}) . \label{fallB}
\ee
We now show  $y_1^\alpha(x) = O_k(r^{-\epsilon})$ by induction. Assume  we know
\be
y_1^\alpha(x)=O_{p-1}(r^{-\epsilon}),
\ee
where $3 \leq p \leq k$. We then have:
\ba
\Db^{\; \; \beta}_\alpha(x)  = \delta_\alpha^\beta+ \partial_\alpha s^\beta(x) +O_{p-2}(r^{-1-\epsilon}), \label{fallDb}
\ea
which in turn implies:
\ba
(\Db^{-1})^{\; \; \beta}_\alpha(x)  = \delta_\alpha^\beta- \partial_\alpha s^\beta(x) +O_{p-2}(r^{-1-\epsilon}), \label{fallDbinv}
\ea
as can be verified from the  inverse matrix formula and Eqns. (\ref{prodf}), (\ref{finv}).  Using (\ref{fallDbinv}) and (\ref{fallB}) in (\ref{partialDb}) one obtains:
\be
\partial_\gamma Y_{\alpha \beta}(x) =  \partial_\gamma \partial_{\alpha} s_{\beta}(x) +O_{p-2}(r^{-2-\epsilon}). \label{partialDb3}
\ee
The  supertranslation terms on both sides simplify and we get:
\be
\partial_\gamma \partial_\beta y_1^\alpha(x) =O_{p-2}(r^{-2-\epsilon}) . \label{partial2y1}
\ee
Integrating with respect to $r$ in similar way as done earlier for Eq. (\ref{ddy}) one finds  $y_1^\alpha(x)=O_{p}(r^{-\epsilon})$ as desired.

Now, recall that the diffeomorphism featuring in the definition (\ref{EEoinv}) is $\phi' \equiv \phi^{-1}$. From appendix \ref{phinvapp} one finds $\phi'$ has the same type of expansion, with leading rotation $(R^{-1})^\alpha_\beta$.  This concludes the specification of $\phi'$ in (\ref{EEoinv}).

We now discuss the $SU(2)$ part.  
Recall that we require $g$  to be such that (\ref{gphEo})  holds.  If (\ref{gphEo}) holds then acting with $\phi_*$ on (\ref{gphEo}) 
implies that the following condition must hold:
\be
g^{-1}(x) \Eo^\alpha  g(x)= \phi_* (\Eo^\beta +  (\even)/r + O(r^{-1-\e})).
\ee
From Eqns. (\ref{phapp}), (\ref{phE}) of appendix \ref{aAapp} together with our considerations above which show that $\phi$ satisfies (\ref{phapp}),
we have that   $ \phi_* (\Eo^\beta +  (\even)/r + O(r^{-1-\e})) = R_\beta^\alpha \Eo^\beta +  (\even)/r + O(r^{-1-\e})$ which implies that:
\be
g^{-1}(x) \Eo^\alpha  g(x)= R_\beta^\alpha \Eo^\beta +  (\even)/r + O(r^{-1-\e}) \label{gEo}. 
\ee
Note that the left hand side expresses the adjoint action of $g(x)$ so that it must equal $R_\beta^\alpha (x) \Eo^\beta$
for some rotation matrix $R_\beta^\alpha (x)$. Conversely from the 2 to 1 map from $SU(2)$ to $SO(3)$, 
$R_\beta^\alpha (x)$ uniquely specifies $g(x)$ upto a sign. Since 
$\lim_{r\rightarrow \infty}R_\beta^\alpha (x)= R_\beta^\alpha$  and since $g(x)$ is continuous in $x^{\alpha}$, 
we have that $\lim_{r\rightarrow \infty} g(x)= \go$ 
with $\go$  being the constant $SU(2)$ rotation satisfying $\go^{-1} \Eo^\alpha \go= R_\beta^\alpha \Eo^\beta$.
From (\ref{gEo}) it is then straightforward to conclude that $g(x)$ must be of the form
\be
g(x)= \go + \lambda(\xh)/r +O_0(r^{-1-\e}), \label{fallg1}
\ee
with $\lambda$ an even function on the sphere.
By comparing derivatives of (\ref{gEo}) and (\ref{fallg1}) one can conclude that the remainder term in (\ref{fallg1}) is $O(r^{-1-\e})$.  By comparing the $r^{-1}(\even)$ term in (\ref{fallg1}) with the one in (\ref{gEo}) one concludes it is of Cartesian degree of differentiability $k$, which in turn implies $\lambda(\xh)$ is $C^k$ as a function on the sphere.  Thus $g(x)$ is an $SU(2)$ rotation as in section \ref{sec7B}. This directly implies $g^{-1}(x)$ is of the same type (see Eq. (\ref{ginv})). Finally from Eq. (\ref{fphi}) one finds that $g'= \phi^{-1}_* g^{-1}$ is also of the form given in \ref{sec7B}, with leading term $\go' \equiv \go^{-1}$. This concludes the characterization of $\autEo$ elements.



\subsection{Asymptotic form of $\phi^{-1}$} \label{phinvapp}
Given the asymptotic form of $\phi \equiv \phi'^{-1}$:
\be
\phi(x)^{\alpha}  =  R^\alpha_\beta (x^\beta+ t^\beta +  s^\beta(\xh) )+O_k(r^{-\e}) , \label{phiapp}
\ee
we want to find the asymptotic form of $\phi' = \phi^{-1}$. 
We first show that (\ref{phiapp}) implies:
\be
\phi'^\alpha(x) =(R^{-1})^\alpha_\beta x^\beta +O_0(1) . \label{phip0}
\ee
From  (\ref{phiapp}) we have that there exists $r_0, M>0$ s.t. $\forall r>r_0$ we have that 
\begin{equation}
| \phi (x) - R(x)| <M  .\label{phirM}
\end{equation}
$\phi$ is a homeomorphism of $\Sigma$
so that  $\phi (B(r_0))$ is closed and $\phi (\partial B(r_0))$ = $\partial \phi (B(r_0))$ where 
$B(r_0)$  is the interior of the sphere of cartesian radius $r_0$  together with its boundary $\partial B(r_0)$.
Since for $r> r_0$ we have that $\phi (x)$ is also in the asymptotic region, there exists $r'_0$ such that 
 $\phi (B(r_0)) \subset B(r'_0)$.  This means that if $|\phi(x)|>r'_0$ then $r>r_0$ so that 
for $|\phi(x)|>r'_0$ we have from equation (\ref{phirM})  that 
$|\phi (x) -R(x) | <M$ which is the same as the statement that for $|x|>r'_0$, we have that 
$|x- R(\phi^{-1}(x))| < M$ which is same as $|R^{-1}(x)- \phi^{-1}(x)| < M$
which means that $\phi^{-1}(x) = R^{-1}(x) + O_0(1)$.

Let $D_\beta^\alpha(x)= \partial_\beta \phi^\alpha(x)$ so that
\be
D_\beta^\alpha(x) = R^\alpha_\gamma( \delta^\gamma_\beta + \partial_\beta s^\alpha(x)) +O_{k-1}(r^{-1-\e}).
\ee
From inverse matrix formula and Eqns. (\ref{prodf}), (\ref{finv}) one finds:
\be
(D^{-1})_\beta^\alpha(x) = (R^{-1})^\gamma_\beta( \delta^\alpha_\gamma - \partial_\gamma s^\alpha(x)) +O_{k-1}(r^{-1-\e}).\label{Dinvc1}
\ee
The inverse derivative formula tells:
\be
 \partial_\alpha \phi'^\beta(x) = (D^{-1})_\alpha^\beta(\phi'(x)). \label{partialphip}
\ee
Substituting (\ref{phip0}) in the RHS of (\ref{partialphip}) and using  that $O_0(|\phi'(x)|^{-1-\e})=O_0(r^{-1-\e}) $ and $\partial_\gamma s^\alpha(\phi'(x))= \partial_\gamma s^\alpha(R^{-1}(x))+O_0(r^{-2})$ by Taylor expansion (see footnote \ref{taylorfn})  one finds:
\be
 \partial_\alpha \phi'^\beta(x) = (R^{-1})^\gamma_\beta( \delta^\alpha_\gamma - \partial_\gamma s^\alpha(R^{-1}(x))) +O_0(r^{-1-\e}) 
\ee
which upon integration leads to:
\be
\phi'^\alpha(x) = (R^{-1})^\alpha_\beta x^\beta - s^\alpha(R^{-1}(x)) + c^\alpha+O_1(r^{-\e}) \label{phip1}
\ee
where $c^\alpha$ are integration constants. The condition $\phi'(\phi(x))=\id$ fixes $c^\alpha=- t^\alpha$. That the remainder in (\ref{phip1}) is $O_k(r^{-\epsilon})$ can be shown by induction using Eq. (\ref{partialphip}):  Assume the remainder is $O_{m-1}(r^{-\e}), 2 \leq m \leq k$ for $\phi'$ on the RHS of (\ref{partialphip}) and compute $\partial_\alpha \phi'^\beta$ from this equation. Using  Eq. (\ref{fphiR}) with $(D^{-1})_\beta^\alpha(x)$ seen as the $C^{k-1}$ function  (\ref{Dinvc1}) and $\phi'$ seen as a $C^{m-1}$ diffeomorphism, one obtains $ \partial_\alpha \phi'^\beta(x) = (R^{-1})^\gamma_\beta( \delta^\alpha_\gamma - \partial_\gamma s^\alpha(R^{-1}(x))) +O_{m-1}(r^{-1-\e})$ which implies the remainder of $\phi'$ is  $O_{m}(r^{-\e})$.  \footnote{The last induction step $m=k$ makes use of the $C^{k+1}$ Cartesian differentiability of $s^\alpha$ (in its role of  $\partial_\gamma s^\alpha$ term in (\ref{Dinvc1})) by the same reason described in footnote \ref{pp1}.}

\subsection{Asymptotic form of $\autEo$ compositions} \label{autgroupapp}

Let $(g,\phi), (g',\phi') \in \autEo$ as in (\ref{fallg}) and $(g'',\phi'') := (g,\phi)(g,\phi)= (g \phi_* g', \phi \phi')$. Definition (\ref{EEoinv}) implies $(g'',\phi'') \in \autEo$ so that from section \ref{symgroup} it has also the asymptotic form:
\ba
\phi''^\alpha(x)=  & = & R^\alpha_\beta(\go'') x^\beta+ t''^\alpha +  s''^\alpha(\xh)+O(r^{-\epsilon}) , \\
g''(x)  & = & \go'' +r^{-1} (\even) + O(r^{-1-\e}),
\ea
where $R^\alpha_\beta : SU(2) \to SO(3)$ denotes  the Adjoint action of $SU(2)$ defined by Eq. (\ref{Rgg}).
From $\phi'' = \phi \phi'$ and using properties as in appendix \ref{Oapp} one finds
\ba
R^\alpha_\beta(\go'') & = & R^\alpha_\mu(\go) R^\mu_\beta(\go') , \label{Rpp}\\
t''^\alpha &=& R^\alpha_\mu(\go) t'^\mu +t^\alpha ,  \label{tpp}\\
s''^\alpha(x) &=&  R^\alpha_\mu(\go) s'^\mu(x) + s^\alpha(R(\go')x) .\label{spp}
\ea
From $g''=g \phi_* g'$ and the properties of appendix \ref{Oapp} one finds $\go''=\go \go'$. This is consistent with (\ref{Rpp}) since $R^\alpha_\gamma(\go) R^\gamma_\beta(\go')=R^\alpha_\beta(\go \go')$. 

From these properties it follows that the set of $\autEo$ elements with trivial asymptotic rotations and translations form a subgroup of $\autEo$. The group $\aut$ is defined as the component connected to identity of this subgroup (one can verify that the connected to identity  property as defined in  footnote \ref{fnconn} is preserved under the group operations).

\subsection{Properties of remainders}\label{Oapp}
  A  $C^p$  field is said to be  $O_n(r^{-\beta})$ ($0 \leq n \leq p \leq k$; $\beta \geq 0$)  if for $m=0,\ldots, n$  the coefficients of the 
$m$-th partial derivatives in the Cartesian chart are bounded by $c r^{-\beta-m}$ for some constant $c$. The case when $n=p$ is denoted by $O(r^{-\beta})$ as  presented in section \ref{sec2A}.  Product of remainders satisfies:
\be
O_m(r^{-\beta}) O_n(r^{-\gamma})=O_{\min(m,n)}(r^{-\beta-\gamma}), \label{prodO}
\ee
as can be  verified by taking derivatives on the left hand side.  We now describe the behaviour under various operations of functions of the form:
\ba
f(x) & = & f_0+ f_1(x)+ O(r^{-1-\epsilon}),  \label{gralf} \\
a(x) & =  & a_2(x)+ O(r^{-2-\epsilon}), \label{grala}
\ea
where $f$ is $C^p$, $a$ is $C^{p'}$,  $0 \leq p,p' \leq k$,  $f_0 =$  constant and 
\be
f_1(x)=g(\xh)/r , \quad a_2(x)=b(\xh)/r^2, \label{f1a2}
\ee
with $g(\xh)$, $b(\xh)$ respectively $C^{p+1}$, $C^{p'+1}$,  functions on the sphere.

We first consider the product of  functions: Given $f(x),a(x)$ and $f'(x)= f'_0+ f''_1(x)+ O(r^{-1-\epsilon})$ as in (\ref{gralf}), it is easy to verify that their product is again of the type (\ref{gralf}), (\ref{grala}) with:
\ba
f(x) f'(x) & = & f_0 f'_0 +(f_0 f'_1(x) +f'_0 f_1(x))+O(r^{-1-\epsilon}),  \label{prodf} \\
f(x) a(x) & = & f_0 a_2(x) +O(r^{-2-\epsilon}). \label{prodfa}
\ea
Next we note that when $f_0 \neq 0$:
\be
1/f(x)= f^{-1}_0 -  f^{-2}_0  f_1(x) + O(r^{-1-\epsilon}) \label{finv}.
\ee
To see (\ref{finv}), use Taylor expansion around $t=1$ of the function $t \mapsto 1/t$ to obtain:   $1/f(x)= f^{-1}_0 -  f^{-2}_0  f_1(x) + O_0(r^{-1-\epsilon})$ . The appropriate  bounds on the derivatives of the remainders can be obtained by induction as follows. Assume $1/f(x)= f^{-1}_0 -  f^{-2}_0  f_1(x) + O_n(r^{-1-\epsilon})$. Writing $\partial_\alpha (1/f(x))=-(1/f(x))^2  \partial_\alpha f(x)$ and using (\ref{prodO}) one concludes that the derivative of the remainder is $O_n(r^{-2-\epsilon})$. This in turn implies the  remainder is $O_{n+1}(r^{-1-\epsilon})$.

More generally one can verify that:
\be
(f(x))^\lambda= f^{\lambda}_0 +\lambda f^{\lambda-1}_0  f_1(x) + O(r^{-1-\epsilon}) ,\label{flam}
\ee
for any  $\lambda \in \reals$ and $f_0 \neq 0$. This can again be shown by induction: Assume  $f^\lambda(x)= f^{\lambda}_0 +\lambda f^{\lambda-1}_0  f_1(x) + O_n(r^{-1-\epsilon})$ ($n=0$ case is obtained by Taylor expansion of $t \mapsto t^\lambda$).  From $\partial_\alpha (f^\lambda(x))= \lambda (f^\lambda(x)) (1/f(x)) \partial_\alpha f(x) $ and using (\ref{prodO}) and (\ref{finv}) one concludes that  the derivative of the remainder is $O_n(r^{-2-\epsilon})$, which in turn  implies the  remainder is $O_{n+1}(r^{-1-\epsilon})$.

We now describe the behaviour of functions (\ref{gralf}), (\ref{grala}) under composition with diffeomorphisms. To simplify the discussion we first consider diffeomorphisms with asymptotic trivial rotational part. Consider a  $C^{l}$, $0 \leq l \leq k$  diffeomorphism with asymptotic form
\be
\phi(x)^{\alpha}  =   x^\alpha+\yo^\alpha(\xh)+O(r^{-\e}) \label{phiO}
\ee
with $\yo^\alpha(\xh)$ a $C^{l+1}$ function on the sphere (the case $l=k$, $\yo^\alpha(\xh)=t^\alpha+s^\alpha(\xh)$ corresponds to the diffeomorphisms of section \ref{symgroup}). 
For $f$ as in (\ref{gralf}) we have:
\be
f(\phi(x))=f_0+ f_1(x)+O(r^{-1-\epsilon}).\label{fphi}
\ee
To show (\ref{fphi}) we need to verify that:
\ba
f(\phi(x)) &= &f_0+ f_1(x)+ O_0(r^{-1-\epsilon}) \label{fphi0}\\
\partial_{\beta_n} \ldots \partial_{\beta_1} f(\phi(x)) & =& \partial_{\beta_n} \ldots \partial_{\beta_1}f_1(x)+ O_0(r^{-(n+1)-\epsilon}), \quad n=1,\ldots, \min(l,p) . \label{fphin}
\ea
($\min(l,p)$ is the differentiability of $f \circ \phi$).    First, notice that argumentation similar to that after Eq. (\ref{phip0}) implies that diffeomorphisms (\ref{phiO})  do not change $O_0(r^{-\beta})$  bounds:
\be
F(x)=O_0(|x|^{-\beta}) \iff F(x)=O_0(|\phi(x)|^{-\beta}), \label{Ophi}
\ee
where   $|\phi(x)| := \sqrt{\phi^\alpha(x) \phi^\alpha(x)}$ and $|x| \equiv r$. 
We thus have:
\be
f(\phi(x))= f_0 + f_1(\phi(x)) + O_0(r^{-1-\epsilon}) . \label{f0}
\ee
Taylor expanding $ f_1(\phi(x))$ around $x$ we find:\footnote{Let $x'^\alpha:=\phi^\alpha(x)$, $r':=|\phi(x)|$ and $\xh'^\alpha=x'^\alpha/r'$. Using that $x' = x+O_0(1)$ it is easy to verify using Taylor approximation that $r'^{-1}=r^{-1}(1+O_0(r^{-1}))$ and $\xh'=\xh+O_0(r^{-1})$. The latter implies $g(\xh')=g(\xh)+O_0(r^{-1})$. Using these expansions for $f_1(x')=r'^{-1}g(\xh')$ one obtains (\ref{tf0}).  Analogous considerations for functions of the form $f_n(x)=r^{-n}g(\xh), n>0$ leads to $f_n(x')=f_n(x)+O_0(r^{-(n+1)})$. \label{taylorfn}}
\be
f_1(\phi(x))= f_1(x)+O_0(r^{-2}). \label{tf0}
\ee
From (\ref{f0}) and (\ref{tf0}) we conclude (\ref{fphi0}).  To show (\ref{fphin}) we use  chain rule formula  to take the $n$-th partial derivative of $f(\phi(x))$,
\be
\partial_{\beta_n} \ldots \partial_{\beta_1} f(\phi(x))= \sum_{m=1}^{n} C^{\gamma_m \ldots \gamma_1}_{\beta_n \ldots \beta_1 }(x)  (\partial_{\gamma_m} \ldots \partial_{\gamma_1} f)(\phi(x)), \label{chain1O}
\ee
where:
\be
C^{\gamma_m \ldots \gamma_1}_{\beta_n \ldots \beta_1 }(x):= \sum \partial_{\beta_{i^{j_m}_{I_m}}} \ldots \partial_{\beta_{i^1_{I_m}}}\phi^{\gamma_m}(x) \cdots \partial_{\beta_{i_{I_1}^{j_1}}} \ldots \partial_{\beta_{i_{I_1}^{1}}} \phi^{\gamma_1}(x), \label{chain2O}
\ee
 where the  sum is over all distinct partitions $I_1,\ldots, I_m$ of $\{1,\ldots,n\}$ with $I_k=\{i_{I_k}^{1},\ldots, i_{I_k}^{j_k}\}$, $i_{I_k}^{1} <i_{I_k}^{2} <\ldots < i_{I_k}^{j_k}$ (thus $j_k$ is the cardinality of $I_k$ and $j_1+\ldots+j_m=n$). The formula  (\ref{chain1O}), (\ref{chain2O}) is a realization to the present  case  of an abstract  chain rule formula described in appendix A of \cite{grosser}.   
 
 From (\ref{gralf}) and (\ref{Ophi}) we have the following rough bound:
 \be
 (\partial_{\gamma_m} \ldots \partial_{\gamma_1} f)(\phi(x)) =O_0(r^{-(m+1)}). \label{rbphi}
 \ee
Using (\ref{rbphi}) and the following rough bounds on  derivatives of  $\phi$: 
\be
\partial \phi (x) =O_0(1),  \quad \partial^n \phi(x)= O_0(r^{-n}) , \;n>1,
\ee
 one can verify that all $m< n$ terms in (\ref{chain1O}) are $O_0(r^{-(n+2)})$ whereas the $m=n$ one is $O_0(r^{-(n+1)})$. It then follows that:
 \be
\partial_{\beta_n} \ldots \partial_{\beta_1} f(\phi(x)) =  (\partial_{\beta_n} \ldots \partial_{\beta_1} f)(\phi(x))+O_0(r^{-(n+2)}),\label{pnf}
 \ee
where we have used that for $m=n$, Eq. (\ref{chain2O}) reduces to: $C^{\gamma_n \ldots \gamma_1}_{\beta_n \ldots \beta_1 }= \partial_{\beta_n} \phi^{\gamma_n} \ldots \partial_{\beta_{1}} \phi^{\gamma_1}$.
Using (\ref{gralf}) for $f$ in the right hand side of (\ref{pnf}) we obtain
 \be
\partial_{\beta_n} \ldots \partial_{\beta_1} f(\phi(x)) =  (\partial_{\beta_n} \ldots \partial_{\beta_1} f_1)(\phi(x))+O_0(r^{-(n+1)-\epsilon}).\label{pnf2}
 \ee
 Finally, Taylor expanding the first term in (\ref{pnf2}) around $x$ we find:\footnote{See footnote \ref{taylorfn}. When $p \leq l$, the condition  $g(\xh)$ being $C^{p+1}$ is used to obtain the  $n=p$ case of (\ref{tfn}). \label{pp1}} 
 \be
(\partial_{\beta_n} \ldots \partial_{\beta_1} f_1)(\phi(x))= (\partial_{\beta_n} \ldots \partial_{\beta_1} f_1)(x)+O_0(r^{-(n+2)}). \label{tfn}
\ee
Using (\ref{tfn}) in (\ref{pnf2}) we obtain the desired result (\ref{fphin}). For later use in appendix \ref{infintesimalapp} we note that the proof goes through if  $p=l=k=\infty$, so that the analogue of equation (\ref{fphi}) still holds in the smooth setting. The only difference is that in this case the  $O_0(r^{-(n+1)-\epsilon})$ bounds in (\ref{fphin}) are in general  of the form $c_n r^{-(n+1)-\epsilon} $ with $n$-dependent constants $c_n$. 

This result can be used to show that the composition of (\ref{gralf}) with a $C^l$ diffeomorphism with nontrivial rotation
\be
\phi(x)^\alpha= R^\alpha_\beta(x^\beta +\yo^{\beta}(\xh)) +O(r^{-\e}) \label{phiRO}
\ee
is given by
\be
f(\phi(x))=f_0+ f_1(R(x))+O(r^{-1-\epsilon}),\label{fphiR}
\ee
where $R(x)^\alpha \equiv R^\alpha_\beta x^\beta$. To see (\ref{fphiR}), write $\phi^\alpha(x) =R^\alpha_\beta \psi^\beta(x)$ with $\psi^\alpha(x):=(R^{-1})^\alpha_\beta \phi^\alpha(x)$. Next, note that $F(x)=O(r^{-\beta}) \iff F(R(x)) =O(r^{-\beta}) $. It follows that $f(R(x))=f_0+f_1(R(x))+O(r^{-1-\epsilon})$. Using (\ref{fphi}) for $f(R(x))$ in place of $f(x)$ and $\psi^\alpha(x)$ in place of $\phi^\alpha(x)$ one obtains (\ref{fphiR}), where it is easily verified that $f_1(R(x))$ satisfies the same conditions of the type (\ref{f1a2}) which are satisfied by $f_1(x)$.
 
We finally show that for $\phi$ as in (\ref{phiRO}) and $a(x)$ as in (\ref{grala}) one has:
\be
a(\phi(x))= a_2(R(x))+ O(r^{-2-\epsilon}) \label{grala2}.
 \ee
This can be seen by writing   $a(x) = \left[ r(x) a(x) \right]\left[ 1/r(x) \right]$, $r(x)\equiv \sqrt{x^\alpha x^\alpha}$, 
and noting that:  The function  $f(x):=r(x) a(x)$ is of the form (\ref{gralf}) (with $f_0=0$), and so it satisfies  (\ref{fphiR});  $1/r(x)$ is also of the form (\ref{gralf}) (with $f_0=0$ and $f_1=r^{-1}$) and so  $1/r(\phi(x))=1/r+O(r^{-1-\epsilon})$. Substituting 
 these expansions in 
\be
a(\phi(x))= \left[r(\phi(x)) a(\phi(x))\right] \left[1/r(\phi(x))\right] \label{rarinv}
\ee
one obtains (\ref{grala2}).

\section{Phase space description of gauge and asymptotic symmetry group} \label{infintesimalapp}
In this appendix we study the classical phase space description of  gauge and asymptotic symmetries.  After reviewing the infinitesimal generators as described in \cite{mcparity}, we study the corresponding finite transformations. As in \cite{beigom,ttparity,mcparity}, we restrict attention to  $C^\infty$ fields. 
\subsection{Infinitesimal transformations} \label{infapp}

In standard  phase space description,  the kinematical gauge transformations are generated by the Gauss and diffeomorphism constraints:
\be
G[\Lambda] := -\int_\Sigma \tr[\Lambda (\partial_a E^a+[A_a,E^a] )] , \;  \quad D[\xi] := \int_{\Sigma}  \tr[E^a \L_{\xi}A_a ] .\label{diffconst}
\ee
The fall-off conditions 
\be 
\Lambda^i = \frac{\lambda^i(\xh)}{r}+O(r^{-1-\e}),  \; \quad \xi^\alpha = S^\alpha(\xh)+O(r^{-\e}) ,
\ee
with $\lambda(\xh)$ even and $S^\alpha(\xh)$ odd  ensure the well-definedness of the constraints as phase space functions. Their action is then given by:
\be
\begin{array}{lllll}
(\Lambda,\xi) \cdot A_a & :=& [\Lambda,A_a] - \L_\xi A_a -\partial_a \Lambda &=& \{  G[\Lambda]+D[\xi], A_a \}, \\
(\Lambda,\xi) \cdot E^a &:=& [\Lambda,E^a]- \L_\xi E^a &=& \{  G[\Lambda]+D[\xi], E^a \} , \label{lieautAE}
\end{array}
\ee
and their commutator evaluates to:
\be
[(\Lambda,\xi),(\Lambda',\xi')]  = (-\L_\xi \Lambda' + \L_{\xi'} \Lambda+[\Lambda,\Lambda'], -[\xi,\xi']), \label{commgen}
\ee
which  corresponds to a representation of the Poisson bracket algebra of the constraints (\ref{diffconst}).  Equations (\ref{lieautAE}), (\ref{commgen}) represent the infinitesimal version of (\ref{autAE}) and (\ref{aap}) respectively.  In the case of  asymptotic symmetries, the fall-off of $SU(2)$ multiplier and shift are:  
\ba
\Lambda^i & = &\LR^i+\frac{\lambda^i(\xh)}{r}+O_2(r^{-1-\e}) \label{falllam} ,\\
\xi^\alpha & = & T^\alpha +R^\alpha + S^\alpha(\xh)+O_2(r^{-\e}) ,\label{fallshift}
\ea
where
\be
\LR^i :=\frac{1}{2} \e_{ijk}\Eo^j_a \L_{R} \Eo^a_k  \label{LR}.
\ee
The phase space action of these generators is given by the same expressions as in the $\aut$ case,  Eq. (\ref{lieautAE}), and the commutator of generators by (\ref{commgen}).  This action is generated via Poisson brackets by well-defined phase space functions representing  the total linear and angular momenta  of the spacetime \cite{ttparity,mcparity}.

\subsection{Finite transformations} \label{expapp}
We now describe the asymptotic form of  the finite transformations associated to the above infinitesimal generators. We proceed in steps:

\subsubsection{Flow equations}
Given a pair $(\Lambda,\xi)$ as in (\ref{falllam}), (\ref{fallshift}) we wish to solve the `flow equation' $d a_t / dt = (\Lambda,\xi) a_t$. Writing $a_t=(g_t,\phi_t)$ and assuming the composition law (\ref{aap}) this translates into:
\ba
\frac{d}{dt} \phi^a_t  & = &  \xi^a (\phi_t) ,\label{odephi} \\
\frac{d}{dt} g_t & =  & \Lambda g_t-\L_\xi g_t. \label{odegt}
\ea
Conversely, it is easy to verify that the flow equations (\ref{odephi}) and (\ref{odegt}) together with (\ref{aap}) yield the infinitesimal transformations of section \ref{infapp} so that (\ref{aap}) is indeed a finite version of these infinitesimal transformations. 

\subsubsection{Asymptotic form of diffeomorphisms (\ref{odephi})}
We show  the diffeomorphisms $\phi_t$ obtained by integrating (\ref{odephi}) have  asymptotic form as in (\ref{fallphi}).  Define the following  vector fields in the asymptotic region:
\be
\xio^\alpha :=T^\alpha +R^\alpha, \quad  Y^\alpha := \xi^\alpha - \xio^\alpha, \quad \delta^\alpha := Y^\alpha- S^\alpha,
\ee
so that
\ba
\xi^\alpha & = & \xio^\alpha+ Y^\alpha \label{xiY}, \\
Y^\alpha & = & S^\alpha +\delta^\alpha ,\\
\delta^{\alpha} &= & O(r^{-\e}). \label{falldelta}
\ea
Let $\phio_t$ be the flow of  $\xio^\alpha$ and define
\ba
\psi_t &:=& \phio^{-1}_t \circ \phi_t \label{defpsit} \\
 Y^\alpha_t &:= &\phio^{-1}_{t \, *} Y^\alpha. \label{defYt}
\ea
It is easy to verify that $\psi_t$ is the flow of the $t$-dependent vector field $Y^\alpha_t$, that is, $\psi_{t=0}(x)=x$ and
\be
\frac{d}{dt} \psi_t^\alpha(x)= Y^\alpha_t(\psi_t(x)). \label{odepsi}
\ee
Indeed, one has:
\ba
\frac{d}{dt} \phio^{-1}_t(\phi_t(x))^\alpha &= &- \xio^\alpha( \phio^{-1}_t (\phi_t(x))) +\frac{\partial (\phio^{-1}_t)^\alpha}{\partial y^\beta}(\phi_t(x))\xi^\beta (\phi_t(x)) \\
& =& - \xio^\alpha( \psi_t(x)) + (\phio^{-1}_{t \, *} \xi)^\alpha(\psi_t(x))\\ 
&=& (\phio^{-1}_{t \, *} Y)^\alpha(\psi_t(x)), \label{odephit} 
\ea
where in going from the first to second line we used the local coordinate expression for the push-forward of vector fields: $(\phi_* X)^\alpha(x)= \partial \phi^\alpha/\partial y^\beta(\phi^{-1}(x)) X^{\beta}(\phi^{-1}(x))$, and in going to the last line we used that $\phio^{-1}_{t \, *} \xio^a=\xio^a$.   

$\phio_t$ is  given by:
\be
\phio_t(x) = R_t(x) +c_t, \label{phio}
\ee
with $R_t(x) \equiv (R_t)_\beta^\alpha x^\beta$  the rotation generated by $R^\alpha$ (so that $\frac{d}{dt}R_t(x)^\alpha =R^\alpha(R_t(x))$) and $c^\alpha_t= (R_t)^\alpha_\beta \int_0^t (R_{t'}^{-1})^\beta_\gamma T^\gamma dt'$ the translation piece.   The nontrivial result we will  show is that:
\be
\psi^\alpha_t(x)= x^\alpha +s^\alpha_t(\hat{x})  + O(r^{-\epsilon}) , \label{fallpsi}
\ee
with $s^\alpha_t(\hat{x})$ odd. From (\ref{fallpsi}), (\ref{phio}), (\ref{defpsit}) the desired result (\ref{fallphi}) directly follows.  Condition (\ref{fallpsi}) can be expressed as:
\ba
 \psi^\alpha_t(x) & =& x^\alpha+s^\alpha_t(\hat{x})+ O_0(r^{-\epsilon}) \label{psi0} \\
\partial_{\beta_n} \ldots \partial_{\beta_1} \psi^\alpha_t(x) & =& \partial_{\beta_n} \ldots \partial_{\beta_1}(x^\alpha+s^\alpha_t(\hat{x}))+ O_0(r^{-n-\epsilon}), \quad n=1,2, \ldots  , \label{falldnphi}
\ea
where $O_0(r^{-\gamma})$ denotes bounded by $c r^{-\gamma}$ for some constant $c$. We  first show (\ref{psi0}) and then provide an induction argument for  (\ref{falldnphi}). The integrated version of (\ref{odepsi}) reads:
\be
\psi_t(x)^\alpha = x^\alpha + \int^t_0 Y_{t'}^{\alpha}(\psi_{t'}(x)) dt' . \label{intpsit}
\ee
Let 
\be
y^\alpha_t(x):= \int^t_0 Y_{t'}^{\alpha}(\psi_{t'}(x)) dt'  .\label{yt}
\ee
Boundedness of $Y^\alpha$ implies boundedness of  $Y^\alpha_t$ which in turn implies implies boundedness of $y^\alpha_t(x)$. Thus $\psi_t(x)$ stays in the asymptotic region and $y^\alpha_t(x)=O_0(r^{0})$:
\ba
\psi^\alpha_t(x)  &=& x^\alpha +y^\alpha_t(x)\\
& =  &x^\alpha+ O_0(r^0). \label{psi1}
\ea
We now use (\ref{psi1}) to determine the asymptotic form of the integrand in (\ref{intpsit}):
\be
Y_t^\alpha(\psi_t(x)) =  \phio^{-1}_{t \, *} S^\alpha( \psi_t(x) ) + \phio^{-1}_{t \, *} \delta^\alpha(\psi_t(x)) . \label{Ytdec}
\ee
For the first term in (\ref{Ytdec}) we have:
\ba
\phio^{-1}_{t \, *} S^\alpha(\psi_t(x)) &  =  & (R_{t}^{-1})^\alpha_\beta S^\beta(R_t(x)+ R_t(y_t(x)) +c_t )  \\
&  =  & (R_{t}^{-1})^\alpha_\beta S^\beta(R_t(x)) +  O_0(r^{-1}), \label{yqs2}
\ea
where we used a Taylor expansion to zeroth order yielding a remainder which is $O_0(r^{-1})$.\footnote{Taylor expansions in (\ref{yqs2}), (\ref{lastpSt}) and (\ref{lastpnSt}) are realized along the lines described in footnote \ref{taylorfn}.}
For later purposes we denote the first term in (\ref{yqs2}) by
\be
S^\alpha_t(x):=(R_{t}^{-1})^\alpha_\beta S^\beta(R_t(x)). \label{defSt}
\ee
The second term in  (\ref{Ytdec}) is $O_0(r^{-\epsilon})$. 
We thus conclude that:
\be
Y_t^\alpha(\psi_t(x)) = S^\alpha_t(x)+ O_0(r^{-\epsilon}) .
\ee
Integrating with respect to $t$ as in (\ref{intpsit}), and using that derivatives with respect to $x$ commute with the $t$ integral we conclude:
\be
\psi^\alpha_t(x)= x^\alpha + s^\alpha_t(\hat{x}) +O_0(r^{-\epsilon})  \label{psios}
\ee
where
\be
s^\alpha_t(\hat{x}) := \int_0^t S^\alpha_{t'}(\xh) dt'.
\ee
We thus recover  Eq. (\ref{psi0}).  We now show (\ref{falldnphi}) for $n=1$. Let
\be
D^\alpha_{\beta \, t}(x) := \partial_\beta \psi_t^\alpha(x) . \label{defL}
\ee
Differentiating (\ref{odepsi}) with respect to $x^\beta$ we find that $D^\alpha_{\beta \, t}(x)$ satisfies the differential equation:\footnote{We assume that, as in the compact manifold case, $\psi_t(x)$ is smooth as a function from $\reals \times \Sigma$ to $\Sigma$. In particular this implies $\partial_{x^\alpha} \partial_t \psi_t(x)= \partial_t \partial_{x^\alpha} \psi_t(x)$ which is what is being used to obtain (\ref{odeL}). Similar considerations are needed when showing  the remaining $n>1$ terms of (\ref{falldnphi}). \label{dxdtcomm}}
\be
\frac{d}{dt} D^\alpha_{\beta \, t}(x) = M^\alpha_{\gamma \, t}(x) D^\gamma_{\beta \, t}(x)  \label{odeL}
\ee
with
\be
M^\alpha_{\gamma \, t}(x)  := \partial_\gamma Y_t^\alpha (\psi_t(x)). \label{defM}
\ee
For fixed $x$ and $\beta$,  (\ref{odeL}) represents an ordinary differential equation for the three-component vector  $v^\alpha(t)=D^\alpha_{\beta \, t}(x)$. The solution with initial condition $D^\alpha_{\beta \, t=0}(x)=\delta^\alpha_\beta$ is then given by:
\ba
D^\alpha_{\beta \, t}(x) & = & {\rm T} \exp ( \int_0^t M_{t'}(x) dt')^\alpha_\beta \\
& = & \delta^\alpha_\beta +  \int_0^t M^\alpha_{\beta \, t'}(x) dt' + \int_0^t \int_0^{t'} M^\alpha_{\gamma \, t'}(x) M^\gamma_{\beta \, t''}(x) dt'' dt' + \ldots. \label{oeM}
\ea
We now describe the asymptotic form of  $M^\alpha_{\gamma \, t}(x)$.  There are two contribution to this matrix:
\be
M^\alpha_{\beta \, t}(x)  =    \partial_\beta (\phio^{-1}_{t \, *} S^\alpha) ( \psi_t(x) ) + \partial_\beta (\phio^{-1}_{t \, *} \delta^\alpha)(\psi_t(x)) . \label{Mdec}
\ee
For the first term in (\ref{Mdec}) we have:
\ba
 \partial_\beta (\phio^{-1}_{t \, *} S^\alpha)( \psi_t(x) ) &= & (R_{t}^{-1})^\alpha_\mu (R_t)^\nu_\beta \partial_\nu S^\mu(R_t(x)+ R_t(y_t(x)) +c_t)  \\
 &= & (R_{t}^{-1})^\alpha_\mu (R_t)^\nu_\beta \partial_\nu S^\mu(R_t(x)) + O_0(r^{-2}) , \\
 &= & \partial_\beta S^\alpha_t(x) +O_0(r^{-2}), \label{lastpSt}
\ea
where we Taylor expanded to zeroth order obtaining a remainder which is $O_0(r^{-2})$  and used (\ref{defSt}) to express the first term in compact form. The second term in (\ref{Mdec}) is  $O_0(r^{-1-\epsilon})$ and we conclude that:
\be
M^\alpha_{\beta \, t}(x) = \partial_\beta S_t^\alpha(x) +O_0(r^{-1 - \epsilon}) . \label{fallM}
\ee 
Using  (\ref{fallM}) in (\ref{oeM}) we obtain (only the linear term in $M$ contributes, higher order ones are already $O_0(r^{-2})$):
\be
D^\alpha_{\beta \, t}(x)  =\delta^\alpha_\beta + \partial_\beta s_t^\alpha(x) +O_0(r^{-1 - \epsilon}) , \label{fallL}
\ee
which corresponds to  $n=1$ of Eq. (\ref{falldnphi}).  We now sketch the argument for the general $n$ case. For $n=1,2,\ldots$ Define the tensors:
\ba
D^{\alpha}_{\beta_n \ldots \beta_1 \, t}(x) & := & \partial_{\beta_n} \ldots \partial_{\beta_1} \psi^\alpha_t(x). \label{defDn} \\
M^\alpha_{\beta_n \ldots \beta_1 \, t}(x) & := & (\partial_{\beta_n} \ldots \partial_{\beta_1} Y_t^\alpha)(\psi_t(x)). \label{defMn}
\ea
In order to avoid notational clutter we are omitting an  `$n$' label in the name of each tensor; tensors associated to different values of $n$ are only distinguished by the number of indices they have. The $n=1$ case corresponds to the  tensors (\ref{defL}), (\ref{defM}).


Differentiating (\ref{odepsi}) $n$ times, commuting $t$ and $x$ derivatives (see footnote \ref{dxdtcomm}) and using chain rule one obtains a system of differential equations in $t$ satisfied by the tensors $D^\alpha_{\beta_n \ldots \beta_1 \, t}(x)$:
\be
\frac{d}{dt} D^\alpha_{\beta_n \ldots \beta_1 \, t}(x) = \sum_{m=1}^{n} C^{\gamma_m \ldots \gamma_1}_{\beta_n \ldots \beta_1 \, t}(x) M^\alpha_{\gamma_m \ldots \gamma_1 \, t}(x), \label{chain1}
\ee
where \cite{grosser},
\be
C^{\gamma_m \ldots \gamma_1}_{\beta_n \ldots \beta_1 \, t}(x):= \sum \, D^{\gamma_m}_{\beta_{i^{j_m}_{I_m}} \ldots\beta_{i^1_{I_m}} \, t}(x) \ldots D^{\gamma_1}_{\beta_{i_{I_1}^{j_1}} \ldots\beta_{i_{I_1}^{1}} \, t}(x), \label{chain2}
\ee
where the sum is over partitions of $\{1,\ldots,n\}$ as described for Eq. (\ref{chain2O}).  For given $n$, Eq. (\ref{chain1}) involves the differentials of order $m \leq n$. The only occurrence of the $n$-th order differential is in the $m=1$ term, for which (\ref{chain2})  becomes:
\be
C^{\gamma}_{\beta_n \ldots \beta_1 \, t}(x) = D^{\gamma}_{\beta_n \ldots \beta_1 \, t}(x) .
\ee 
All remaining $m>1$ sumands of (\ref{chain1}) involve differentials of order strictly less than $n$. If we collectively denote these terms by:
\be
N^\alpha_{\beta_n \ldots \beta_1 \, t}(x) := \sum_{m=2}^{n} C^{\gamma_m \ldots \gamma_1}_{\beta_n \ldots \beta_1 \, t}(x) M^\alpha_{\gamma_m \ldots \gamma_1 \, t}(x),\label{defN}
\ee
equation (\ref{chain1}) takes the form:
\be
\frac{d}{dt} D^\alpha_{\beta_n \ldots \beta_1 \, t}(x) = M^\alpha_{\gamma \, t}(x) D^{\gamma}_{\beta_n \ldots \beta_1 \, t}(x) + N^\alpha_{\beta_n \ldots \beta_1 \, t}(x) . \label{odeD}
\ee
For fixed $x$, and $\beta_n \ldots \beta_1$, Eq. (\ref{odeD}) represents an inhomogenous ordinary linear differential equation in $t$ for the three dimensional vector $v^\alpha(t):=D^\alpha_{\beta_n \ldots \beta_1 \, t}(x) , \alpha=1,2,3$. Since the source term $N^\alpha_{\beta_n \ldots \beta_1 \, t}(x)$ depends on the differentials of order smaller than $n$, the solutions can be obtained  iteratively.  For $n>1$, the general solution to (\ref{odeD}) with initial condition $D^\alpha_{\beta_n \ldots \beta_1 \, t=0}(x)=0$ is given by:
\be
D^\alpha_{\beta_n \ldots \beta_1 \, t}(x) = D^\alpha_{\mu \, t}(x) \int_{0}^{t} (D^{-1})^\mu_{\nu \, t'}(x) N^\nu_{\beta_n \ldots \beta_1 \, t'}(x)   dt' , \label{solD}
\ee
where $(D^{-1})^\alpha_{\beta \, t}(x)$ is the inverse matrix of $D^\alpha_{\beta \, t}(x)$ (the invertibilty of this matrix follows from the fact that it is the differential of a diffeomorphism). In order to determine the asymptotic form of  (\ref{solD}) we need to study the fall-offs of (\ref{defN}). This requires knowledge of the fall-offs of the differentials of $\psi_t$ up to order $n-1$ (which we assume are given) as well as knowledge of the fall-offs of (\ref{defMn}). For the latter there are two contributions:
\be
 M^\alpha_{\beta_n \ldots \beta_1 \, t}(x) = (\partial_{\beta_n} \ldots \partial_{\beta_1}  (\phio^{-1}_{t \, *} S^\alpha) ( \psi_t(x) ) +(\partial_{\beta_n} \ldots \partial_{\beta_1}  (\phio^{-1}_{t \, *} \delta^\alpha) ( \psi_t(x) ) . \label{Mndec}
\ee
For the first term in (\ref{Mndec}) we have:
\ba
(\partial_{\beta_n} \ldots \partial_{\beta_1}  \phio^{-1}_{t \, *} S^\alpha) ( \psi_t(x) ) &= & (R_{t}^{-1})^\alpha_\mu (R_t)^{\nu_n}_{\beta_n}\ldots  (R_t)^{\nu_1}_{\beta_1} \partial_{\nu_n} \ldots \partial_{\nu_1}  S^\mu(\phio_t(\psi_t(x)))  \\
 &= &\partial_{\beta_n} \ldots \partial_{\beta_1}  S^\alpha_t(x) +O_0(r^{-n-1}), \label{lastpnSt}
\ea
where we Taylor expanded to zeroth order obtaining a remainder which is $O_0(r^{-n-1})$ and used (\ref{defSt}) to express the first term in compact form. The second term in (\ref{Mndec}) is $O_0(r^{-n-\epsilon})$ and so it follows that:
\be
 M^\alpha_{\beta_n \ldots \beta_1 \, t}(x) = \partial_{\beta_n} \ldots \partial_{\beta_1}  S^\alpha_t(x)+O_0(r^{-n-\epsilon})  . \label{fallMn}
\ee
Going now to  (\ref{defN}),   one can verify that the highest order term is given by the $m=n$ sumand, for which (\ref{chain2}) becomes the product of $n$ first order differentials:
\be
C^{\gamma_n \ldots \gamma_1}_{\beta_n \ldots \beta_1 \, t}(x):=  D^{\gamma_n}_{\beta_n \, t}(x) \ldots D^{\gamma_1}_{\beta_1 \, t}(x).
\ee
Using that  $D^{\gamma}_{\beta \, t}(x)= \delta^\gamma_\beta+O_0(r^{-1})$  and (\ref{fallMn})  we conclude:
\be
N^\alpha_{\beta_n \ldots \beta_1 \, t}(x) =  \partial_{\beta_n} \ldots \partial_{\beta_1}  S^\alpha_t(x)+O_0(r^{-n-\epsilon}) . \label{falloffN}
\ee
Using (\ref{falloffN}) in (\ref{solD}) one obtains the $n$-th condition in Eq. (\ref{falldnphi}).

\subsubsection{Step  3: Integrating the flow equation (\ref{odegt})}

A solution to (\ref{odegt}) is given by:
\be
g_t= (\phi_t)_*( {\rm T} e^{ \int_{0}^t (\phi^{-1}_{t'})_* (\Lambda) d t'}), \label{gt}
\ee
where ${\rm T}$ denotes the ordered matrix product so that
\be
\frac{d}{dt} {\rm T} e^{ \int_{0}^t (\phi^{-1}_{t'})_* (\Lambda) d t'} = (\phi^{-1}_{t})_*(\Lambda) {\rm T} e^{ \int_{0}^t (\phi^{-1}_{t'})_* (\Lambda) d t'} . \label{Tordered}
\ee
We assume that (\ref{gt}) is the unique solution to (\ref{odegt}) with initial condition $g_{t=0}=\idtwo$. 
It is then straightforward to verify that $a_t=(g_t, \phi_t)$ 
 satisfies the one-parameter subgroup property $a_t a_s = a_{t+s}$ 
with the product rule given by Eq. (\ref{aap}).


Next, we argue that the explicit expression (\ref{gt}) satisfies the analogue of the fall-offs described in Eq. (\ref{fallg}). 
Consider first the $SU(2)$-valued function arising from the time ordered exponential in (\ref{gt}) 
satisfying (\ref{Tordered}). For $\Lambda$ and $\xi^a$ as in (\ref{falllam}),
(\ref{fallshift}) and the smooth version of composition property (\ref{fphi}) described in section \ref{Oapp}  we obtain
\be
(\phi^{-1}_{t})_*\Lambda  =\Lo + (\even) r^{-1} + O(r^{-1-\epsilon}) .
\ee
 Note that
here and in the following equations, the  terms  $(\even) r^{-1} + O(r^{-1-\epsilon})$ 
are in general $t$-dependent. Let
\be
h_t := e^{-t \Lo} {\rm T} e^{ \int_{0}^t (\phi^{-1}_{t'})_* (\Lambda) d t'} \label{ht}.
\ee
It satisfies
\be
\frac{d}{dt} h_t = \tilde{\Lambda}_t h_t 
\ee
with
\be
\tilde{\Lambda}_t = (\phi^{-1}_{t})_*(e^{-t \Lo} (\Lambda-\Lo) e^{t \Lo}) = (\even) r^{-1} + O(r^{-1-\epsilon}) , \label{tildelam}
\ee
where again we used the composition property (\ref{fphi}). The expression for $h_t$ can then be written as:
\be
h_t = {\rm T} e^{ \int_{0}^t \tilde{\Lambda}_{t'} d t'}.
\ee
Writing down the explicit power series 
defining the ordered matrix product and using the fall-off in (\ref{tildelam}) one finds
\be
h_t= \idtwo +(\even) r^{-1} + O(r^{-1-\epsilon}) .
\ee
From Eqns. (\ref{ht}), (\ref{gt}), and composition property (\ref{fphi})  we find:
\be
g_t = e^{t \Lo} + (\even) r^{-1} + O(r^{-1-\epsilon}), 
\label{gtasymp}
\ee
which is what we wanted to show.\footnote{Note that  $\Lo= \LR = \theta$ (see equation (\ref{LR})) so that the rotation of the fiducial flat triad $\Eo^a$ 
by   $e^{t \Lo}$ is exactly `undone' by the leading order part of $\phi_t$.\label{goundo}}

\section{Supplementary material for section \ref{sec8}} \label{appsec8}

\subsection{Finiteness of $\alpha(a,E)$} \label{finitealphaapp}
From the fall-offs of $g$ and its derivative:
\be
g  =  \go+ (\even) r^{-1}+O(r^{-1-\e})  ,   \quad \partial_a g = (\odd) r^{-2}+O(r^{-2-\e}), \label{falldg}
\ee
we find:
\ba
\partial_a g g^{-1} & = & \partial_a g (\go^{-1} + (\even) r^{-1}+O(r^{-1-\e}))  \nn  \\
& = & \partial_a (g \go^{-1}) + (\odd) r^{-3}+O(r^{-3-\e}) ,\label{partialgginv}
\ea
where we used that $\partial_a \go^{-1} =0$.  Using (\ref{falldg}), (\ref{partialgginv}) and the fact that $a \cdot E^a$ satisfies the fall-off conditions (\ref{fallE}),
\be
a \cdot E^\alpha = \Eo^\alpha +(\even) r^{-1}+O(r^{-1-\e}) ,\label{fallaE}
\ee
 the asymptotic form of the first term in (\ref{rho}) is found to be:
\be
\tr[(a \cdot E^\alpha) \partial_\alpha g g^{-1}] =  \partial_\alpha \tr[\Eo^\alpha  (g \go^{-1})] +(\odd) r^{-3}+O(r^{-3-\e}) ,\label{1st}
\ee
where we used  $\partial_\alpha \Eo^\alpha=0$ to bring $\Eo^\alpha$ inside the derivative.  

We now focus on the total derivative  term in (\ref{rho}). Using that $\tr[E^a]=\tr[\Eo^a]=0$ together with, (\ref{falldg}), (\ref{fallaE}), we find:
\be
\tr[(E^\alpha - \Eo^\alpha) g  \go^{-1}] =(\even) r^{-2}+ O(r^{-2-\e}),  \label{EEo}
\ee
from which we obtain the following  asymptotic form for the second term in (\ref{rho}):
\be
-\partial_\alpha \tr[E^\alpha g \go^{-1}]  = - \partial_\alpha \tr[\Eo^\alpha  (g \go^{-1})] + (\odd) r^{-3}+O(r^{-3-\e}) .\label{2nd}
\ee
Adding (\ref{1st}) and (\ref{2nd}) we find
\be
\rho(a,E) =(\odd) r^{-3}+O(r^{-3-\e}) .
\ee
so that $\alpha(a,E)$ is finite.

We conclude with two observations  related to Eq. (\ref{EEo}). First, we note that using (\ref{EEo}), $\alpha(a,E)$ can  alternatively be written as,
\be
\alpha(a,E)= \int_\Sigma \tr[a \cdot E^a \, \partial_a g g^{-1}]  -\oint_\infty dS_a \tr[\Eo^a g \go^{-1}] . \label{alpha3}
\ee
The form (\ref{alpha3}) makes it explicit that the `additional' surface term is sensitive only to the limiting value of $E^a$.  Second, we note that for $\Eb^a \in \E$ we have
\be
\tr[\Eb^\alpha g  \go^{-1}] =(\even) r^{-2}+ O(r^{-2-\e}),  \label{sfcEb}
\ee
so that
\be
\oint_\infty dS_a \tr[\Eb^a g  \go^{-1}] =0. \label{intsEb}
\ee
Eq. (\ref{sfcEb}) follows from the same argument given for Eq. (\ref{EEo}). Eq. (\ref{intsEb}) implies that if one uses formula (\ref{alpha2}) with $\Eb^a \in \E$, one recovers the phase factor for `barred' electric fields, Eq. (\ref{alpha}). This in turn implies Eq. (\ref{alphaEEb}).

\subsection{Eq. (\ref{phasesaap})} \label{phasecompapp}

Let $a=(g,\phi)$, $a'=(g',\phi')$ and 
\be
a''=(g'',\phi''):= a a' = (g \phi_* g', \phi \phi'). \label{app}
\ee
The phase associated to $a''$ is 
\be
\alpha(a'',E) = \int_\Sigma \rho(a'',E),
\ee
with $\rho$ in (\ref{rho}) evaluated on the group element (\ref{app}):
\ba
\rho(a'',E) & = & \tr[ a \cdot (a' \cdot E^a) \partial_a(g \phi_* g') \phi_* g'^{-1} g^{-1} ]- \partial_a  \tr[ E^a g'' \go''^{-1}] \\
& = & \tr[ a \cdot (a' \cdot E^a) \partial_a g g^{-1}] + \phi_* \tr[a' \cdot E^a \partial_a g' g'^{-1}]- \partial_a  \tr[ E^a g'' \go''^{-1}]. \label{rhopp}
\ea
We now focus on the total derivative term in (\ref{rhopp}). Using the expansion (\ref{falldg}) for $g$ and $g'$ and the fact that  $\phi_* \go'=\go'$ (since $\go'$ is a constant) we find:
\ba
\go'' & = & \go \go' \\
g'' & = & g \go'+ \go \phi_* g' -  \go  \go'+ (\even) r^{-2}+ O(r^{-2-\e})  \label{gppgg} \\
g'' \go''^{-1} & =& g \go^{-1} +\go  (\phi_* g' \go'^{-1})\go^{-1}  -\idtwo + (\even) r^{-2} + O(r^{-2-\e}). \label{gohpp}
\ea
The  relevant terms appearing in $\tr[ E^a g'' \go''^{-1}]$ are then:
\be
\tr[E^\alpha g \go^{-1}] = \tr[a' \cdot E^\alpha \,  g \go^{-1}] + (\even) r^{-2} + O(r^{-2-\e}) , \label{Eggo}
\ee
where the equality follows from Eq. (\ref{sfcEb}) together with the fact that $a' \cdot E^a - E^a \in \E$,
and
\ba
\tr[\go^{-1} E^\alpha \go  \phi_* g' \go'^{-1} ] &= &   \phi_* \tr[ \phi^{-1}_* (\go^{-1} E^\alpha \go) g' \go'^{-1} ]  \nn \\
&= & \phi_* \tr[  E^\alpha g' \go'^{-1} ] +(\even) r^{-2} + O(r^{-2-\e}), \label{tdrhopp}
\ea
where in the last equality we again used Eq. (\ref{sfcEb}) since,
\be
\phi^{-1}_* (\go^{-1} E^\alpha \go) = \Eo^\alpha +(\even) r^{-1}+ O(r^{-1-\e}) .
\ee
From (\ref{gohpp}), (\ref{Eggo}) and (\ref{tdrhopp}) it follows that
\be
\tr[ E^a g'' \go''^{-1}] = \tr[a' \cdot E^\alpha \,  g \go^{-1}] +\phi_* \tr[  E^\alpha g' \go'^{-1} ] +X^a, \label{trEgpp}
\ee
where,
\be
X^a = (\even) r^{-2}+ O(r^{-2-\e}), \label{X}
\ee
collects the remainder terms arising in (\ref{gohpp}), (\ref{Eggo}) and (\ref{tdrhopp}).

Using Eq. (\ref{trEgpp}) in (\ref{rhopp}), and recalling the definition of $\rho$ (Eq. (\ref{rho})), Eq. (\ref{rhopp}) becomes:
\be
\rho(a'',E) = \rho(a' \cdot E,a) + \phi_* \rho(a',E) +\partial_a X^a ,\label{rhoppfinal}
\ee
 Integrating (\ref{rhoppfinal}) over $\Sigma$ we find that the total derivative term in (\ref{rhoppfinal}) give zero contribution since $\oint_{\infty} dS_a X^a=0$ for $X^a$ as in (\ref{X}). Finally, using $\int_{\Sigma} \phi_* \rho(a',E)  = \int_{\Sigma}  \rho(a',E) = \alpha(a',E)$ we obtain (\ref{phasesaap}).

\subsection{Properties of the rank sets}\label{rsapp}
In this appendix we show that the rank sets (\ref{V0}), (\ref{V1}), (\ref{V2}) as well as the intersections given in Eq. (\ref{cond3}) can be decomposed into finite union of semianalytic submanifolds. In \cite{mm1} this result followed from the fact that  the sets can be described in terms of semianalytic functions, together with compactness of $\Sigma$. The rank sets here are also described in terms of semianalytic functions (see Eqns. (6.4)-(6.9) of \cite{mm1}). Below we adapt the proof to the present case by dividing $\Sigma$ into an `inside' compact region and an `outside' non-compact region where $E^a$ is of rank 3. 

In the following we work with a given fixed  electric field $E^a$ of the type described in section \ref{sec2}. The fall off conditions (\ref{fallE}) allow us to find  $r_0$ such that $E^a$ is of rank 3 for points outside the 2-sphere $S_{r_o}$ (with respect to the Cartesian chart $\{x^\alpha \}$).  Let us fix $r>r_0$, and denote by $\Sigma_r$ the compact, `inside' region so that in the Cartesian chart $\Sigma \setminus \Sigma_r =\{ \vec{x} \in \reals^3 : (x^1)^2+(x^2)^2+(x^3)^2>r^2 \}$. We can describe $\Sigma_r$ in terms of a semianalytic function $f_r$ such that $\Sigma_r=\{ f_r(x) =0 \}$ and $\Sigma \setminus \Sigma_r = \{  f_r(x) > 0 \}$. We define $f_r$ by  
\be
f_r(x)= \left\{
\begin{array}{cll}
(|\vec{x}|^2-r^2)^m  & \text{for}  &  |\vec{x}|>r  \\
0 &  & \text{elswhere}
\end{array} \right. 
\ee
where the first line refers to the Cartesian chart $\{x^\alpha \}$ with  $|\vec{x}| \equiv  \sqrt{(x^1)^2+(x^2)^2+(x^3)^2}$, and  $m>k$.

It follows that $V^E_2= \{f_r > 0 \} \cup (\{f_r = 0 \} \cap V^E_2  )$ and $V^E_{n}= \{f_r = 0 \} \cap V^E_n $ for $n=0,1$. Thus, all rank sets and their intersections (\ref{cond3}) are of the form $X=\{ f_r=0 \} \cap_{i=1}^n \{ f_i s_i 0 \} $ for some given semianalytic functions $f_i : \Sigma \to \reals, i=1,\ldots n$ and choices of symbols $s_i \in \{ =, <, > \}$.  We now adapt the proof given in appendix B of \cite{mm1}  to show that such $X$ is a finite unions of submanifolds. In the following the index  $\alpha$ stands for $\alpha=r,1,\ldots,n$ so that $X=\cap_\alpha \{f_\alpha  s_\alpha 0 \}$ with $s_r = ``=''$.

 Let $\{ \chi_I, U_I \}$ be a semianalytic atlas of $\Sigma$ compatible with the functions $\{f_\alpha \}$.  That is,  each $U_I$ admits a semianalytic partition compatible with the functions $f_\alpha$. We recall that a  semianalytic partition means a decomposition of the form:
\be
U'_I:=\chi_I(U_I)=\cup_{\sigma^I} V^I_{\sigma^I}, 
\ee
where $\sigma^I:=\sigma^I_1,\ldots,\sigma^I_{n_I}$ is a sequence of $n_I$ symbols $<,>$ or $=$ and   $V^I_{\sigma^I}$  is defined in terms of  analytic functions  $h^I_{i}, i=1,\ldots n_I$, on $U'_I$ by   $V^I_{\sigma^I} := \cap_{i=1}^{n_I} \{h^I_i  \sigma^I_i 0 \}$. The compatibility of the functions $f_\alpha$ with the partition means there exists analytic functions   $\{ f^I_{\alpha \sigma^I} : U'_I \to \reals\}$  such that
\be
f_\alpha \circ \chi^{-1}_I |_{V^I_\sigma}=f^{I}_{\alpha \sigma}|_{V^I_\sigma} .
\ee

 Thus, on each local chart the set of interest  takes the form:
 \be
\chi_I(X \cap U_I)= \cup_{\sigma} \left( \{ \cap_\alpha f^I_{\alpha \sigma} s_\alpha 0 \} \cap\{ h^I_1(x) \sigma_1 0 \} \ldots \cap  \{ h^I_{n_I}(x) \sigma_{n_I} 0 \} \right). \label{zls}
 \ee
The sets featuring in the union (\ref{zls}) can be realized as sets of a new  partition of $U'_I$ defined in terms of the functions $\{ \{h^I_i \} ,\{ f^I_{\alpha \sigma^I} \} \}$. By  proposition A.9 of \cite{lost} it follows that  every $x \in U_I$ has  an open neighborhood $U^x_I$ such that
\be
\chi_I(X \cap U^x_I) =  \text{finite union of analaytic submanifolds}.
\ee
This in turn implies that $X \cap U^x_I$ is a finite union of \emph{semianalytic} submanifolds of $\Sigma$.  Now, the  (uncountable) collection of such open sets $\{U^x_I \}$ cover $\Sigma$ and in particular $\Sigma_r$. Since the latter is compact, it follows there exists a \emph{finite} subfamily  $\{W_\beta \} \subset \{U^x_I \}$ that covers $\Sigma_r$. Thus  $\Sigma_r \subset \cup_{\beta} W_\beta$. Since $X \subset \Sigma_r$ it follows that
\be
X= \cup_{\alpha} (X \cap W_\alpha) = \text{finite union of semianalytic submanifolds}.
\ee


\begin{thebibliography}{99}
\bibitem{alm2t}
  A.~Ashtekar, J.~Lewandowski, D.~Marolf, J.~Mourao and T.~Thiemann,
  ``Quantization of diffeomorphism invariant theories of connections with local degrees of freedom,''
  J.\ Math.\ Phys.\  {\bf 36} (1995) 6456
  [gr-qc/9504018]

\bibitem{leecarlo}
  C.~Rovelli and L.~Smolin,
  ``Spin networks and quantum gravity,''
  Phys.\ Rev.\ D {\bf 52}, 5743 (1995)
  [gr-qc/9505006]

\bibitem{baezspinnet}
  J.~C.~Baez,
  ``Spin network states in gauge theory,''
  Adv.\ Math.\  {\bf 117}, 253 (1996)
  [gr-qc/9411007]

\bibitem{aajurekarea} 
  A.~Ashtekar and J.~Lewandowski,
  ``Quantum theory of geometry. 1: Area operators,''
  Class.\ Quant.\ Grav.\  {\bf 14}, A55 (1997)
  [gr-qc/9602046]


\bibitem{abhayaflat}
A.~Ashtekar,
``New perspectives in canonical gravity'' , Bibliopolis (1988)

\bibitem{kos} 
  T.~A.~Koslowski,
  ``Dynamical Quantum Geometry (DQG Programme),''
  arXiv:0709.3465 [gr-qc].
  

\bibitem{hanno} 
  H.~Sahlmann,
  ``On loop quantum gravity kinematics with non-degenerate spatial background,''
  Class.\ Quant.\ Grav.\  {\bf 27}, 225007 (2010)
  [arXiv:1006.0388 [gr-qc]]

\bibitem{koshanno} 
  T.~Koslowski and H.~Sahlmann,
  ``Loop quantum gravity vacuum with nondegenerate geometry,''
  SIGMA {\bf 8}, 026 (2012)
  [arXiv:1109.4688 [gr-qc]]




\bibitem{mm1}
 M.~Campiglia and M.~Varadarajan,
  ``The Koslowski--Sahlmann representation: gauge and diffeomorphism invariance,''
  Class.\ Quant.\ Grav.\  {\bf 31}, 075002 (2014)
  [arXiv:1311.6117 [gr-qc]]

\bibitem{arnsdorf} 
  M.~Arnsdorf and S.~Gupta,
  ``Loop quantum gravity on noncompact spaces,''
  Nucl.\ Phys.\ B {\bf 577}, 529 (2000)
  [gr-qc/9909053].

\bibitem{itp} 
  T.~Thiemann and O.~Winkler,
  ``Gauge field theory coherent states (GCS) 4: Infinite tensor product and thermodynamical limit,''
  Class.\ Quant.\ Grav.\  {\bf 18}, 4997 (2001)
  [hep-th/0005235]

  \bibitem{me}
  M.~Varadarajan,
  ``The generator of spatial diffeomorphisms in the Koslowski--Sahlmann representation,''
  Class.\ Quant.\ Grav.\  {\bf 30} (2013) 175017
  [arXiv:1306.6126 [gr-qc]]

\bibitem{mm2}
M.~Campiglia and M.~Varadarajan
``The Koslowski-Sahlmann representation:
Quantum Configuration Space''
arXiv:1406.0579 [gr-qc]


\bibitem{fs} 
  ``Spin 1/2 From Gravity,''  J.~L.~Friedman and R.~D.~Sorkin,

  Phys.\ Rev.\ Lett.\  {\bf 44}, 1100 (1980).

\bibitem{arnsdorfspin}
  M.~Arnsdorf and R.~S.~Garcia,
  ``Existence of spinorial states in pure loop quantum gravity,''
  Class.\ Quant.\ Grav.\  {\bf 16}, 3405 (1999)
  [gr-qc/9812006]
\bibitem{rt} 
  T.~Regge and C.~Teitelboim,
  ``Role of Surface Integrals in the Hamiltonian Formulation of General Relativity,''
  Annals Phys.\  {\bf 88}, 286 (1974).

\bibitem{beigom} 
  R.~Beig and N.~\'o Murchadha,
``The Poincare group as the symmetry group of canonical general relativity,''
  Annals Phys.\  {\bf 174}, 463 (1987)  


\bibitem{ttparity} 
  T.~Thiemann,
  ``Generalized boundary conditions for general relativity for the asymptotically flat case in terms of Ashtekar's variables,''
  Class.\ Quant.\ Grav.\  {\bf 12}, 181 (1995)

\bibitem{mcparity}
M.~Campiglia,
``Note on the  phase space of asymptotically flat gravity in Ashtekar-Barbero variables''



\bibitem{ttbook} 
  T.~Thiemann,
  ``Modern canonical quantum general relativity'',
  Cambridge, UK: Cambridge Univ. Pr. (2007) 

\bibitem{acz} 
  A.~Ashtekar, A.~Corichi and J.~A.~Zapata,
  ``Quantum theory of geometry III: Noncommutativity of Riemannian structures,''
  Class.\ Quant.\ Grav.\  {\bf 15}, 2955 (1998)
  [gr-qc/9806041]

\bibitem{lost} J.~Lewandowski, A.~ Oko\l\'ow, H.~Sahlmann, T.~Thiemann, ``Uniqueness of the diffeomorphism invariant state
    on the quantum holonomy-flux algebras'', Comm. Math. Phys. \textbf{267}
    703-733  (2006)

\bibitem{rendall} 
  A.~D.~Rendall,
  ``Comment on a paper of Ashtekar and Isham,''
  Class.\ Quant.\ Grav.\  {\bf 10}, 605 (1993)



\bibitem{aajurekhoop} 
  A.~Ashtekar and J.~Lewandowski,
  ``Representation theory of analytic holonomy C* algebras,''
  gr-qc/9311010.


\bibitem{glimmjaffe}
J.~Glimm and A.~M.~Jaffe,
  ``Quantum Physics. A Functional Integral Point Of View,''
  New York, USA: Springer (1987)

\bibitem{josedon} 
  D.~Marolf and J.~M.~Mourao,
  ``On the support of the Ashtekar-Lewandowski measure,''
  Commun.\ Math.\ Phys.\  {\bf 170}, 583 (1995)
  [hep-th/9403112]

\bibitem{velhinho} 
  J.~M.~Velhinho,
  ``On the structure of the space of generalized connections,''
  Int.\ J.\ Geom.\ Meth.\ Mod.\ Phys.\  {\bf 1}, 311 (2004)
  [math-ph/0402060]

  \bibitem{aajurekproj} 
  A.~Ashtekar and J.~Lewandowski,
  ``Projective techniques and functional integration for gauge theories,''
  J.\ Math.\ Phys.\  {\bf 36}, 2170 (1995)
  [gr-qc/9411046]





  \bibitem{alrev} 
  A.~Ashtekar and J.~Lewandowski,
  ``Background independent quantum gravity: A Status report,''
  Class.\ Quant.\ Grav.\  {\bf 21}, R53 (2004)  

\bibitem{kp2} 
S. G. Krantz  and H. R. Parks, 
 ``The Implicit Function Theorem: History, Theory, and Applications'',
Birkh{\"a}user (2012)



  \bibitem{lang} 
S. Lang,  ``Introduction to Differentiable Manifolds'', Springer (2002)



\bibitem{leepositive} 
  L.~Smolin,
  ``Positive energy in quantum gravity,''
  arXiv:1406.2611 [gr-qc]


\bibitem{ttqsd} 
  T.~Thiemann,
  ``Quantum spin dynamics (QSD),''
  Class.\ Quant.\ Grav.\  {\bf 15}, 839 (1998)
  [gr-qc/9606089]

\bibitem{aloklqg} 
  A.~Laddha,
  ``Hamiltonian constraint in Euclidean LQG revisited: First hints of off-shell Closure,''
  arXiv:1401.0931 [gr-qc]

\bibitem{aloku11} 
  A.~Henderson, A.~Laddha and C.~Tomlin,
  ``Constraint algebra in loop quantum gravity reloaded. I. Toy model of a $U(1)^3$ gauge theory,''
  Phys.\ Rev.\ D {\bf 88}, no. 4, 044028 (2013)
  [arXiv:1204.0211 [gr-qc]]
\bibitem{aloku12} 
  A.~Henderson, A.~Laddha and C.~Tomlin,
  ``Constraint algebra in loop quantum gravity reloaded. II. Toy model of an Abelian gauge theory: Spatial diffeomorphisms,''
  Phys.\ Rev.\ D {\bf 88}, no. 4, 044029 (2013)
  [arXiv:1210.3960 [gr-qc]]
\bibitem{meu11} 
  C.~Tomlin and M.~Varadarajan,
  ``Towards an Anomaly-Free Quantum Dynamics for a Weak Coupling Limit of Euclidean Gravity,''
  Phys.\ Rev.\ D {\bf 87}, no. 4, 044039 (2013)
  [arXiv:1210.6869 [gr-qc]]

\bibitem{meu12}
  M.~Varadarajan,
  ``Towards an Anomaly-Free Quantum Dynamics for a Weak Coupling Limit of Euclidean Gravity: Diffeomorphism Covariance,''
  Phys.\ Rev.\ D {\bf 87} (2013) 4,  044040
  [arXiv:1210.6877 [gr-qc]]

\bibitem{sandipan1} 
  S.~Sengupta,
  ``Quantum geometry with a nondegenerate vacuum: a toy model,''
  Phys.\ Rev.\ D {\bf 88}, 064016 (2013)
  [arXiv:1306.6013 [gr-qc]]

\bibitem{sandipan2} 
 S.~Sengupta,
  ``Asymptotic Flatness and Quantum Geometry,''
  Class.\ Quant.\ Grav.\  {\bf 31}, 085005 (2014)
  [arXiv:1309.5266 [gr-qc]

\bibitem{menewrep} 
  M.~Varadarajan,
  ``Towards new background independent representations for loop quantum gravity,''
  Class.\ Quant.\ Grav.\  {\bf 25}, 105011 (2008)
  [arXiv:0709.1680 [gr-qc]]

\bibitem{grosser}
M. Grosser,  M. Kunzinger, M. Oberguggenberger, R. Steinbauer,
``Geometric Theory of Generalized Functions with Applications to General Relativity'', Springer (2001)



\bibitem{aalog}
A.~Ashtekar,  ``Logarithmic ambiguities in the description of spatial infinity'',
 Foundations of physics 15 no. 4, 419-431 (1985)

\end{thebibliography}
\end{document}